\documentclass[apj]{emulateapj-rtx4}
\usepackage{natbib}

\newcommand  \sii   {[\ion{S}{2}]}

\newcommand  \ha    {H$\alpha$} 

\newcommand  \hii   {\ion{H}{2}} 
\newcommand  \hi    {\ion{H}{1}}
\newcommand  \nii   {[\ion{N}{2}]}
\newcommand  \um    {$\mu$m}
\newcommand  \ysoai {J014353.94$-$743224.71}
\newcommand  \ysoafi {J014349.20$-$743200.63}
\newcommand  \ysobi {J014942.43$-$743704.76}
\newcommand  \ysobii {J014929.21$-$743916.48}
\newcommand  \aebebi {J014918.59$-$743753.93}
\newcommand  \ysocfi {J015635.59$-$741701.10}
\newcommand  \ysocfii {J015654.01$-$741526.13}
\newcommand  \ysoei {J020649.68$-$744359.95}
\newcommand  \ysoefi {J020622.18$-$744254.43}
\newcommand  \ysoefii {J020631.11$-$744247.84}
\newcommand  \ysofi {J021440.18$-$742127.03}
\newcommand  \ysoffi {J021434.94$-$742339.94}
\newcommand  \ysogi {J021526.72$-$740432.77}
\newcommand  \ysogfi {J021525.25$-$740600.82}
\newcommand  \ysogali {J022152.32$-$744537.83}
\newcommand  \ysogalii {J022152.99$-$744534.94}

\shorttitle{Young Massive Stars in the Magellanic Bridge}
\shortauthors{Chen et al.}

\begin{document}

\title{{\it Spitzer} View of Massive Star Formation in the Tidally Stripped Magellanic Bridge}

\author{C.-H.~Rosie~Chen\altaffilmark{1,2}, 
 Remy~Indebetouw\altaffilmark{1,3},
 Erik~Muller\altaffilmark{4,5}, 
 Akiko~Kawamura\altaffilmark{4,5},
 Karl~D.~Gordon\altaffilmark{6},
 Marta~Sewi{\l}o\altaffilmark{7},
 Barbara~A.~Whitney\altaffilmark{8},
 Yasuo~Fukui\altaffilmark{5}, 
   Suzanne~C.~Madden\altaffilmark{9},
   Marilyn~R.~Meade\altaffilmark{8},
   Margaret~Meixner\altaffilmark{6},
   Joana~M.~Oliveira\altaffilmark{10},
   Thomas~P.~Robitaille\altaffilmark{11},
   Jonathan~P.~Seale \altaffilmark{6,7},
   Bernie~Shiao\altaffilmark{6},
 and 
   Jacco~Th. van Loon\altaffilmark{10}
}
\email{rchen@mpifr-bonn.mpg.de}
\altaffiltext{1}{Department of Astronomy, University of Virginia, Charlottesville, VA 22904}
\altaffiltext{2}{Current address: Max Planck Institute for Radio Astronomy, D-53121 Bonn, Germany}
\altaffiltext{3}{National Radio Astronomy Observatory, Charlottesville, VA 22903}
\altaffiltext{4}{National Astronomical Observatory of Japan, Mitaka, Tokyo 181-8588, Japan}
\altaffiltext{5}{Department of Astrophysics, Nagoya University, Furocho, Chikusaku, Nagoya 464-8602, Japan}
\altaffiltext{6}{Space Telescope Science Institute, Baltimore, MD 21218}
\altaffiltext{7}{Department of Physics and Astronomy, Johns Hopkins
  University, Baltimore, MD 21218}
\altaffiltext{8}{Department of Astronomy, University of Wisconsin-Madison, Madison, WI 53706}
\altaffiltext{9}{CEA, Laboratoire AIM, Irfu/SAp, Orme des Merisiers, F-91191 Gif-sur-Yvette, France}
\altaffiltext{10}{Astrophysics Group, Lennard-Jones Laboratories, Keele University, ST5 5BG, UK}
\altaffiltext{11}{Max Plan ck Initutute for Astronomy, D-69117 Heidelberg, Germany}

\begin{abstract}

The Magellanic Bridge is the nearest low-metallicity, tidally stripped 
environment, offering a unique high-resolution view of physical 
conditions in merging and forming galaxies.
In this paper we present analysis of candidate massive young stellar 
objects (YSOs), i.e., {\it in situ, current} massive star formation (MSF) 
in the Bridge using {\it Spitzer} mid-IR and complementary optical and 
near-IR photometry.
While we definitely find YSOs in the Bridge, the most massive are 
$\sim10 M_\odot$, $\ll45 M_\odot$ found in the Large Magellanic Cloud (LMC).
The intensity of MSF in the Bridge also appears decreasing, as the most 
massive YSOs are less massive than those formed in the past. 
To investigate environmental effects on MSF, we have compared properties 
of massive YSOs in the Bridge to those in the LMC.
First, YSOs in the Bridge are apparently less embedded than in the LMC:
81\% of Bridge YSOs show optical counterparts, compared to only 56\% of 
LMC sources with the same range of mass, circumstellar dust mass, 
and line-of-sight extinction.
Circumstellar envelopes are evidently more porous or clumpy in the Bridge's 
low-metallicity environment.
Second, we have used whole samples of YSOs in the LMC and the Bridge to 
estimate the probability of finding YSOs at a given \hi\ column density, N(HI).  
We found that the LMC has $\sim3\times$ higher probability than the Bridge 
for N(HI) $>10\times10^{20}$~cm$^{-2}$, but the trend reverses at 
lower N(HI).  
Investigating whether this lower efficiency relative to HI is due
to less efficient molecular cloud formation, or less efficient cloud
collapse, or both, will require sensitive molecular gas
observations.

\end{abstract}


\section{Introduction}
 
The dependence of star formation on environment is fundamental in 
both the nearby and distant universe.
Star formation is often enhanced in galaxies undergoing interaction 
or merger \citep{LT78}.
This enhancement can produce global starbursts such as Arp 220,
though more frequently observed are local concentrations of star 
forming regions \citep{BNetal03}.
Although the overall star formation rate (SFR) can be enhanced, 
it is not clear how such physical conditions affect the star or cluster 
mass distribution.
The higher pressure and density environment of mergers might 
encourage the preferential formation of {\it massive} stars and 
clusters \citep[e.g.,][]{EE97}.
On the other hand, the increased turbulence in the interstellar medium
(ISM)  can result in larger gas dispersions, hampering the formation
of giant molecular clouds and hence the massive clusters formed within 
\citep[e.g.,][]{SC97}.
As massive stars and clusters are the energy source of the ISM and 
in turn affect the evolution of their host galaxies, it is important to 
understand their formation in a variety of environments that are 
different from our Galaxy.

In addition to the influence of dynamical interactions, the formation 
of massive stars may depend on metallicity.
A lower dust abundance and greater permittivity to ultraviolet 
radiation of the ISM is expected to affect pre-formation gas dynamics,
as well as cooling and feedback from massive young stellar objects
(YSOs) \citep{Poetal95}.
It is thus critical to understand massive star formation in a 
low-metallicity, dynamically disturbed environment, to 
interpret similar situations in the early  universe.
Furthermore, to understand the detailed physics of the process;
the geometric, morphological and temporal relationships of the molecular 
and dust components with the forming stellar population,
we must spatially resolve the relevant structures, i.e., molecular 
clouds and individual young massive stars.

Located between the Large and Small Magellanic Clouds (LMC and SMC)
at a distance of $\sim$ 50--60 kpc \citep[e.g.,][]{HHH03,TVetal10}, 
the Magellanic Bridge (hereafter the Bridge) is the closest tidal 
system, and one of the few where clusters and interstellar structures can be resolved and 
studied in detail.
The Bridge was first identified in an \hi\ survey \citep{Hietal63}
and its production has been suggested to be the result of a recent close encounter
between the LMC and SMC $\sim$ 200 Myr ago \citep[e.g.,][see also 
\citealt{BGetal10}]{GSF94}.
The Bridge's tidal environment together with its significantly low metallicity 
$\sim$ 1/5--1/8 $Z_\odot$ \citep{RWetal99,LJetal05} provide an excellent 
laboratory to study massive star formation under such physical conditions. 
Furthermore, the Bridge's high \hi\ mass $\sim 1.5\times10^8 M_\odot$ 
and substantial \hi\ column density (N(\hi )) up to $\sim 3\times10^{21}$ 
cm$^{-2}$ \citep{Muetal03b} make it a promising site to search for 
newly formed massive stars.
This high \hi\ mass also qualifies the Bridge as a potential region
to develop into a dwarf galaxy \citep[\hi\ mass ranging from 
$10^6-10^9 M_\odot$,][]{MF99}, providing insight into their  
development and evolution as well.

Several studies have found evidence of stars in the Bridge 
less than $\sim$100 Myr old, which if the 200--300 Myr formation timescale
is correct must have formed {\em in situ} in tidal gas.
\citet{HJ07} analyzed optical color-magnitude diagrams to derive
a star formation history beginning $\sim 200-300$ Myr ago, with two distinct 
episodes $\sim$ 160 and 40 Myr ago.
Studies specifically targeting massive stars provide 
more compelling evidence of {\em in situ} star formation.
The handful 
of large \ha\ shells and several small \hii\ regions \citep{MJ86,MP07} 
in the Bridge are most likely formed by massive stars.
One of the large \ha\ shells, DEM\,S\,171, has been suggested to be 
ionized by one or more O-type stars or blown by a supernova explosion 
\citep{MJ86,Gretal01}, though for other \ha\ shells and regions the 
underlying stellar population is not known.
A population of blue stars are also identified in the Bridge using 
broadband $BV$ or $BVR$ photometry, with estimated ages ranging from 
as young as $\sim$ 10--25 Myr to $\sim 100$ Myr \citep{Iretal90,DI91,DB98}.
Nevertheless, sufficient uncertainty in the timescales, and no clear
association with natal gaseous material, leave open the possibility 
that these massive stars actually formed in the SMC body and were 
stripped out along with the gas.  
Pre-main-sequence stars identified in the Bridge would require an order 
of magnitude shorter timescale and it would be very hard to argue 
against {\em in situ} formation.
A recent near-infrared (IR) $JHK_s$ survey of the Bridge finds 
Herbig Ae/Be (HAeBe) candidates with ages possibly down to $\sim 2$ 
Myr, but this color-selected candidate list is of moderate reliability 
\citep[only $\sim 40$\%  likely bona-fide HAeBe,][]{Nietal07} 
and requires spectroscopic confirmation.

Recent {\it Spitzer Space Telescope} imaging observations in the mid-IR 
have enabled the detection of individual massive YSOs in the LMC and SMC 
\citep[e.g.,][]{CYetal05,WBetal08,GC09,BAetal07,Sewiloetal13}.
Follow-up {\it Spitzer} spectroscopic observations further confirms 
a $> 95$\% reliability rate in identifying massive YSOs in the LMC using
our method based on examination on multi-wavelength spectral energy 
distributions (SEDs) and images \citep{CCetal09,GC09,Seetal09}.
In this paper we present a similar inventory of massive YSOs in the Bridge,
compare their properties and distribution to the molecular clouds,
and probe a causal relationship between the initial condition 
(gas) and the end product (stars) in the most direct way.
With the knowledge of current and recent stellar content and 
expected stellar energy feedback, it is then possible to assess
if the star formation is triggered and its relative strength
to that from spontaneous processes, and further estimate how the 
star formation efficiency (SFE) of a molecular cloud varies with time.
Comparisons between the SFEs of clouds in the Bridge to 
those in a variety of metallicities and galactic environments
such as the Galaxy or the LMC
then allow us to probe the effect of environment on massive star 
formation.

\begin{figure}
\epsscale{0.8}
\plotone{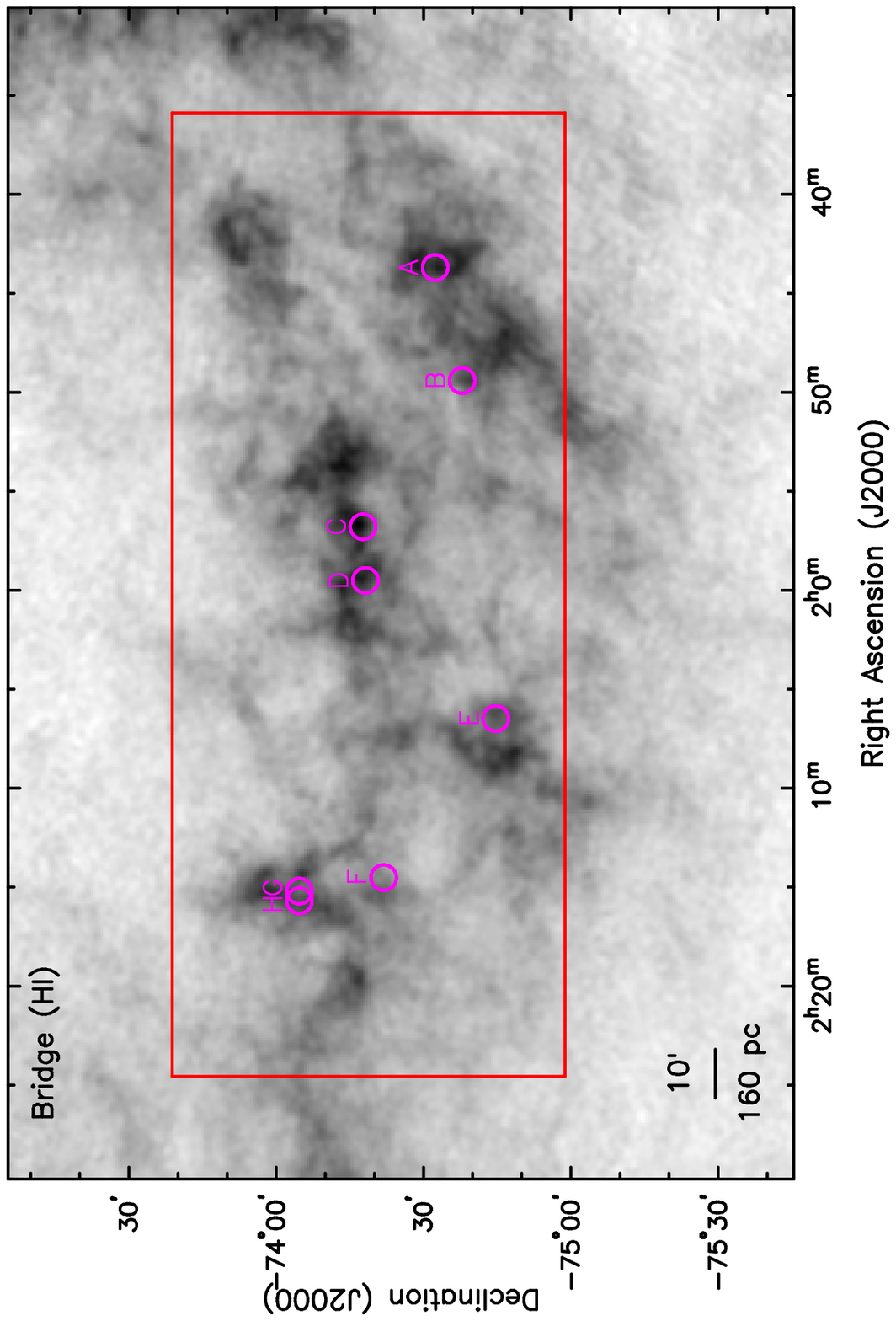}
\caption{N(\hi ) map of the Bridge \citep{Muetal03b} overlaid with positions
  of Molecular Clouds A--H (magenta circles) cataloged in \citet{MNetal06}.   The box outlines 
  the high N(\hi ) area that is fully mapped in the {\it Spitzer} bands from 3.6 to 160 
  \um .     
 \label{fig:hiimg}}
\end{figure}

As part of the the {\it Spitzer} survey of the SMC 
\citep[SAGE-SMC,][]{GKetal11}, the high N(\hi )  
portion of the Bridge (where N(\hi ) = 2--27$\times10^{20}$ 
cm$^{-2}$  with an average  = 10$\times10^{20}$ cm$^{-2}$) 
has been mapped fully in the {\it Spitzer} bands from 3.6 to 
160 \um\ (Figure~\ref{fig:hiimg})\footnote{The region 
of our study is located in the western part of the continuous stellar 
bridge between the LMC and SMC \citep{Iretal90}.   It also appears
to be extending from the SMC Body and Wing and hence has been 
referred as ``the SMC Tail'' \citep[e.g.,][]{GKetal09}.
For simplicity we call this region the Bridge.}.
Molecular clouds have been detected via CO J=1-0 emission 
\citep{Muetal03a,MNetal06}, providing an excellent opportunity 
to investigate if the formation of massive YSOs depends on physical 
conditions of the clouds.
To study the current massive star formation in the Bridge,
we have used {\it Spitzer} mid-IR observations and archival catalogs 
and data in the optical and near-IR wavelengths.
The paper is organized as follows: the observations and data 
reduction are described in Section 2, the identification of YSO
and HAeBe candidates is reported in Section 3, the derivation 
of physical properties of these candidates is detailed in Section 4, 
the properties of massive star formation in the Bridge is discussed
in Section 5, and a summary is given in Section~6.

\section{Observations and Data Reduction}

YSOs are primarily identified by their IR excess, requiring analysis of 
multiple colors, or equivalently the spectral energy distribution (SED) 
over as wide a wavelength range as possible from optical to IR.  
Our primary datasets are {\it Spitzer} mid-IR imaging, but to extend 
the wavelength coverage and improve angular resolution, we include data 
from several archival ground-based optical and near-IR surveys.

\subsection{{\it Spitzer} IRAC and MIPS Observations}

\begin{figure}
\epsscale{1.0}
\plotone{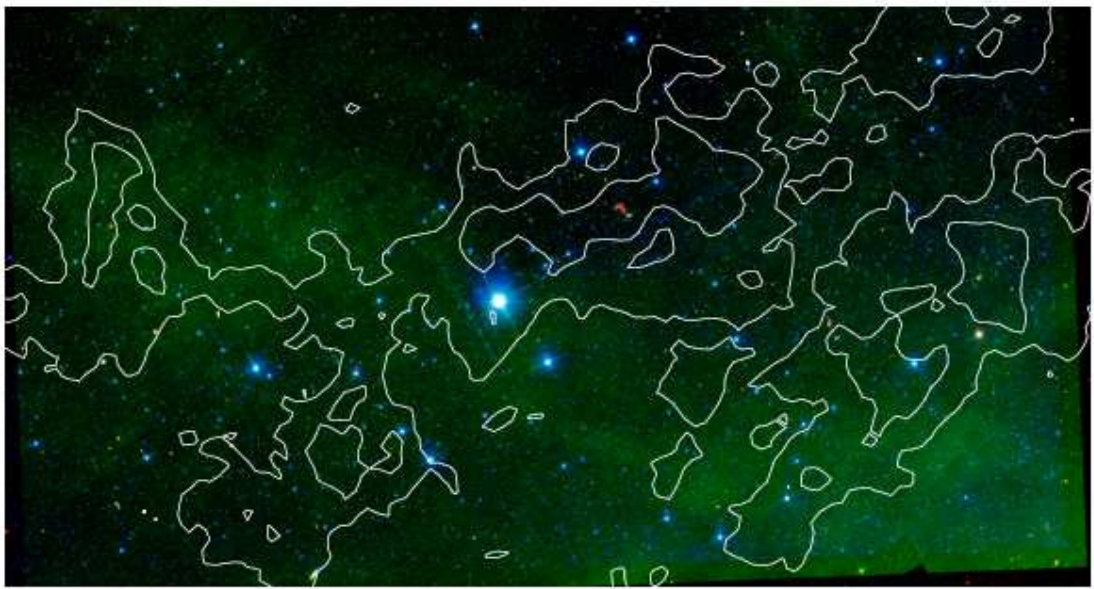}
\caption{
 Color composite of the Bridge with {\it Spitzer} 3.6, 8.0, and 24 \um\ images mapped in 
 blue, green, and red, respectively.  Contours show N(\hi ) = 11 and 18 $\times 10^{20}$~cm$^{-2}$ 
 (40 and 67\% of the peak value, respectively) in white lines.
 \label{fig:mbimg}}
\end{figure}

The {\it Spitzer} observations of the Bridge were obtained as part 
of the Legacy Program ``Surveying the Agents of Galaxy Evolution in 
the Tidally-Stripped, Low-Metallicity Small Magellanic Cloud''
\citep[SAGE-SMC;][]{GKetal11}.
These observations included images taken at 3.6, 4.5, 5.8, and 
8.0 \um\ bands with the InfraRed Array 
Camera (IRAC) and at 24, 70, and 160 \um\ bands with
the Multiband Imaging Photometer for {\it Spitzer} (MIPS).
The details of data processing are given in \citet{GKetal11}. 
The final mosaics have exposure times 48s per pixel at each of the four 
IRAC bands, and 60, 30, and $\sim 9$s per pixel at MIPS 24, 70, and 160 
\um\ bands, respectively.
The $180' \times 80'$ area of the Bridge covered by all IRAC and MIPS 
bands was analyzed here; it includes all molecular clouds detected
in the Bridge. 
Figure~\ref{fig:mbimg} shows the color composite of the analyzed 
field made with images in the 3.6, 8.0, and 24 \um\ bands, to 
demonstrate the distribution of different emission.
The stellar emission is depicted in the 3.6 \um\ image,
the polycyclic aromatic hydrocarbon (PAH) emission is traced in 
the 8.0 \um\ image, and the dust continuum emission dominates 
the 24 \um\ image 
\citep{LD01,DL07}.

The IRAC and MIPS photometry of point sources in the Bridge
are available from SAGE-SMC point source catalogs.
These catalogs are intended for photometric measurements with 
consistent quality for point sources, at the expense of excluding 
sources that are not well fitted by the point-spread-functions (PSFs).
With IRAC's resolution of $\sim 2''$, or $\sim$ 0.6 pc at the 
Bridge's distance, massive YSOs that have formed compact \hii\ 
regions or are superimposed on large-scale, diffuse dust features 
associated with \hii\ complexes can appear slightly extended 
or irregular compared to the IRAC PSF and be excluded from the point source
catalog.
A more complete list can be created by relaxing the point source 
criteria, but this comes at the expense of significant numbers 
of unreliable sources (e.g., knots of structured diffuse emission) that 
must be culled by manual inspection.   
The process that we used to produce a more complete list 
is described in detail in \citet{CCetal09,CCetal10}, and outlined below.  
Candidate point sources are found with \texttt{daofind} 
\citep{ST87} using relatively inclusive point source criteria.
The fluxes of these sources were then measured using the IRAF aperture 
photometry package \texttt{apphot} with a source aperture of 3\farcs6 
(3 pixels) radius and an annular background aperture extending across 
radii of 3\farcs6--8\farcs4 (3-7 pixels), and applied an aperture correction  
provided in the IRAC Instrument Handbook\footnote{\url{http://irsa.ipac.caltech.edu/data/SPITZER/docs/\\
irac/iracinstrumenthandbook/}}.
This IRAC catalog is then merged with the SAGE-SMC catalog of MIPS 
point sources allowing for 1$''$ positional differences.  All of our sources 
should be unresolved at the poorer spatial resolution of MIPS than IRAC, 
and hence present in the MIPS point source catalog if detected.

\subsection{Additional Data Sets}

To construct multi-wavelength SEDs for sources in the {\it Spitzer} 
catalog, we have expanded it by adding photometry from optical and 
near-IR surveys covering the Bridge, i.e., $BRI$ photometry from the 
Super COSMOS Sky Surveys \citep[SSS;][]{HNetal01a} and $JHK_s$ 
photometry from the Two Micron All Sky Survey \citep[2MASS;][]{SMetal06}.
As the photometric limit of the 2MASS catalog is relatively shallow
($K_s \sim 14.5$) and the long exposure 2MASS catalog, 2MASS 6x, 
only covers about one third of the area we analyze, the deeper 
(by $\sim 2$ mag) point source catalog from the InfraRed Survey 
Facility \citep[IRSF;][]{Kaetal07} is also used to match those 
IRAC sources without 2MASS counterparts.
The data sets are merged by allowing a 1$''$ position error for 
matching {\it Spitzer} sources with optical or near-IR sources. 

To examine the large-scale distribution of gas in the Bridge, 
we have used the \ha\ images from the Super COSMOS \ha\ survey 
\citep{Parkeretal05} 
to examine dense ionized gas and the \hi\ map from the Australia 
Telescope Compact Array (ATCA) and Parkes surveys \citep{Muetal03b}
to examine the neutral gas.

\section{Identification of Massive YSOs}

\subsection{Selection of Massive YSO Candidates}

Owing to the presence of circumstellar dust and hence the excess IR 
emission, YSOs are positioned in redder parts of color-magnitude 
diagrams (CMDs) than normal stars without circumstellar dust, 
such as main-sequence stars.
However, redder sources include not only YSOs, but also background 
galaxies and evolved stars such as asymptotic giant branch (AGB) 
or post-AGB stars, and these contaminants exist in non-negligible numbers.
Figure~\ref{fig:cmds} displays the [8.0] versus ([4.5]$-$[8.0]) CMD 
of all sources detected in the Bridge.
The prominent vertical branch centered at 
$([4.5]-[8.0]) \sim 0.0$ is composed mostly of main-sequence, giant, and 
supergiant stars.
Also plotted in Figure~\ref{fig:cmds} are expected loci from models for 
Galactic C- and O-rich AGB stars \citep{Gr06}.
These loci are only from models for a stellar luminosity of 3000 $L_\odot$
and can shift vertically from 1.2 to $-3.3$ mag for the luminosity 
range $1\times10^3
-6\times10^4 L_\odot$ reported for AGB stars \citep{Po93}.

The initial selection of massive YSO candidates was done using 
two color-magnitude criteria $([4.5]-[8.0]) \ge 2.0$ and $[8.0] < 
14 - ([4.5]-[8.0])$, which have been demonstrated in 
\citet{CCetal09} and \citet{GC09} as effective criteria to exclude 
contaminants such as galaxies and evolved stars.
Applying these two criteria to our catalog produces 60 YSO candidates 
in the Bridge. 
As a comparison, applying the same criteria to the SAGE-SMC catalog 
(that have stringent criteria on selecting point sources) produces 
25 YSO candidates.
Note that the initial list of YSO candidates includes a non-negligible 
fraction of small dust features, obscured evolved stars, and bright 
background galaxies that need to be excluded.
This requires examining all candidates closely to assess their nature.
Following the same procedure outlined in \citet{CCetal09}, we examine 
SSS $BRI$, 2MASS $JHK_s$, and \ha\ images to better resolve these IR 
sources and their environments.
We also use multi-wavelength SEDs from $B$ band to 70 \um\ constructed 
from the catalog described in Section 2.2.
In the next section we discuss how we use these images and SEDs to 
assess whether these candidates are truly YSOs.

\subsubsection{Identification of Contaminants}
\label{contaminants}

\begin{figure}
\epsscale{0.8}
\plotone{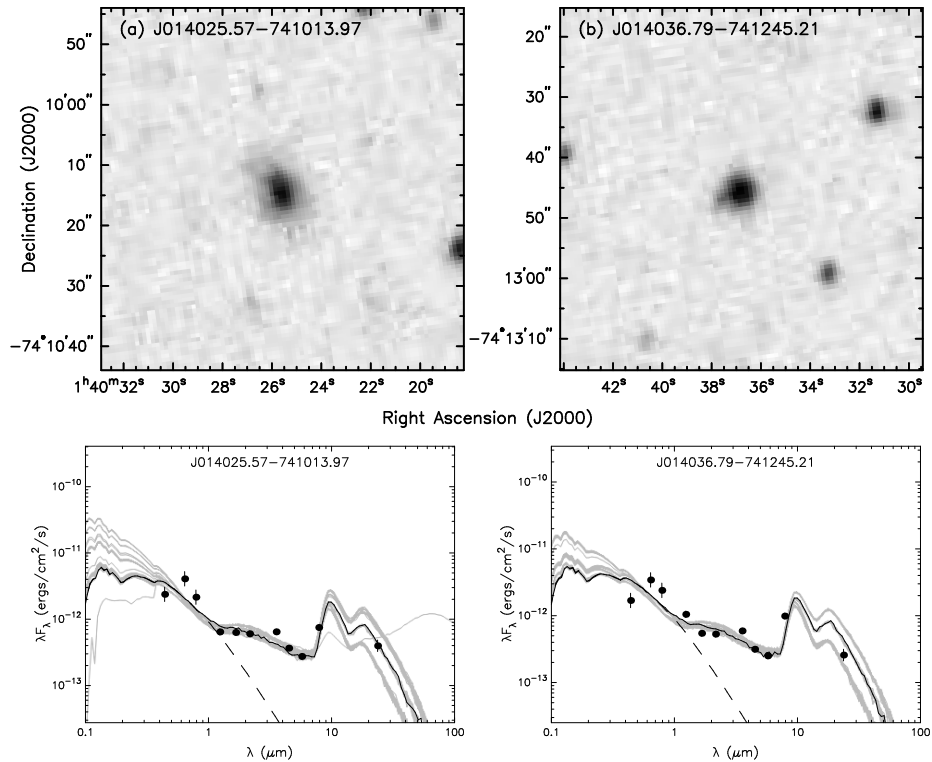}
\caption{SSS $B$ images and SEDs of two CMD-selected YSO candidates that
 are most likely galaxies.  The observed SEDs of these two sources 
 (filled circles), are not well fitted by YSO models (black and grey lines).
 Instead, they have SEDs similar to galaxies and are resolved into a 
 spiral galaxy (a) or appear extended (b).
 \label{fig:gal}}
\end{figure}

Background galaxies, if resolved, can be identified from their 
morphologies.
17 of the CMD-selected YSO candidates are resolved into galaxies
in high-resolution optical $BRI$ and near-IR $JHK_s$ images, and another 
17 candidates show elongated emission extended beyond the point 
sources, most likely galaxies more distant than the resolved ones. 
Figure~\ref{fig:gal} shows $B$-band images and SEDs of two examples from 
sources in these two categories, a resolved galaxy and a galaxy candidate 
with extended morphology.
The SEDs of the first group of resolved sources with galaxy-like morphology
are not well reproduced by YSO models 
but resemble late-type galaxies, i.e., characterized by two 
broad humps with one over optical and near-IR range coming 
from stellar emission and the other over mid- to far-IR range coming from dust 
emission.
The second group of 17 sources have similarly double-peaked SEDs, poorly fit 
by any of our YSO models.  Although we do not have the definitive 
evidence of spatially-resolved imaging, and these 
sources could also be main-sequence stars with nearby dust, or more evolved 
(Class III) YSOs, we classify these additional sources, for a total of 34, 
as background galaxies. \label{galaxysed}

Unlike YSOs found in the LMC \citep[e.g.,][]{CCetal09,GC09}, 
background galaxies appear to be the main and only
contaminants in the list of CMD-selected YSO candidates.
In our LMC studies such lists include a significant number of 
small dust clumps, local peaks of large-scale dust filaments, 
and evolved stars.
The Bridge does not have bright \hii\ complexes nor the associated
photo-dissociation regions (PDRs), and hence little large-scale
diffuse dust emission where dust clumps and peaks are frequently
found. 
The Bridge also has a relatively young stellar population of $\le$
200-300 Myr \citep[][see also \citealt{Noeletal13} and 
\citealt{Bagherietal13} reporting older stars possibly stripped off 
from the SMC]{HJ07} and it is thus reasonable that it has few extreme 
evolved stars (redder than our color selection $([4.5]-[8.0]) \ge 2.0$).

\subsubsection{Massive YSOs and Their Classification}

The results of our examination of 60 YSO candidates are given in 
Table~\ref{ysoclass}, which includes source name, ranking of the 
brightness at 8 \um , magnitudes in the {\it Spitzer} bands from
3.6 to 70 \um , source classification, and remarks.
Photometry taken from available optical and near-IR catalogs 
in $BRIJHK_s$ bands are listed in Table~\ref{ysoclassb}.
The identification of YSOs and non-YSOs is shown in the [8.0] versus 
([4.5]$-$[8.0]) CMD in Figure~\ref{fig:cmds}.
As aforementioned, among the list of 60 YSO candidates, 34 background 
galaxies are identified.
After excluding these sources, 26 YSO candidates remain.
Since these are most likely bona fide YSOs, we will simply call them 
YSOs in the rest of the paper.
Among these 26 YSOs, 20 are from the SAGE-SMC point source catalog
and the other 6 are from our IRAC point source catalog produced 
using more relaxed criteria of the PSFs.

\begin{figure}
\epsscale{0.8}
\plotone{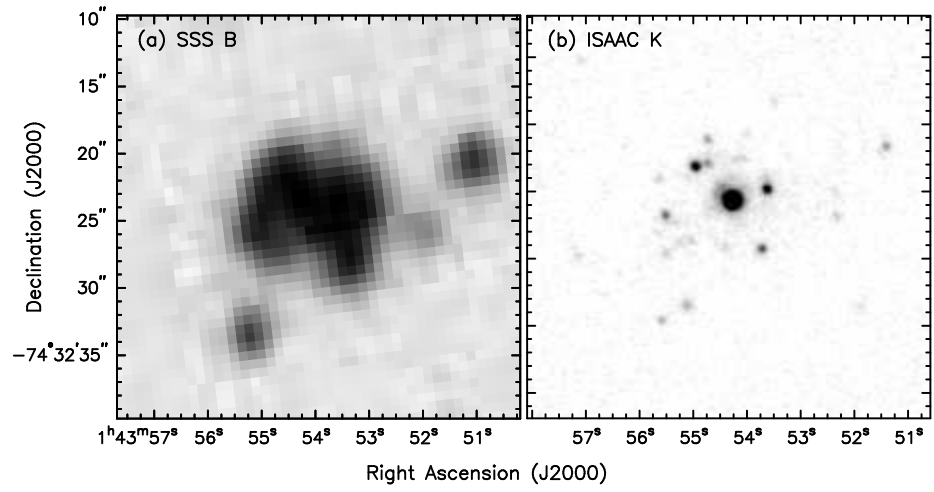}
\caption{High resolution SSS $B$ and ISAAC $K$ images of YSO \ysoai .
 This YSO is resolved into a multiple system. 
 \label{fig:ysoa}}
\end{figure}

We note that two of the YSOs, \ysoai\ and \ysogali , were 
previously identified as galaxies LEDA 2816287 and LEDA 248457 respectively
based on their {\it IRAS} colors. 
Their radial velocities, 2,199 and 26,087 km~s$^{-1}$ respectively, 
were estimated from \ha ,  \nii , and \sii\ lines in 
low-spatial and -spectral resolution spectra taken as part of the PSCz 
redshift survey of {\it IRAS} galaxies \citep{SWetal00}. 
These velocity estimates are less reliable since these two 
are not single but multiple sources.
Source \ysoai\ is resolved into at least 5 point-like sources of 
different ($B-R$) colors within a 4\farcs5-radius circle in
high-resolution SSS $B$ and $R$ images at a pixel scale of  
0\farcs67~pixel$^{-1}$.
These 5 sources have the clear morphology of a stellar cluster 
surrounded by fainter stars
in the archival $JHK_s$ images taken with the ISAAC telescope in 
the European Southern Observatory (ESO) at a superb pixel scale of 
0\farcs15~pixel$^{-1}$ (Figure~\ref{fig:ysoa}).
In addition, our {\it Spitzer} IRS spectrum of this source shows typical 
PAH emission features of a YSO.
The other source \ysogali\ is $\le$ 2\farcs0 from another YSO 
\ysogalii\ as revealed in all images except at the {\it Spitzer} MIPS'
poorer resolution.
These two neighboring sources have SEDs that are fitted relatively 
well by YSO models (Figure~\ref{fig:fit}) but quite different from 
galaxies discussed in Section \ref{galaxysed}.
Given the above considerations, we thus reclassify sources 
\ysoai\ and \ysogali\ as YSOs, not background galaxies.

Using the classification scheme proposed in \citet{CCetal09}, 
we have further categorized the 26 YSOs in the 
Bridge into Types I, II, and III based on their SEDs and 
surrounding interstellar environment that are expected as
a result of evolution of massive YSOs.
Type I YSOs have large circumstellar envelopes that dominate
the radiation; thus, their SEDs show a steep rise from the near-IR 
to 24 \um\ and beyond. 
They are generally not visible at wavelengths shorter than $K_s$ 
band but brightens up toward longer wavelengths; they are often
found in dark clouds.
Type II YSOs show their stellar cores and circumstellar disks 
after the envelopes have dissipated; thus, their SEDs exhibit 
a low peak in the optical and a higher peak at 8--24 \um .
They are faint in the optical but bright in the near- to mid-IR 
up to 8 \um , and then fading at 24 \um .
Type III YSOs have largely exposed their stellar cores and possessed
only remnant circumstellar material; thus, their SEDs peak in the
optical and show only modest dust emission in the near- to mid-IR.
They are bright in the optical and fading at longer wavelengths;
they are often associated with small \hii\ regions. 

This ''Type'' classification is straightforward for YSOs that are 
unresolved by IRAC,
but more complicated for those in multiple systems or complex surroundings.
Nine of the 26 YSOs are resolved by the SSS and 2MASS images
into multiple sources within the IRAC PSF (e.g. the aforementioned 
YSO \ysoai), and another 5 appear more extended than the PSF of SSS 
and 2MASS that need higher-resolution images such as the on-going
VISTA survey of the Magellanic Clouds System \citep[VMC;][]{Cionietal11}
to identify multiple components.
YSOs are often found in dark clouds, dust columns, or \hii\ regions.
These interstellar features can be identified in high-resolution optical images, 
but the mid-IR emission from these dust features 
can be blended with that of the YSOs in the {\em Spitzer} images, 
especially in the 24 \um\ band 
at a resolution $\sim 6''$, increasing the uncertainty of the YSO's 
classification.

Our classification of the 26 YSOs and remarks on the multiplicity
and association with dark cloud, dust column, and \hii\ region are
given in Table~\ref{ysoclass}.
For YSOs in multiple systems or complex interstellar surroundings,
the classification has larger uncertainties.

\subsection{Selection of Fainter YSO Candidates}
\label{fainteryso}

\begin{figure}
\epsscale{0.8}
\plotone{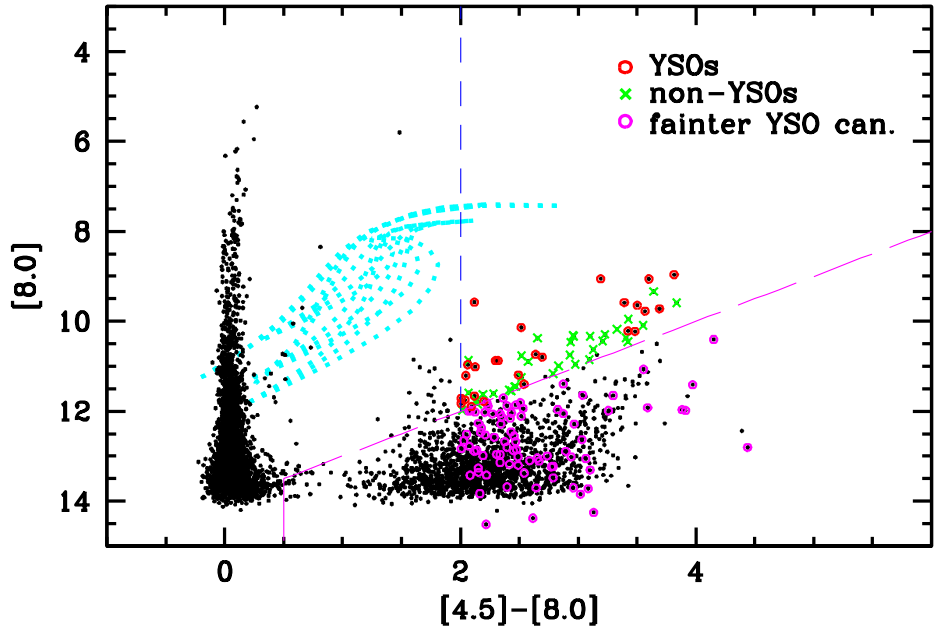}
\caption{
 [8.0] vs.\ [4.5]$-$[8.0]
 CMD of all sources detected in the Bridge.  
 Expected loci from AGB stellar models \citep{Gr06} are marked with filled 
 cyan squares; only models for a stellar luminosity of 3000 $L_\odot$ are 
 plotted and hence these loci can shift vertically from 1.2 to $-$3.3 
 mag for the luminosity range of AGB stars.  
 The criterion to exclude normal
 and AGB stars is shown in short-dashed lines and that to exclude background
 galaxies in long-dashed lines. 
 Sixty YSO candidates are found in the upper
 right wedge that has the minimum contamination from stars and background
 galaxies.  These candidates have been through detailed examination using
 multi-wavelength images and SEDs.  Candidates that are most likely YSOs are
 marked with additional red open circles and non-YSOs with green crosses. 
 In addition to these YSOs, the same examination procedures are carried out 
 on 1028 sources in the corresponding lower wedge to identify lower mass 
 or more evolved YSOs.  Candidates that are most likely YSOs are marked with
 additional magenta open circles.  
  \label{fig:cmds}}
\end{figure}

The lower part of the [8.0] versus ([4.5]$-$[8.0]) CMD 
(Figure~\ref{fig:cmds}) is populated by numerous background galaxies
\citep{HPetal06} as well as YSOs that have masses 
lower than $\sim 4 M_\odot$, and more evolved YSOs with 
reduced circumstellar dust and infrared excess \citep{WBetal04a,RTetal06}.
The primary selection above excludes these YSOs in order to reduce the 
number of background galaxies.

Among the eight molecular clouds detected in the Bridge 
\citep{MNetal06}, 
the most massive one, Molecular Cloud C does not have sources 
that meet our initial YSO selection criteria, but a cluster of eight 
red sources fainter than the color-magnitude cut bounded by 
$[8.0] \ge 14 - ([4.5]-[8.0])$.
To assess the nature of these sources, we examined their 
multi-wavelength SEDs and images.
One of them was resolved into a spiral galaxy in the archival 
ISAAC images, and another has a galaxy-like SED as discussed in 
Section \ref{galaxysed}; these two are most likely background galaxies.
The remaining six red sources have SEDs similar to YSOs.
Furthermore, our IRS spectra of the two brightest 24 \um\ sources 
among them show PAH features or red continuum with a rising slope, 
typical of YSOs (Indebetouw et al.\ in preparation).
Given the above considerations and that these six sources
are in a molecular cloud, they are most likely YSOs.

We have further expanded the examination of SEDs and images on 
all sources within the same lower wedge in the CMD, to assess 
if they are YSOs.
Among 1028 sources found in this wedge, 192 have photometric
measurements less than four bands to effectively distinguish 
their nature;
753 are resolved into galaxies in high-resolution 
SSS or 2MASS images, or have SEDs that are not well reproduced 
by YSO models but resemble late-type galaxies, or active galactic 
nuclei, i.e., flat from optical to far-IR or obscured in the 
optical due to the viewing angle with respect to its dust torus 
\citep{FAetal05,HEetal05,RRetal05}. 
Furthermore, resolved galaxies with similar SEDs are found in 
close vicinities of most of these 753 sources, suggesting 
that they are most likely bona fide galaxies. 
After excluding the above 192 and 753 sources, 83 sources remain.
These sources are likely YSO candidates, though a fraction 
of them could be star-forming (dwarf) galaxies having SEDs 
similar to YSOs but unresolved in the SSS or 2MASS images.
Note that such confusion is much less serious in our primary CMD-selected 
YSOs because galaxies would have to be quite nearby to be that bright and 
hence are usually resolved in the SSS or 2MASS images.
Given that these 83 fainter sources contain a fraction of contaminants 
and to distinguish them from the primary high-confidence sample, 
we will call them fainter YSO candidates in 
the rest of the paper.
The results of identifying these fainter YSO candidates 
are shown along with the primary YSOs in the [8.0] versus 
([4.5]$-$[8.0]) CMD (Figure~\ref{fig:cmds}). 

Finally, we remind that our both sets of selection criteria 
excluded YSOs with [4.5]$-$[8.0] $<$ 2.0, i.e., those that have 
much less circumstellar dust owing to even lower $M_\star$ 
or later evolutionary stages than their redder counterparts.
Identifying YSOs in this bluer part of the CMD is more 
challenging as the contamination from evolved stars such 
as AGB or post-main-sequence Be stars is significant 
\citep{Boyeretal11,Bonanosetal10}, and often requires 
additional data such as light curves or spectral lines 
to distinguish YSOs from contaminants 
\citep[e.g.,][]{deWitetal03,deWitetal05}.
\citet{Sewiloetal13} proposed a statistical approach to 
select YSO candidates in the SMC from 5 sets of CMD 
criteria that include sources outside our selection criteria. 
They calculated each source's location in 
the color-magnitude space with respect to contaminants,
determined a ``CMD score'' as a measure of confidence 
level for this source being a non-contaminant, and compared
SEDs of such candidates with YSO models to assess the 
likelihood of them being bona fide YSOs.
Owing to brighter magnitude cuts to alleviate contamination 
from galaxies, their list of YSO candidates with high 
confidence recovers subsets of 9 and 4 in our lists of 
26 primary YSOs and 83 fainter YSO candidates, respectively. 
Nonetheless, they identify three additional YSO candidates 
with colors bluer than our selection criteria: Y954, Y969, 
and Y970 \citep[nomenclature from][]{Sewiloetal13}.
These three candidates are in a corner defined by 
1.5 $<$ [4.5]$-$[8.0] $<$ 2.0 and 10.0 $<$ [8.0] $<$ 12.0:
Y969 has an SED similar to our YSOs, whereas Y954 and Y970
have SEDs consistent with B[e] supergiants but of fainter
luminosities \citep{Bonanosetal10}, requiring additional 
data to assess their nature as Be stars or YSOs.
The much smaller number of YSO candidates with [4.5]$-$[8.0] 
$<$ 2.0 than YSO candidates with redder colors reported 
in the \citet{Sewiloetal13} study lends credence that our 
criteria select the majority of YSOs in the Bridge.

\section{Determining YSO Properties from Model Fits of SEDs}

\subsection{Modeling the SEDs}
\label{modelfitting}

To infer the probable range of physical parameters for a YSO,
we compare the observed SED with those from a model grid
and select the best-fit models using $\chi^2$ minimization.
We use the large grid from \citet{RTetal06} that includes 
20,000 pre-calculated dust radiative transfer models,
each containing a central star (a photospheric emitter of 
pre-main-sequence or main-sequence spectrum) surrounded by 
a flared circumstellar disk and a flattened rotating envelope 
with bipolar cavities.
The best-fit models for observed SEDs is determined using the 
code from \citet{RTetal07} that requires fluxes and their errors
of a YSO as the input.
The fluxes of a SED are converted from magnitudes in Tables 
\ref{ysoclass} and \ref{ysoclassb}.
When calculating uncertainties associated with fluxes,
we include errors from measurement (in Tables \ref{ysoclass} 
and \ref{ysoclassb}) and absolute flux calibration 
(i.e., 10\% in each band from $B$ to $K_s$,
5\% in each band from 3.6 to 8.0 \um , 10\% in 24 \um , and 20 \% in 
70 \um ; \citealt{HNetal01a}; \citealt{SMetal06}; \citealt{Kaetal07};
IRAC Data Handbook; MIPS Data Handbook), and estimate the total
uncertainty of a flux using the quadratic sum of these two errors.
The 26 YSOs in our sample have been analyzed with SED fitting. 
As the models are calculated for single YSOs, comparisons are plausible
to the 17 YSOs that appear single or are clearly the dominant sources 
within the IRAC PSF.
The remaining 9 YSOs have multiple sources only resolved at some 
wavelengths.
Although the total fitted luminosity of 
each group or multiple is robust, differences in color between the members 
may indicate different evolutionary stages; we discuss cases of such sources 
in \S\ref{cldA} and \S\ref{cldB}.

The results of SED fitting to the 26 YSOs are shown in Figure~\ref{fig:fit},
with the best-fit and acceptable models overplotted on the SED of each YSO.
The 26 YSOs are shown in the following order: first the 17 ``single'' YSOs 
arranged by order of increasing Types from our empirical classification 
and within each Type by order of increasing [8.0] magnitude, and then 
the 9 ``multiple'' YSOs arranged simply by order of increasing [8.0] magnitude.
The best-fitting model for each source is determined by the minimum $\chi^2$
($\chi_{\rm min}^2$), but there is typically 
a range of models which are nearly as consistent with the data, i.e. $\chi^2$ 
not significantly greater. 
We have used a cutoff of $\chi^2 - \chi_{\rm min}^2 \le 3$ per datapoint 
for these acceptable models.  
The figure shows that uncertainty often results from parts of the SED at 
longer or shorter wavelengths than can be constrained by our data.
Table~\ref{sedfits} lists the results of model fits for the 26 YSOs 
which are tabulated in the same order as Figure~\ref{fig:fit}.
This table includes the source name, [8.0] magnitude, and type 
from our empirical classification, and weighted averages and standard
deviations of selected physical parameters: 
central stellar mass
($M_\star$), total luminosity ($L_{\rm tot}$), envelope accretion
rate ($\dot{M}_{\rm env}$), disk mass ($M_{\rm disk}$), central
stellar age ($\tau_\star$), foreground extinction ($A_V$), and 
inclination angle.
The derived $\dot{M}_{\rm env}$ and $M_{\rm disk}$ have been 
scaled by a factor of 3.5 to account for the higher gas-to-dust ratio
in clusters in the Bridge, which is likely to be representative for the 
YSO-forming regions, than the Galactic value used in YSO models 
\citep[][]{GKetal09}.
These averages and standard deviations show a possible 
range of these physical parameters; they are calculated from 
best-fit and acceptable models using the inverse square of 
$\chi^2$ as the weight.
For each accepted model, the evolutionary stage is determined 
using $\dot{M}_{\rm env}/M_\star$ and $M_{\rm disk}/M_\star$
ratios as defined in \citet{RTetal06}, i.e., Stage I -- 
$\dot{M}_{\rm env}/M_\star > 10^{-6}$ yr$^{-1}$; Stage II -- 
$\dot{M}_{\rm env}/M_\star < 10^{-6}$ yr$^{-1}$ and 
$M_{\rm disk}/M_\star > 10^{-6}$; and Stage III -- 
$\dot{M}_{\rm env}/M_\star < 10^{-6}$ yr$^{-1}$ and 
$M_{\rm disk}/M_\star < 10^{-6}$.
The range of the evolutionary stage, Stage Range, is the weighted 
standard deviation of the stages determined from each of the acceptable 
models for a YSO.

The model SEDs in general fit well with the observed SEDs of the 17 
''single'' YSOs (Figure~\ref{fig:fit}),
though discrepancies are seen in a number of the Type II YSOs.
Several Type II YSOs have lower observed than modeled 
4.5 \um\ fluxes. 
This 4.5 \um\ brightness dip is most likely a result of 
unaccounted contribution from PAH emission at 
3.3, 6.2, 7.7, and 8.6 \um\ in the other three IRAC bands.  
Since PAH emission 
features were not included in these pre-calculated models,
in SED fits they are compensated by an increase in small 
grain continuum emission.
However, our detailed analysis for YSOs in the LMC with and without 
correction of PAH emission demonstrates that this effect does not 
alter the derived physical parameters more than their quoted 
uncertainties \citep{CCetal10}. 
PAH emission has been 
detected in 87\% of IRS spectra of 277 YSOs in the LMC and 80\% 
of IRS spectra of 5 YSOs in the Bridge (\citealt{Seetal09}; 
Indebetouw et al.\ in preparation).
Bright PAH emission is also likely responsible for the YSOs with 
observed lower than modeled fluxes at both 4.5 and 24 \um\ in SEDs.

Second, a few Type II YSOs have observed fluxes brighter than modeled 
fluxes at optical wavelengths.
As massive YSOs evolve, compact \hii\ regions typically form,
but their small sizes $\sim 0.1$--1 pc, or $\sim 0.4$--4$''$ in the Bridge,
can only be resolved in high-resolution \ha\ images.
Such cases have been demonstrated in massive YSOs in the LMC using
{\it HST} or 4~m MOSAIC \ha\ images 
\citep[e.g.,][]{CCetal09,Vaetal09,CCetal10}.
The optical $BRI$ photometry were adopted from the SSS catalog based
on broadband images taken with a CCD camera of a 0\farcs67 pixel$^{-1}$
scale and a resolution of $\sim 1\farcs5$ \citep{HNetal01b}.
This would be inadequate to resolve central stars from surrounding compact 
\hii\ regions of size $\lesssim 2''$.
Bright nebular emission such as \ha\ would have raised the fluxes
in $R$ and resulted in deviation from the dust 
radiative transfer models.

The most robust physical parameters derivable from model fitting are 
the total luminosity and circumstellar dust mass; the correspondingly 
derived stellar mass and evolutionary state are robust as well when
single-YSO models are applicable. 
There is excellent agreement found between those stellar luminosities and masses 
inferred from SED fits of YSOs and from ionizing fluxes of corresponding 
ultra-compact \hii\ regions in the LMC \citep{CCetal09,CCetal10}.
For YSOs whose SEDs show discrepancies in the optical wavelengths, 
total luminosity can remain reliable for an SED relatively well reproduced
in the mid-IR segment, as shown by YSOs 052207.3$-$675819.9 in N\,44 and 
054004.40$-$694437.6 in N\,159 \citep{CCetal09,CCetal10}.
In the models, circumstellar dust is distributed in a disk and a rotating
flattened envelope.  Although the relative distribution between those 
two components is difficult to constrain without (sub)millimeter photometry, 
the total mass of circumstellar dust is relatively robust.  
In cases where compact \hii\ regions already formed around YSOs,
the dust may be considered circumstellar in the SED fits rather than the more
likely interstellar origin, 
and the inferred $\dot{M}_{\rm env}$ and 
$M_{\rm disk}$ may be overestimated.  This is most likely in those sources
with observed optical emission greater than the model, which as previously 
noted we believe to be likely optical nebular contamination.

\subsection{Evolutionary Stage of YSOs}

\subsubsection{Comparisons Between Empirical ''Type'' and 
Model-Derived ''Stage'' Classifications}

There is not yet a well-defined 
classification system for massive YSOs, since neither their formation 
mechanisms nor their circumstellar mass distribution is well-known.
In the \citet{RTetal06} study, we proposed a 
``Stage'' classification based on the {\it physical}
quantities $\dot{M}_{\rm env}$ and $M_{\rm disk}$, derived from the best-fitting
models for a given source.  Solar-mass YSO classifications have traditionally
been based on the infrared spectral index, or infrared color excess.
However, extension of the scheme to massive YSOs must be carefully parameterized, 
because a change in the central source's photospheric temperature and luminosity 
changes the emitted spectrum of heated circumstellar dust independently of the 
{\em mass} of circumstellar dust \citep{WBetal04b}.  Using the model-derived 
ratio of envelope to central source mass takes into account these effects in a 
physically self-consistent way.

At the distance of the Bridge, 
multiple sources or small clusters would be unresolved by {\it Spitzer};
thus, the circumstellar dust geometry may be complex and not correspond 
very well with these single-YSO models.
As the physical conditions and structure of the surrounding ISM change 
as YSOs evolve, we proposed an empirical classification ``Type'' that 
uses both SEDs and immediate environmental morphology on sub-parsec scale 
to assess the evolutionary states of massive YSOs in the Magellanic system 
\citep{CCetal09}.
There is also uncertainty in the ``Type'' classification as it assumes 
that the environmental morphology corresponds tightly to the evolutionary 
state of a single or dominant source.
Comparisons between the two schemes can help to better understand the 
amount and distribution of dust around massive YSOs in the Bridge 
and to obtain a comprehensive picture of their evolution. 

The analysis of 17 ''single'' YSOs is used to compare the two 
classifications Type and Stage.
As listed in Table~\ref{sedfits}, types and stages are not 
overwhelmingly correlated.
Upon close examination, among the seven Types II or II/III YSOs (relatively evolved) 
that have inferred Stage $\sim 1.0$--1.2 (relatively unevolved), five show observed 
optical fluxes higher than model SEDs, implying presence of nebular 
emission and hence \hii\ regions likely to have formed.
The mid-IR dust emission from unresolved \hii\ regions 
included in the YSO SED is interpreted by the models
as circumstellar disks or envelopes even though it is unlikely to accrete 
onto the sources \citep[however see e.g.,][for a discussion of accretion 
of ionized gas in massive protostars]{KE07}.
When excluding these five Types II or II/III YSOs, there is a
rough trend between types and stages as Types I and I/II YSOs
have Stages $\le 1.6$ and Types II and II/III have Stages $\sim 2.0$.

Neither Type III nor Stage III YSOs are found in the Bridge, unlike
in LMC \hii\ complexes \citep{CCetal09,CCetal10}.
This may be partly due to the lack of high-resolution \ha\ images
that are used in the Type scheme to discriminate Type III from 
Type II.
Our exclusion of unresolved, extended sources with double-peaked 
galaxy-like SEDs (\S\ref{contaminants}) also may exclude some 
Type III YSOs, but this 
effect should not be different between the LMC and Bridge.
On the other hand, it is possible that the Type/Stage III phase
is shorter or exhibiting different observational properties 
compared to those in the LMC.  Stage III corresponds to transitional 
and debris disks, and a faster dust disk dissipation time at lower metallicity
(harder stellar field, lower interstellar dust-to-gas ratio) is not 
implausible.
Further comparisons between YSOs in the Bridge and the LMC 
are discussed in Section \ref{ysometallicity}.

\subsubsection{Evolutionary Stages of Faint YSO Candidates}

The fainter YSO candidates were selected from the lower part of 
the [8.0] versus ([4.5]$-$[8.0]) CMD (Figure~\ref{fig:cmds}) 
that is populated by YSOs that are more evolved or have lower
masses than those in the upper part of the CMD.
To infer the probable ranges of physical parameters of these
fainter YSO candidates, we have compared their SEDs to YSO
models.
The results of model fits of the 83 fainter candidates are given
in Table~\ref{fysofits}, including the source name, weighted average 
and standard deviation of selected physical parameters: 
$M_\star$, $L_{\rm tot}$, $\dot{M}_{\rm env}$, $M_{\rm disk}$, 
Stage,  $\tau_\star$, $A_V$, and inclination angle.
The weighted average and standard deviation are calculated using 
the same procedure described in Section \ref{modelfitting}.
As shown in Figure~\ref{fig:haebe_hist}, fainter YSO candidates
have more advanced evolutionary states with 77\% being Stage II
(weighted average stage $=$1.5--2.5) and 3\% being Stage III 
(weighted average stage $\ge 2.5$), more evolved than our primary 
YSO sample with 46\% being Stage II and zero Stage III.
These results are consistent with expected for populations in 
these two parts of the CMD.

\subsubsection{Evolutionary Stages of HAeBe Candidates}

As YSOs have evolved from the embedded phases and revealed 
their stellar photospheres, their circumstellar envelopes 
might have mostly dissipated and only remnant disks remain.
Thus the near-IR excess is likely more prominent than mid-IR
excess.
HAeBe are young stars of intermediate masses that are frequently 
identified by their near-IR excess \citep[e.g.][]{HLetal92}.
Since their mass range overlap with our YSO sample, we examine
whether some of them might be previously identified as HAeBe 
candidates and if so, how their evolutionary stages are compared
to our YSOs.
Using the IRSF $JHK_s$ catalog, \citet{Nietal07} selected 
$\sim$ 200 HAeBe candidates in the Bridge based on their 
near-IR excess in $JHK_s$ CMDs.
Such candidates are of modest reliability, as near-IR excess 
can originate from not only circumstellar dust around 
young stars, but also interstellar dust near 
main-sequence and giant stars, or background galaxies 
\citep[e.g.,][]{Naetal05}.  
In this list of HAeBe candidates in the Bridge, \citet{Nietal07}
estimated $\sim$ 60\% contaminants.

With our multi-wavelength SEDs and images, we have demonstrated 
that background galaxies can be identified, as can main-sequence
or giant stars obscured by cool ($\sim 30$ K) interstellar dust, 
opposed to YSOs or evolved stars with warm ($\ge 100$ K) 
circumstellar dust.
To assess the nature of the HAeBe candidates in the Bridge, we
constructed their multi-wavelength SEDs by matching them with 
our photometric catalog.
When matching sources, we allow a 1$''$ error margin.
Among the 203 HAeBe candidates in the \citet{Nietal07} list,
199 are in our working field, and 191 of them have matches
in our catalog.
We compare SEDs of these 191 sources to YSO models as well 
as stellar models from \citet{BH05} for stellar temperature 
$T_\star <$ 10,000 K and from \citet{CK04} for $T_\star \ge$
10,000 K; these stellar models are included in the SED
fitting code \citep{RTetal07}. 
SEDs of main-sequence or giant stars should be well reproduced 
by stellar models reddened by interstellar dust without obvious
mid-IR excess in the IRAC or MIPS bands, for example HAeBe candidate 
J014226.34$-$741432.13 shown in 
Figure~\ref{fig:haebe}.
SEDs of stars with (warm) circumstellar dust would show IRAC 
or MIPS fluxes higher than stellar models reddened by 
interstellar dust and be relatively well reproduced by YSO models, 
such as HAeBe candidate J014744.02$-$742551.95 in Figure~\ref{fig:haebe}.
Based on these comparisons, we have identified 96 HAeBe candidates
in the \citet{Nietal07} list that show IRAC flux excess.
These sources are most likely HAeBe stars, though a fraction of them 
could be more evolved, such as Be stars that also have warm 
circumstellar dust.

The results of model fits of the 96 HAeBe candidates are given
in Table~\ref{aebefits}, including the source name, weighted average 
and standard deviation of selected physical parameters: 
$M_\star$, $L_{\rm tot}$, $\dot{M}_{\rm env}$, $M_{\rm disk}$, 
Stage,  $\tau_\star$, $A_V$, and inclination angle.
The weighted average and standard deviation are calculated using 
the same procedure described in Section \ref{modelfitting}.
84\% of these HAeBe candidates have more advanced evolutionary 
states (Figure~\ref{fig:haebe_hist}), with 55\% being Stage III 
and 29\% being Stage II, more 
evolved than the Stages I and II (with the maximum weighted average Stage 
$=$2.3) 
inferred for most of the YSOs selected using mid-IR CMDs.
Note that the list of HAeBe candidates from \citet{Nietal07} does 
not overlap with our YSO list, contrasted with LMC \hii\ complex N\,159 
where lists of HAeBe candidates and {\em Spitzer}-selected YSOs do overlap
\citep{CCetal10}.
This disjointness of the two samples in the Bridge is mostly attributed to bluer 
near-IR color cuts \citet{Nietal07} used to select HAeBe candidates than
were used in the LMC.
These bluer color cuts were based on the argument in the \citet{deWitetal03} study that the amount of circumstellar dust around HAeBe stars at low 
metallicity would be much less and hence results in smaller NIR color excess 
and bluer colors, though they also found that all their bluer candidates 
have ambiguous natures as their light curves and spectral features are 
also consistent with post main-sequence Be stars.
If the near-IR color cuts that had been used in N\,159 \citep{Naetal05}
were applied to select HAeBe candidates in the Bridge, three candidates 
indeed would overlap with our YSOs and have evolutionary state of Stage II.
The reliability of a list of NIR-selected 
HAeBe candidates can be greatly improved using
multi-wavelength SEDs, but definitive identification of these 
more evolved YSOs, and understanding the evolution of circumstellar 
disks, requires spectroscopic and/or variability study.

\subsection{Masses of YSOs, Fainter YSO Candidates, and HAeBe Candidates}

\begin{figure}
\epsscale{1.1}
\plotone{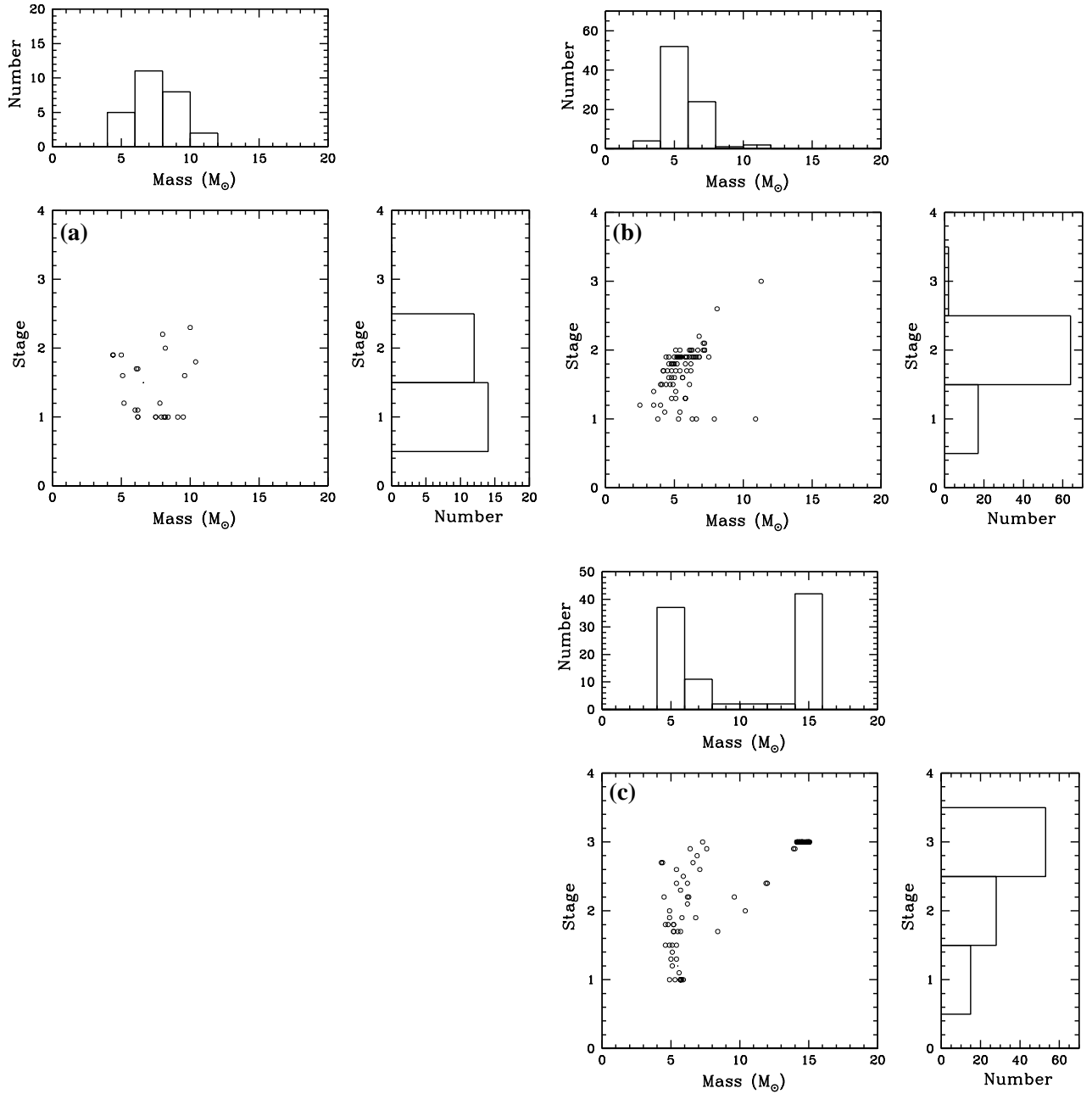}
\caption{Stage versus mass inferred from SED fits for (a) 26 YSOs, (b) 
 83 fainter YSO candidates, and (c) 96 HAeBe candidates in the Bridge. 
 Histograms of mass and Stage are plotted on the sides.
 These three types of sources show different dominant populations of Stages:
 YSOs, fainter YSO candidates, and HAeBe candidates have their dominant populations 
 in Stages I, II, and III, respectively, consistent with expectations for these  sources.
 YSOs and faint YSO candidates have masses ranging 4-11 $M_\odot$. 
 Between the two YSO samples, fainter YSO candidates have masses mostly in the lower 
 end of the range, though the two most massive ones have masses $\sim 11 M_\odot$, 
 comparable to those in our primary YSO sample.
 HAeBe candidates show a bimodal distribution of masses, with the high-mass 
 ($\sim 15 M_\odot$) population all having more evolved evolutionary stage (Stage III).  
 \label{fig:haebe_hist}}.
\end{figure}

The mass estimates, i.e., the $\chi^2$-weighted average mass $\bar{M}_{\ast}$
from the best and acceptable fits, of 26 YSOs in the Bridge are given 
in Table~\ref{sedfits}.
The mass estimates would be reliable for the 17 ``single'' YSOs as their 
SEDs can be properly approximated by single-YSO models.
By contrast, the other nine have SEDs from multiple YSOs that 
might not be well reproduced by single-YSO models; their mass estimates 
have larger uncertainties.
The results of SED fits, as illustrated in Figure~\ref{fig:haebe_hist}a, 
show that ten YSOs in the Bridge have $\bar{M}_{\ast} \ge 8 M_\odot$; 
these are most likely bona fide massive YSOs.
The remaining 16 YSOs have $\bar{M}_{\ast} < 8 M_\odot$; these 
are likely intermediate-mass YSOs.
The Bridge does not contain embedded YSOs as massive as active regions 
in the LMC, which contain 
O-type embedded YSOs with masses 17--45 $M_\odot$ 
\citep[e.g.,][]{CCetal09,CCetal10}.
The most massive embedded YSOs in the Bridge are 
$\sim 10 M_\odot$, corresponding to $\sim$ B2 stars.

The mass estimates from the best and acceptable fits of 83 fainter 
YSO candidates in the Bridge are listed Table~\ref{fysofits}.
Their masses are mostly in the lower end of the mass range 
(Figure~\ref{fig:haebe_hist}b): 80 ($= 96$\%) of the fainter 
YSO candidates have masses $< 8 M_\odot$ with the peak of the 
distribution falling in the 4--6 $M_\odot$ mass bin.
That most sources are in this low-mass bin is expected for sources 
in the fainter part of the CMD.  
However, three of these fainter sources have masses $\ge 8 M_\odot$, 
with two of them upto $\sim 11 M_\odot$, comparable to the most massive 
in our primary sample 
The fainter [8.0] magnitudes (compared to our primary YSO sample) of 
these three massive YSOs are likely attributed to a more edge-on
viewing angle, 57--86$\degr$ as inferred from the SED fitting 
(Table~\ref{fysofits}).
Among these three fainter YSO candidates with masses $\ge 8 M_\odot$, 
the most massive two are found in Cloud C and the third one in Cloud A.

The mass estimates from the best and acceptable fits of 96 HAeBe
candidates in the Bridge are listed Table~\ref{aebefits}.
The masses of the HAeBe candidates show a bimodal distribution, 
with 50 candidates in the mass range 4--10 $M_\odot$ and 46 
in 10--16 $M_\odot$ (Figure~\ref{fig:haebe_hist}c).
Candidates in the lower-mass peak have a mass range comparable 
to the embedded YSOs, implying that the most massive stars formed in 
the Bridge in the last several Myr are $\sim 10 M_\odot$.
Candidates in the higher-mass end (14--16 $M_\odot$, corresponding
to B1V) are all Stage III sources with ages $\gtrsim 3$ Myr, 
though a fraction of them are likely much older Be stars, 
of spectral type B1 and often found in clusters 
with ages ranging from $\gtrsim 3$ to 30 Myr \citep{GE97}.
Comparisons between these mass ranges of YSOs and HAeBe candidates show 
that the Bridge has not been actively producing O-type stars  
in the last several Myr, and that the intensity of {\em massive} star 
formation appears to have {\em decreased with time}, 
as the most massive stars formed
$\gtrsim$ 3 Myr ago are $\sim 15 M_\odot$, while those formed 
currently are only $\sim 10 M_\odot$.

\subsection{Comparisons of YSO Properties at Different Metallicity}
\label{ysometallicity}

It has been suggested that a lower dust abundance and greater 
permittivity to ultraviolet radiation of the ISM is expected to affect 
pre-formation gas dynamics, as well as cooling and feedback from 
massive YSOs \citep{Poetal95}.
The Bridge and the LMC have different metallicities, 1/5--1/8 and
1/3 $Z_\odot$ respectively, and thus provide an excellent opportunity
to examine the metallicity effect on massive star formation.
We compare the properties of YSOs in the Bridge to those from our 
studies of two LMC \hii\ complexes, N\,44 and N\,159 
\citep{CCetal09,CCetal10}.
These two complexes have 41 YSOs in the mass range overlapping with 
YSOs in the Bridge, i.e., 4--11 $M_\odot$.
Comparisons among them show that YSOs in the Bridge appear less 
embedded, as 81\% (=21/26) of them show optical counterparts,
while only 56\% (=23/41) in these two LMC complexes.
The difference is even larger when taking into account that SSS 
$BRI$ images of the Bridge are much shallower than our 4~m MOSAIC 
deep $UBVI$ images of LMC complexes, allowing fewer optical counterparts to be detected.
The higher frequency of optical counterparts implies a smaller 
extinction in YSOs in the Bridge, likely due to its lower dust
content in either the molecular clouds or circumstellar envelopes,
or both.

\begin{figure}
\epsscale{1.0}
\plotone{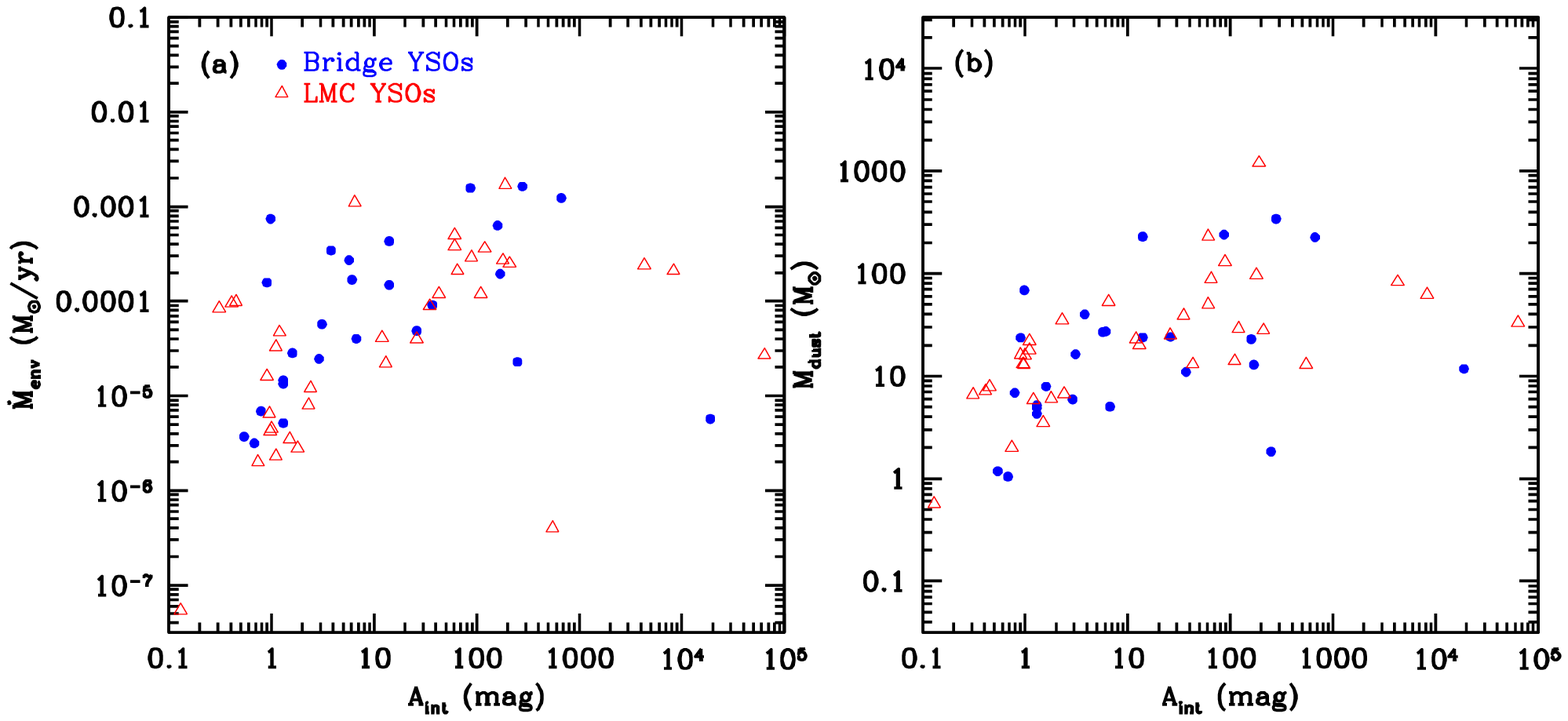}
\caption{(a) $A_{\rm int}$ versus $\dot{M}_{\rm env}$ and (b) $A_{\rm int}$
 versus $M_{\rm dust}$ of YSOs in the Bridge (filled circles).  For comparison, 
 YSOs in two LMC \hii\ complexes \citep[open triangles;][]{CCetal09,CCetal10} 
 are also plotted.  $\dot{M}_{\rm env}$ and $M_{\rm dust}$ in this figure are
 not scaled with respective gas-to-dust ratios for the LMC and Bridge so that
 the comparisons among parameters are all based on dust masses.
 YSOs in the Bridge show a similar distribution in the inferred amount of circumstellar
 dust compared to in the LMC, even though the optical counterparts are more frequently
 detected in the former than the latter.
 \label{fig:extinct}}
\end{figure}

We investigate several possible causes of the higher frequency of optical 
counterparts for YSOs in the Bridge.
First we examine if the extinction difference is mostly from 
molecular clouds by comparing foreground extinction $A_V$ (fitted in our 
SED modeling) in these YSOs.
We find similar values in the Bridge and the LMC complexes,
i.e., $A_V = 3.4\pm 8.9$ and $3.5\pm7.0$ respectively, suggesting
that molecular clouds are not the main cause in the extinction
difference.
To search for differences in circumstellar dust, 
we compare the central source mass $M_\star$, 
line-of-sight extinction due to dust in the circumstellar 
envelope $A_{\rm int}$,
the envelope accretion rate $\dot{M}_{\rm env}$, and total amount 
of circumstellar mass $M_{\rm dust}$. $A_{\rm int}$ is determined 
from the entire SED while $\dot{M}_{\rm env}$ and $M_{\rm dust}$ 
mostly depend on mid-IR emission.
As shown in Figure~\ref{fig:extinct}, there is no obvious difference 
in these parameters between YSOs in the Bridge and those in LMC \hii\ 
complexes. 
Note that since $A_{\rm int}$ depends on dust mass, not dust$+$gas mass,
$\dot{M}_{\rm env}$ and $M_{\rm dust}$ in this figure are thus not 
scaled with respective gas-to-dust ratios for the LMC and Bridge 
so that comparisons among these parameters can be all on dust masses.
If circumstellar material had a smooth spherical distribution, then 
the circumstellar line-of-sight extinction $A_{\rm int}$ would
be directly related to the circumstellar dust mass $M_{\rm dust}$ 
and derived envelope accretion rate $\dot{M}_{\rm env}$.  However, 
for a nonspherical (disk-like or clumpy/irregular) circumstellar 
dust distribution, the line-of-sight extinction is primarily related 
to the viewing angle \citep[inclination angle for a disk-like geometry, 
presence or absence of a dense clump on the line of sight for a clumpy 
geometry, see e.g.,][]{Inetal06}.  The derived dust mass however is primarily 
related to the total mid to far-infrared emission, and less sensitive to 
the geometry.   Finally, the presence of a bright optical counterpart 
corresponds to how clumpy overall the distribution is (or for a disk, the 
size of the bipolar openings) -- a more porous distribution of the same amount 
of dust will allow more short-wavelength light to scatter out and be observed.
What we see in this comparison is that Bridge and 
LMC massive YSOs have similar circumstellar dust masses,
that the presence or absence of a line-of-sight clump is not particularly different, 
but that overall the envelopes of Bridge YSOs are likely more porous or clumpy
than in LMC YSOs.
This may also imply that these YSOs have a shorter timescale 
for dispersal of circumstellar material, and more rapid YSO evolution.

\section{Massive Star Formation in the Magellanic Bridge}

A causal relationship between the physical conditions of interstellar
environment and formation of massive stars is not easy to establish
since once massive stars are formed, their strong UV radiation and
fast stellar winds ionize and disperse the natal clouds and subsequently
alter the physical conditions of the ambient ISM.   
Embedded massive YSOs, on the other hand, have not had time to 
significantly affect their surrounding medium beyond parsec scales,
providing an excellent chance to probe issues on massive star formation.
We use the massive YSOs found in the Bridge to investigate issues such 
as the relationship between star formation properties and interstellar 
conditions and the progression of star formation in the tidal environment.

\subsection{Interstellar Environments and Star Formation Properties}

\begin{figure}
\plotone{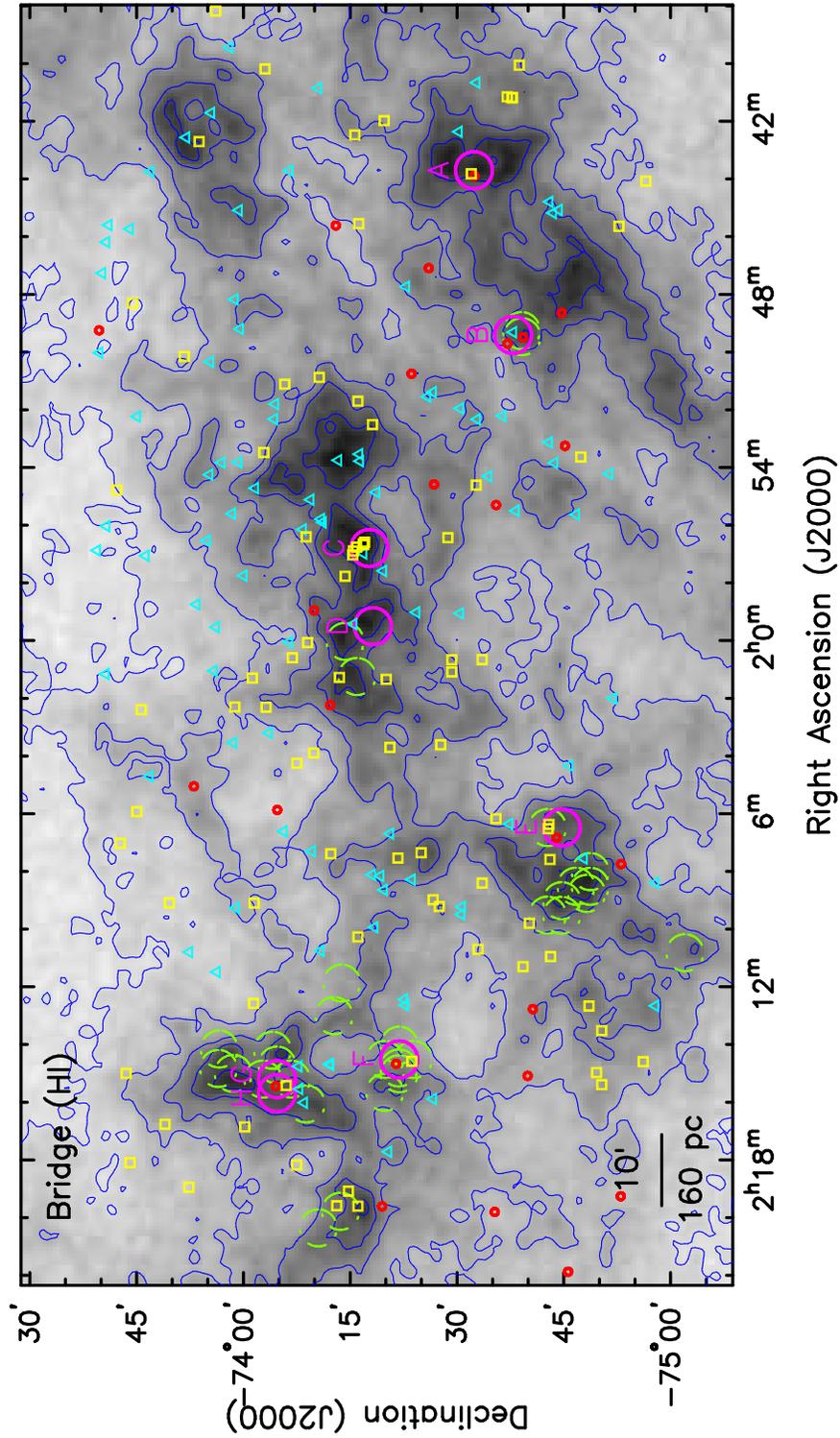}
\caption{Distribution of embedded YSOs (red circles), fainter YSO candidates
 (yellow squares), HAeBe candidates (cyan triangles) with respect to the 
 interstellar environment of the Magellanic Bridge.  The grey scale shows
 the \hi\ map of the Bridge, overlaid with contours (blue lines) in
 4n$\times10^{20}$~cm$^{-2}$ (n$=$1,2,...).  Molecular clouds A--G cataloged
 in \citet{MNetal06} are labeled and marked in large magenta circles; 
 these were part of the NANTEN CO survey of the Bridge (large green circles; 
 Fukui et al.\ in preparation). 
 \label{fig:yso_pos}}
\end{figure}

How massive stars are formed in a region and whether there is 
a dependence on environmental factors such as metallicity and 
tidal interaction are crucial to understand star formation across 
the near and far universe.
To study these questions, one of the most direct ways is to examine 
the relationship between massive YSOs and their natal environment,
particularly the molecular environment as the dense clumps are where
stars are formed.
We thus examine the properties of massive YSOs in the Bridge and 
their relation to molecular clouds. 
Eight molecular clouds have been detected toward the Bridge from Mopra,
SEST, and NANTEN CO J=1-0 surveys \citep{Muetal03a,MNetal06}.
This census of molecular clouds is not complete; these low-metallicity 
clouds require brightness temperature sensitivities of $\lesssim$10 mK 
to detect, so the search was
limited to regions selected primarily based on bright {\it IRAS} 
100 \um\ emission 
(regions observed are marked in Figure~\ref{fig:yso_pos}, and 
the search criteria described in detail in \citealt{MNetal06} and 
Fukui et al.\ in preparation).
These {\it IRAS} sources have a significant fraction of contaminants 
such as background galaxies (recall Section \ref{contaminants} on culling 
out such contaminants from the list of YSO candidates), or diffuse 
dust regions unresolved with {\it IRAS}' 2$'$-resolution ($\sim$ 30 pc 
in the Bridge).
Furthermore, as the majority of the YSOs in the Bridge are intermediate 
masses and thus too faint for {\it IRAS} surveys, it is not surprising 
that we find numerous embedded YSOs not covered by previous CO surveys 
and hence not associated with detected CO.

Despite the small size (eight) and incompleteness of 
the existing CO cloud sample, all but one have associated embedded 
YSOs; the only one without YSOs, Cloud D, has the faintest CO emission 
with a marginal 2-$\sigma$ detection \citep{MNetal06}.
This tight spatial correlation between the molecular clouds and YSOs 
implies that star formation happens quickly once these clouds are formed.
By contrast, star formation does not appear to happen swiftly in 
the large group (36) of massive clouds in the molecular ridge of the LMC 
as very few of them show massive star formation activity in the past 
$\gtrsim$ 10 Myr or host massive YSOs \citep{Inetal08}.
The eight Bridge clouds exhibit an extensive range of massive stars 
formed in the past $\gtrsim$ 10 Myr and at present; we discuss the propagation 
of star formation in these clouds individually in Section \ref{indivclouds}.

\begin{figure}
\plotone{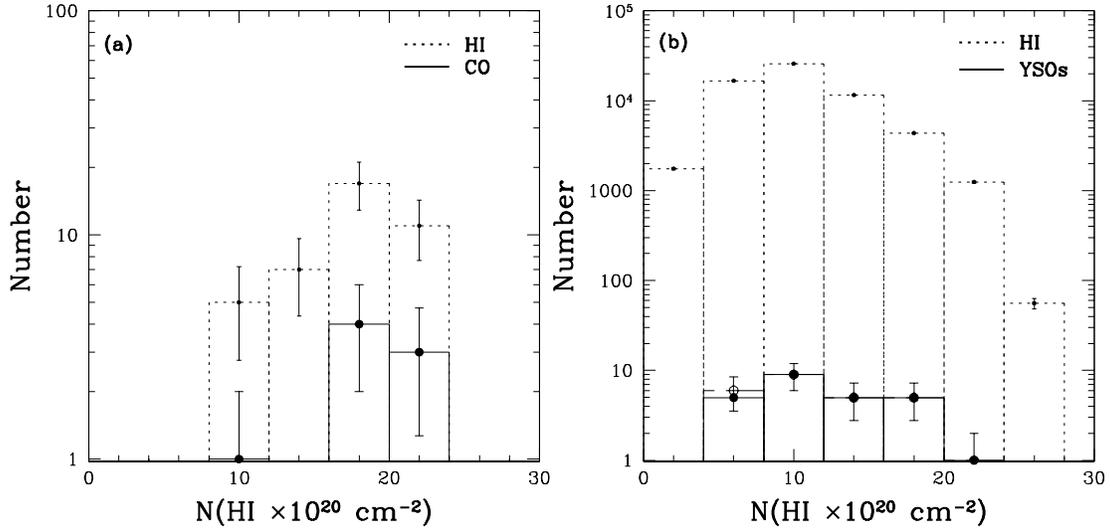}
\caption{(a) Histograms of N(\hi ) at 44 positions mapped in CO surveys 
 of the Bridge (dotted lines; Fukui et al.\ in preparation) and at which eight molecular 
 clouds are detected \citep[solid lines;][]{MNetal06}. 
 Despite the sparsity of CO observations in the Bridge, it shows that the 
 chance of detecting CO is larger at higher N(\hi ). 
 (b) Histograms of all \hi\ resolution elements (dotted lines) and those
 containing YSOs (solid lines) in the Bridge. 
 \label{fig:hi_mcbyso}}
\end{figure}

To better understand the formation mechanisms of the entire YSO sample
in the Bridge, we compare the distribution of YSOs to \hi\ emission,
as N(\hi ) indicates where the bulk of atomic gas is,
and the ATCA \hi\ survey has mapped the entire Bridge we studied \citep{Muetal03b}.
As shown in Figure~\ref{fig:yso_pos}, the N(\hi ) in the Bridge ranges
up to 2.7$\times10^{21}$~cm$^{-2}$; compared to the LMC and SMC that have 
maximum N(\hi )$=$ 8.8 and 14.3 $\times10^{21}$~cm$^{-2}$ respectively 
\citep{KSetal03,SSetal99}, the N(\hi ) distribution of the Bridge 
overlaps with the lower end of the LMC and SMC.
Molecular clouds (and stars forming from them)
tend to be found in regions with higher N(\hi ), as shown in 
the LMC and SMC \citep{WTetal09,Leetal07,Muetal10}, and also in the small 
sample of molecular clouds in the Bridge (Figure~\ref{fig:hi_mcbyso}).
Thus it is not unexpected that the majority, $>$ 70\%, of the YSOs in 
the Bridge are found in N(\hi ) $\ge 8\times10^{20}$ cm$^{-2}$ 
(Figures~\ref{fig:yso_pos}). 
We estimate the mass of molecular material that may be
associated with such a column of \hi:
at first glance, this N(\hi ) appears to correspond to a very low
A$_V < 0.1 $ for a gas-to-dust-ratio N(\hi )/A$_V \sim 10$--$32 \times 
10^{21}$ cm$^{-2}$~mag$^{-1}$ found in the nearby SMC \hii\ region N\,83 
\citep{Leetal09} or the 30$\times$10$^{21}$ cm$^{-2}$~mag$^{-1}$ 
measured in the Bridge \citep{GKetal09}.
This is much lower than A$_V \sim$ 0.5--1 needed for the surface 
H$_2$ formation rate to effectively balance ultraviolet photodestruction
\citep{HBB01}. 
However, considering the \hi\ map's 98$''$-resolution ($\sim$ 29 pc 
in the Bridge)
and \hi :H$_2 =$ 4:1 \citep{DT93}, 
N(\hi ) $\sim 8\times10^{20}$ cm$^{-2}$ could hide a 5-pc-sized molecular 
cloud with gas surface density $\sim 6.7 \times 10^{21}$ cm$^{-2}$, 
corresponding to A$_V \sim$ 0.2--0.7 for the aforementioned 
gas-to-dust-ratio.
Such a molecular cloud would have a mass $\sim 1300 M_\odot$. 

\begin{figure}
\plotone{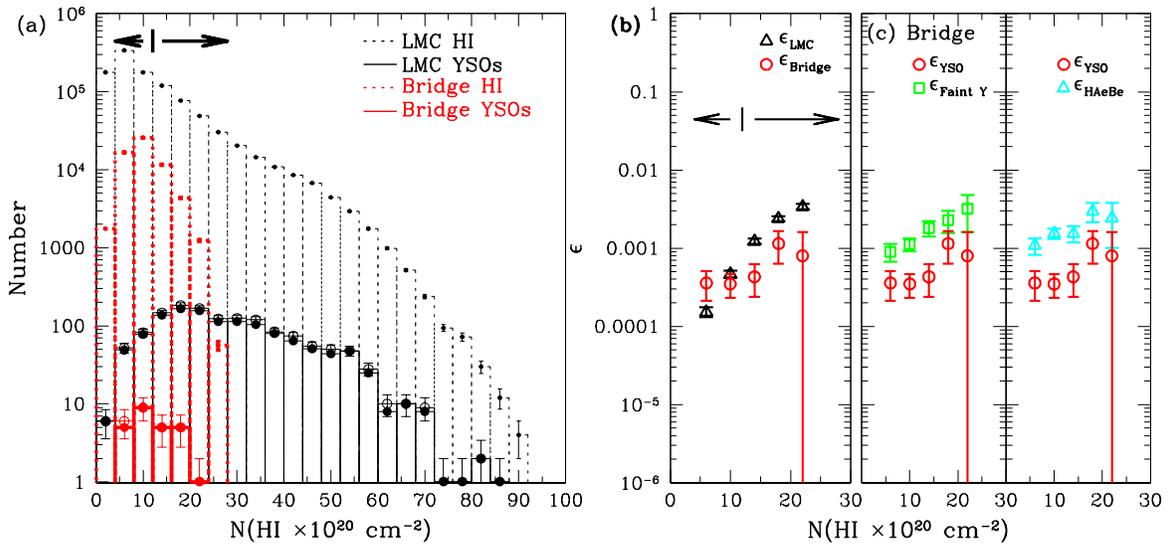}
\caption{(a) Histograms of all \hi\ resolution elements in the LMC (black dotted
 lines) and those containing YSOs \citep[black solid lines,][]{GC09}. 
 Similar histograms for the Bridge are shown as red dotted and solid lines. 
 (b) $\epsilon$ as a function of N(\hi ) for the LMC (black 
 triangles) and the Bridge (red circles).  Error bars are marked.  The Bridge
 has $\epsilon \sim $1/3 the LMC at N(\hi ) $\ge 12\times10^{20}$ cm$^{-2}$.
 However, at lower N(\hi ), the Bridge's $\epsilon$ is comparable or even 
 larger than the LMC.
 (c) $\epsilon$ as a function of N(\hi ) for young massive stars at different 
 evolutionary stages in the Bridge: YSOs (red circles), faint YSO candidates 
 (green boxes) and HAeBe candidates (cyan triangles).  The two older 
 populations show a mild decrease with decreasing N(HI), while YSOs exhibit 
 a steep decrease in low N(HI).
  \label{fig:hi_mcblmc}}
\end{figure}

To quantify the SFE in the Bridge, we examine $\epsilon$,
the fraction of \hi\ resolution elements containing YSOs 
as a function of N(\hi ). 
This is equivalent to the probability of finding YSOs at 
a given N(\hi ).
Figure~\ref{fig:hi_mcbyso} shows histograms of the Bridge's 
\hi\ resolution elements, i.e., 30$''$~pixel$^{-1}$, and YSOs, 
as well as $\epsilon$ as a function of N(\hi ).
To examine clustering of YSOs, an additional histogram of YSOs
is made to count those in the same \hi\ pixel as one YSO.
Figure~\ref{fig:hi_mcbyso} (open compared to filled circles) shows 
that the clustering is minimal since only one \hi\ pixel in the 
Bridge contains two YSOs.
$\epsilon$ of the Bridge appears to be higher at higher N(\hi ), 
though there is no obvious linear correlation.
To assess if star and cloud formation is affected by metallicity 
or galactic environment, we further compare $\epsilon$ between 
the Bridge and LMC.
Histograms of the LMC's \hi\ resolution elements in the same pixel 
scale of 30$''$ and YSOs \citep{KSetal03,GC09} are shown
in Figure~\ref{fig:hi_mcblmc}, along with $\epsilon$ as a function
of N(\hi ).
As in the Bridge,
only a small fraction, $\lesssim 10$\%, of \hi\ pixels in the LMC
contain multiple YSOs, so clustering on scales $<30''$ doesn't 
affect the result there either.
Comparisons of $\epsilon$ show that the Bridge is $\lesssim 1/3$ 
the LMC in their overlapping bins of N(\hi ) $=$ 12--24 $\times 
10^{20}$ cm$^{-2}$.
However, at lower N(\hi ), the Bridge shows a flatter slope 
so that its $\epsilon$ is even up to $\sim 2$ times the LMC.
To investigate if this flatter slope is a result of
local variations within the LMC, we have also estimated $\epsilon$ for seven 
regions across the LMC with the same angular size as the Bridge, 
and 1--1.5 times the Bridge's average N(\hi ).
We found that all these regions may have $\epsilon$ higher or 
lower than the average $\epsilon$ of the entire LMC, but their 
slopes are always rising.
The Bridge's flatter slope does indeed appear unique.

We examine the YSO properties in the LMC and the Bridge
to investigate the possible causes of different $\epsilon$ 
in these two systems.
The YSOs in the Bridge have inferred masses of 4--10 $M_\odot$,
not as massive as those up to $45M_\odot$ found in the LMC
\citep[e.g.,][]{CCetal09,CCetal10}; as will be discussed
in Section 5.3, the dearth of YSOs $\ge 10 M_\odot$ in the
Bridge cannot be explained by stochastic effects.
Furthermore, Bridge YSOs are sparsely distributed 
(Figure~\ref{fig:yso_pos}), in sharp contrast to LMC YSOs that are 
usually in groups and clusters \citep[e.g.,][]{CCetal09,CCetal10,GC09}.
The lack of massive clusters in the Bridge is consistent with
the simulations of \citet{KMT09b}, in which cluster-forming 
molecular clouds are scarce at that density and metallicity.
On the other hand, the reverse trend appears to suggest a different 
dominant mechanism of star formation at the lower N(\hi ) regime.
The sparse distribution of YSOs and low masses of molecular clouds, 
10$^3$--10$^4 M_\odot$, in the Bridge \citep{MNetal06},
as well as the rapid star formation implied by the tight spatial 
correlation between molecular clouds and YSOs, are consistent with 
star formation through colliding flows \citep{HFetal06}.
In the \citet{HFetal06} simulations, Taurus-type molecular clouds 
\citep[masses $\sim 1.5\times10^4 M_\odot$,][]{Pietal10} are formed 
in colliding \hi\ flows and followed by nearly instantaneous star 
formation.
The comparisons in $\epsilon$ and star formation properties between 
the LMC and the Bridge indicate that at the higher N(\hi ) regime,
the lower metallicity of the Bridge is likely responsible for 
lower efficiencies of star and molecular cloud formation. 
At the lower N(\hi ) regime, Bridge's dynamic tidal environment 
may enhance distributed star formation.

Finally, we have further examined if the flatter slope 
seen in $\epsilon$ calculated using the Bridge YSOs also appears 
in fainter YSO candidates or HAeBe candidates.
Figure~\ref{fig:hi_mcblmc}c illustrates $\epsilon$ determined 
for all three kinds of young massive stars in the Bridge.
Unlike YSOs showing a flatter slope in $\epsilon$ toward low N(\hi ), 
fainter YSO candidates and 
HAeBe candidates exhibit a monotonically mild decrease.
This mild decrease might be attributed to quick dissipation 
of some clouds so that objects once formed in high 
N(HI) now have lower N(HI) surroundings.
In this case, the dissipation timescale has to be shorter
than the age of fainter YSO candidates and HAeBe candidates,
i.e., a few Myr (Tables~\ref{fysofits}--\ref{aebefits}).
We note that such a short dissipation timescale, if proven, 
is consistent with expectations from star formation through 
colliding flows.

\subsection{Star Formation in Individual Molecular Clouds}
\label{indivclouds}

To determine the mode of massive star formation, examine 
its progress in space and time, and assess if some might be 
triggered in the Bridge, we compare the underlying distributions 
of massive stars and YSOs in all eight molecular clouds detected 
from previous CO surveys \citep[Figure~\ref{fig:yso_pos}:][]{Muetal03a,MNetal06}.
Each cloud is discussed individually in the following subsections, 
except that Cloud H is part of Cloud G and hence is included in 
the discussion of Cloud G.
As aforementioned in Section 5.1 that the majority of YSOs in 
the Bridge were not covered in previous CO surveys; we have 
carried out a new CO survey of all YSOs and will discuss the 
results in a forthcoming paper.

\subsubsection{Molecular Cloud A}
\label{cldA}

\begin{figure}
\epsscale{0.8}
\plotone{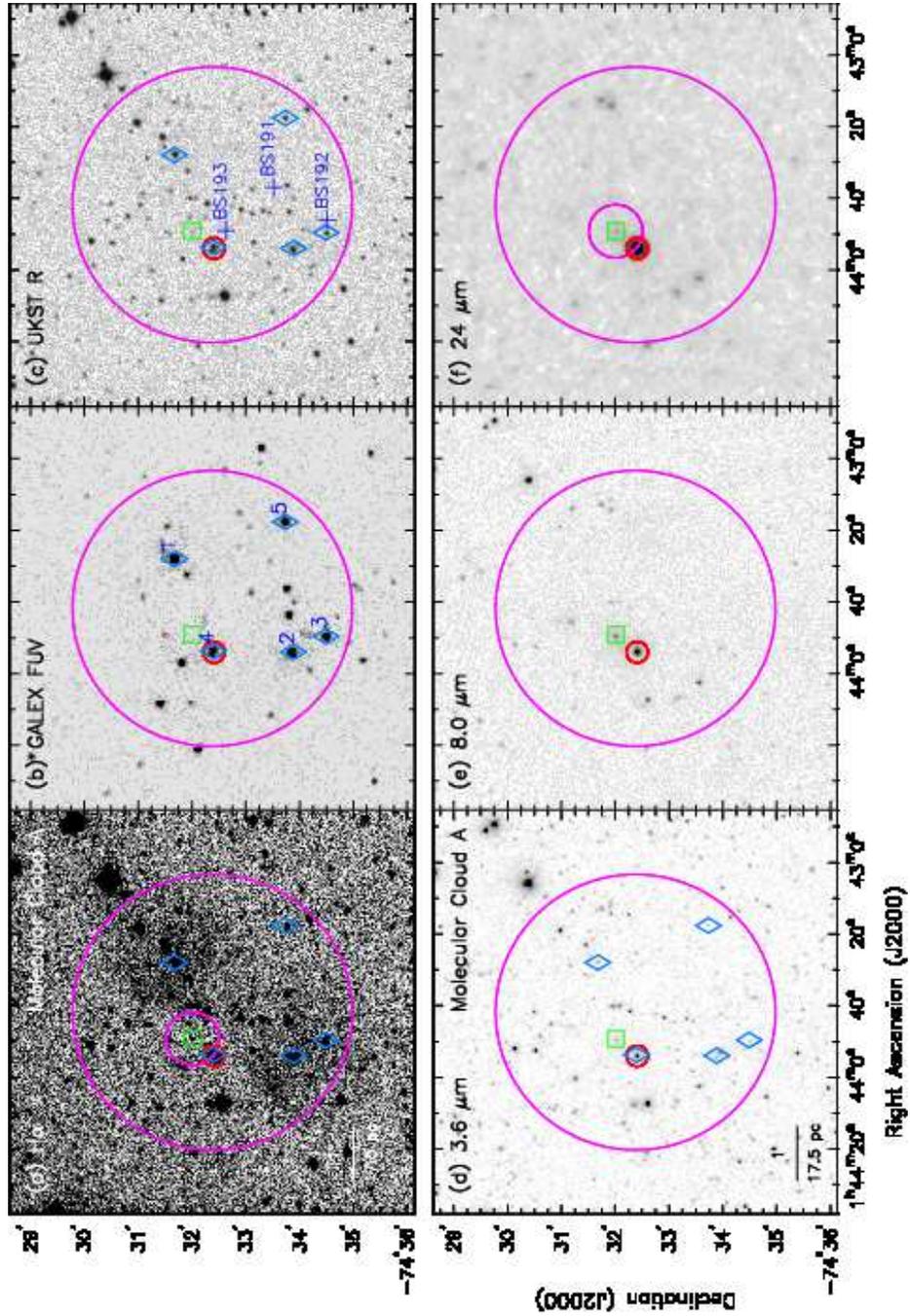}
\caption{(a-f) UKST \ha, {\it GALEX} FUV, UKST $R$, and {\it Spitzer} 3.6,
 8, and 24 \um\ images of Molecular Cloud A.  Positions of embedded YSOs
 (red circles),
 fainter YSO candidates (green squares), candidate HAeBe stars (cyan triangles), and
 five brightest FUV sources (blue diamonds) are marked.  The centers of stellar
 associations and clusters from \citet{BS95} are marked in blue pluses and
 their names are labeled. The position of 
 the CO J=1-0 detection is marked in large magenta circle with a size of
 5\farcm6, twice the beam size of the NANTEN telescope \citep{MNetal06}.
 High-resolution ASTE CO J=3-2 observations have been made toward Clouds A, B, C,
 E, G \citep{Muetal13}; the position of each CO 3-2 peak
 is marked with a smaller magenta circle 1\farcm0 in size.
 \label{fig:yso_coa}}
\end{figure}

Star formation has been occurring around Molecular Cloud A, as it is
located within a large stellar association BS191 and is associated
with two compact stellar clusters BS192 and BS193 \citep[nomenclature
from][]{BS95}.  The evidence of massive stars formed in the last 10
Myr is revealed by the presence of two faint, circular \hii\ regions,
each with size $\sim 100''$ or $\sim 27$ pc shown in the \ha\ image in
Figure~\ref{fig:yso_coa}a.  It has been suggested that the \hii\
regions are unlikely ionized by the aforementioned association and
clusters based on spatial separation \citep{MP07}.  Indeed, far-
and near-ultraviolet (FUV and NUV) images of the Bridge from the 
{\it GALEX} All-Sky Survey reveal three bright blue stars near the \hii\
region centers (Figure~\ref{fig:yso_coa}b).  These three stars are
most likely the ionizing stars as they are the brightest and also the
bluest within the \hii\ regions.  They have $FUV =$ 14.37, 14.39, and
15.28, and ($FUV-NUV$) $= -0.28$, $-0.17$, and $-0.28$, respectively
({\it GALEX} Data Release 6); these magnitudes and colors are
consistent with one to two B2V stars or up to one B1V star at 55 kpc,
the distance of the Bridge, with little extinction, i.e., $FUV =$
15.0, 14.2, to 13.9, respectively, and the same ($FUV-NUV$) $= -0.30$.
Thus, these \hii\ regions appear to be ionized by single or double B2
or B1 stars, each with stellar mass $\sim 10$, up to $13 M_\odot$.

The current massive star formation in Molecular Cloud A is revealed by
our {\it Spitzer} observations.  Two YSOs are identified,  an
embedded YSO \ysoai\ and a fainter YSO candidate
\ysoafi\ (Figure~\ref{fig:yso_coa}).  Both YSOs have mass estimates
$\sim 8 M_\odot$ (Tables~\ref{sedfits} and \ref{fysofits}), but show 
different stellar environments.  The embedded YSO is in a compact 
multiple system containing massive stars of similar masses at different 
apparent evolutionary stages.  As shown in Figures~\ref{fig:cmds} 
(Section 3.1.2) and \ref{fig:yso_coa}, the YSO, dominating the NIR 
and MIR light, is young with an estimated age of 0.11$\pm$0.06 Myr
(Table~\ref{fysofits}), while the FUV bright star, dominating the UV
and blue light, has $FUV =$ 15.75 and ($FUV-NUV$) $= -0.11$ that are
consistent with those of a B2-3V star (mass $\sim 8-10 M_\odot$), and
hence has at least reached the zero-age main-sequence (ZAMS) and is
older than a few Myr \citep{BM94}.  In contrast, the fainter YSO 
in the cloud appears to have only one massive source.  The fainter YSO
has an age estimate of 2.6$\pm$1.2 Myr, much more evolved than the
embedded YSO but still young enough to possess circumstellar dust 
and hence less likely reaching the ZAMS.  
This fainter YSO is at
the CO peak of Cloud A revealed by $\sim 20''$-resolution ASTE CO
observations \citep{Muetal13}.

Our examination of massive star and YSO populations in Molecular Cloud
A shows that a handful of early B (B1-3) stars in small groups formed
some 5--10 Myr ago.  Current (most recent few Myr) star formation
might have been triggered by the expansion of \hii\ regions as the
YSOs are found near the edge of the \hii\ regions.  Furthermore, the
most massive star formed in the current generation is $\sim 8
M_\odot$, less massive than $\lesssim 13 M_\odot$ in the previous
generation; this is consistent with the trend observed in Galactic and
LMC \hii\ regions that the being-triggered second generation  
is less massive the first
\citep{Poetal09,Fletal10,CCetal10}.  If the current star formation is 
indeed triggered and given that Cloud
A's low mass of $1\times10^3 M_\odot$ \citep{MNetal06} is similar to
the Taurus Cloud forming mostly low-mass stars, the star formation
mode in Cloud A most likely will remain as distributed, small groups 
of B stars rather than a large OB association.

\subsubsection{Molecular Cloud B}
\label{cldB}

\begin{figure}
\epsscale{0.8}
\plotone{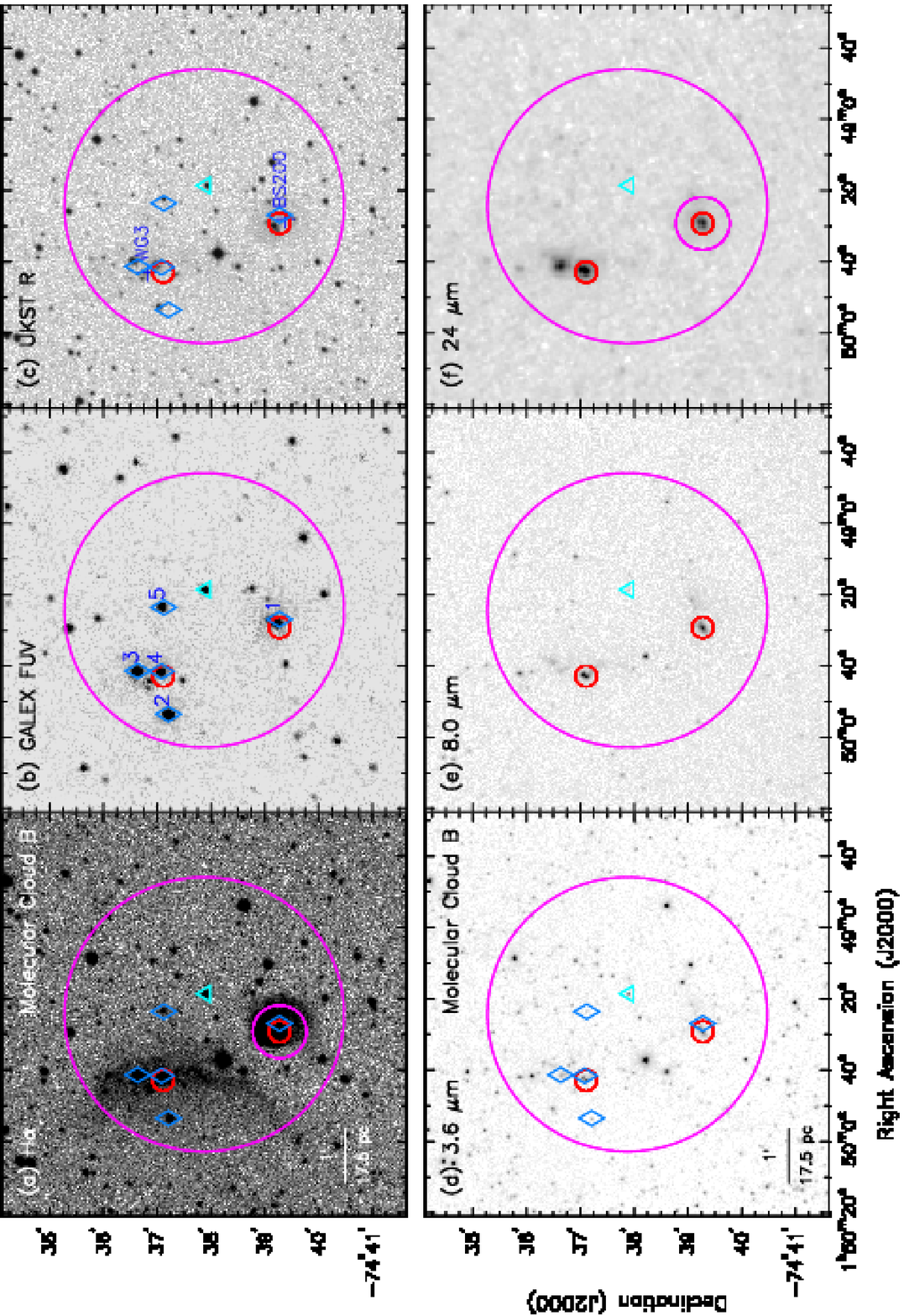}
\caption{Molecular Cloud B, with the same images and annotations as Fig~\ref{fig:yso_coa}.
 \label{fig:yso_cob}}
\end{figure}

Star formation has been occurring in Molecular Cloud B as it contains
a stellar association BS200 and a compact cluster WG3, and both are 
associated with nebular emission \citep{BS95}.  
The \ha\ image of Cloud B shows an arc filament overlapping with WG3 
and a compact bright \hii\ region spatially coincident with BS200
(Figure~\ref{fig:yso_cob}a).  It has been suggested that the ionizing
stars of these \hii\ regions are likely from these stellar association
and cluster \citep{MP07}, and the {\it GALEX} UV data reveal their nature and
location more precisely (Figure~\ref{fig:yso_cob}b).  The brightest
star is at the center of the compact bright \hii\ region coincident with 
BS200; it has
$FUV = 13.71$ and ($FUV-NUV$) $= -0.10$, consistent with a B1V star at
55 kpc.  The second and third brightest stars are in the interior and
within the arc filament respectively and have $FUV = 14.33$ and 14.51
and ($FUV-NUV$) $= -0.38$ and $-0.34$, consistent with a B1-2V star.
This spatial arrangement of UV stars, optical cluster, and \ha\ arc
shown in Figure~\ref{fig:yso_cob}a-c suggests that the second
brightest UV star has formed an asymmetric \hii\ region with bright
emission (arc) resulting from its expansion into the denser ISM
surrounding the compact cluster WG3.  A fraction of the \ha\ emission
in the arc may come from the third brightest UV star, apparently a
member of WG3, but unlikely from other cluster members since they are
much fainter in the UV and hence less massive.  Thus, the \hii\
regions in Cloud B, like in Cloud A, are ionized by single or double
B2 or B1 stars each with stellar mass $\sim 10$, up to $13 M_\odot$.

The current massive star formation in Molecular Cloud B is represented
by two embedded YSOs, \ysobi\ and \ysobii, and a HAeBe candidate,
\aebebi . 
Both YSOs have mass estimates
$\sim 8 M_\odot$ (Table~\ref{sedfits}), and both are in compact
($\lesssim 5''$, or $\sim 1.3$ pc) multiple systems that have massive
stars of similar masses at different evolutionary stages.  As shown in
Figure~\ref{fig:yso_cob}, YSO \ysobi, dominating the MIR light, is
young with an estimated age $\sim$ 0.1 Myr (Table~\ref{sedfits}),
while the companion UV bright star is the ionizing star of the compact
\hii\ region and hence at least a few Myr old.  The other YSO \ysobii\
has an estimated age $\sim$ 0.1 Myr, and the companion is a
non-embedded, UV bright star with $FUV =$ 15.51 and ($FUV-NUV$) $=
0.10$, comparable to a B3V but with a color redder by $\sim 0.3$ mag.
Note that this redder color could result from unresolved redder stars
in the vicinity, but would not be caused by extinction since the NUV
filter overlaps the 2200 \AA\ absorption peak such that
$E(FUV-NUV)/A_V=-0.17$, making an extincted source fainter and {\it
  bluer} \citep{CCM89}.  Lastly, the HAeBe candidate has estimated
mass $\sim 11 M_\odot$ and age $\sim$ 1.0 Myr (Table~\ref{aebefits}),
falling between the ages of MIR-selected YSOs and UV-bright ionizing
stars.

Our examination of massive stellar and YSO populations in Molecular Cloud
B shows that a few small groups of early B (B1-3) stars formed
3--10 Myr ago.  YSO \ysobii\ is found in the \ha\ arc, possibly
triggered by expansion of the older arc-shaped \hii\ region.  
Comparisons in their masses demonstrate the trend generally seen in 
triggered star formation, i.e., the most massive star formed in the 
current generation, $\sim 8 M_\odot$, is less massive than $\sim 13 M_\odot$
in the previous generation.  In contrast, the other YSO \ysobi\ is
located in the CO peak of Cloud B revealed by ASTE observations, in
which star formation just began in the last few Myr (as evidenced by
the bright compact \hii\ region).  Given that Cloud B has higher mass
of $3\times10^3 M_\odot$ \citep{MNetal06}, it is possible that the
compact system around YSO \ysobi\ can eventually form a more massive
cluster.

\subsubsection{Molecular Cloud C}

\begin{figure}
\epsscale{0.8}
\plotone{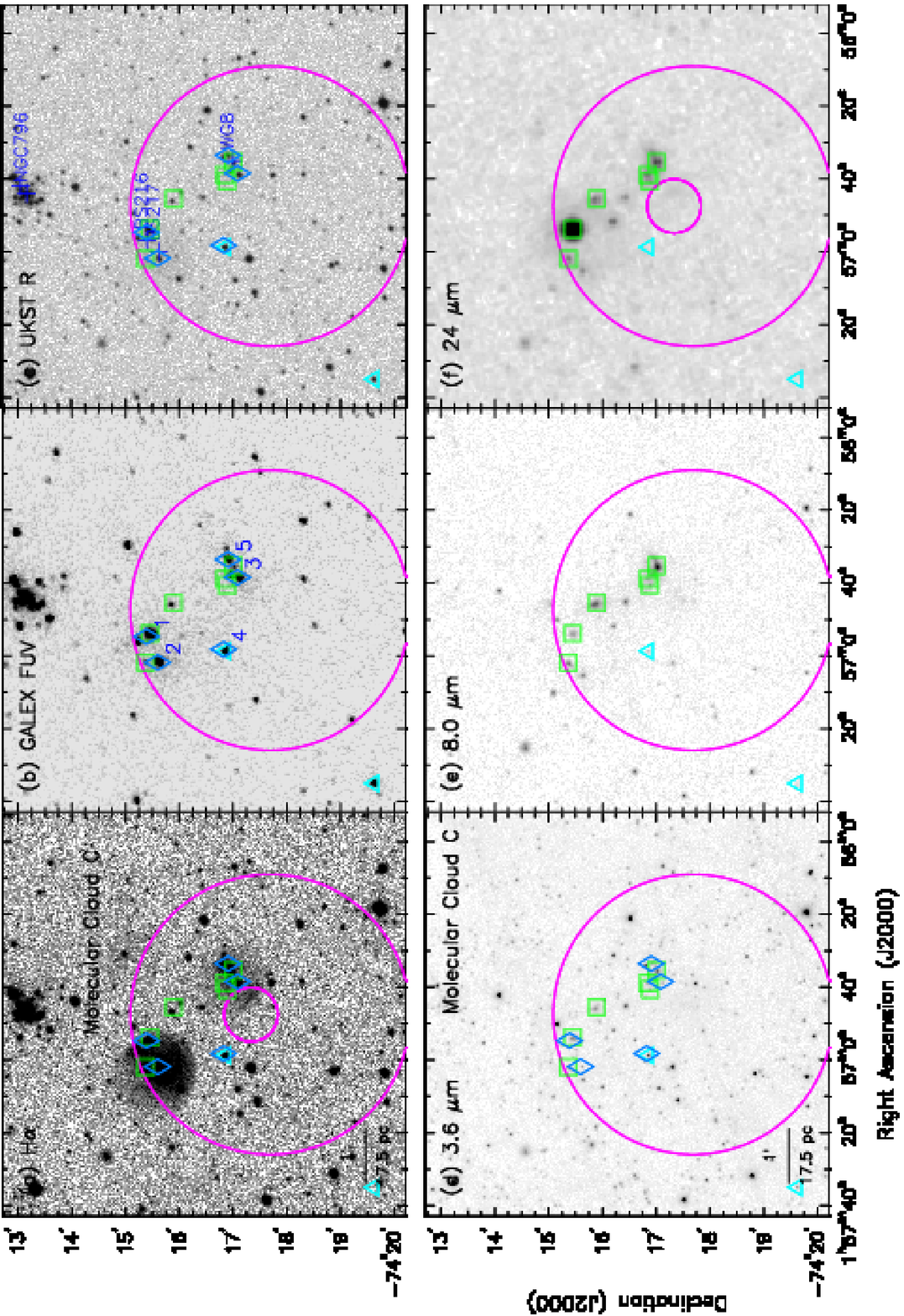}
\caption{Molecular Cloud C, with the same images and annotations as Fig~\ref{fig:yso_coa}.
 \label{fig:yso_coc}}
\end{figure}

Molecular Cloud C appears to be located along a string of active star
formation: from the compact blue cluster NGC\,796 at 55$''$ ($\sim 15$
pc) north to the cloud, to a large association BS217 containing a
compact cluster BS216 in the northern part of the cloud, to a small
association WG8 near the center of the cloud
(Figure~\ref{fig:yso_coc}c).  Four \ha\ blobs have been identified
within and north of Cloud C \citep{MP07}.  The north blob has
little diffuse emission and is spatially coincident with stars in
NGC\,796 (Figure~\ref{fig:yso_coc}a--c); indeed, the spectroscopic
study by \citet{Ahetal02} revealed a small equivalent width (3.89 \AA)
of \ha\ emission and suggested an age of $\gtrsim 6-10$ Myr.  The
other three \ha\ blobs appear to be typical \hii\ regions.
The bright compact \hii\ region at the northeast of the cloud, associated 
with BS216,
contains the two brightest UV stars in the cloud;
these stars have $FUV=14.20$ and 14.48 and ($FUV-NUV$) $= -0.10$ and
$-0.27$ respectively, consistent with B1-2V stars.  The second brightest 
compact
\hii\ region in the south near the cloud center, associated with WG8, 
contains the next
two UV-brightest stars; they have $FUV=14.83$ and 15.57, and
($FUV-NUV$) $= -0.11$ and $-0.10$ respectively, consistent with $\sim$
B2V stars.  Finally, the very faint \hii\ region at the northwest of the cloud 
appears to center on a blue star of $FUV=18.10$ and ($FUV-NUV$) 
$= -0.07$, consistent with a B6V star.  These three \hii\ regions are ionized 
by small groups of early B stars or a single mid B star.

Despite the string of active star formation, there is no embedded MIR-bright 
YSO but there is a group of six fainter YSO candidates found in 
Molecular Cloud C (Figure~\ref{fig:yso_coc}).  
As discussed in Section \ref{fainteryso}, these six candidates are most likely 
bona fide YSOs based on examination of their multi-wavelength SEDs and 
images, and further confirmed by IRS spectra of the two with the brightest 
24 \um\ emission among them, \ysocfi\ and \ysocfii.
All six YSOs are found within or near the edge of the two bright \hii\ regions: 
three in the south-central \hii\ region,
two in the northeast \hii\ region and one near its edge
(Figure~\ref{fig:yso_coc}a).  The three YSOs in south-central \hii\
region have lower masses, two $\sim 4-5 M_\odot$ and one $\sim 8
M_\odot$; this group have younger ages with two $\lesssim 0.1$ Myr 
and one $\sim 2$ Myr.  The other three YSOs in and near the 
northeast \hii\ region have higher masses, all three $\sim 7-11 M_\odot$, 
including the two most massive ones among the fainter YSO candidates; 
this group have older ages with one $\gtrsim 0.1$ Myr and two $\sim 3$ Myr 
(Table~\ref{fysofits}).

Compared to other Bridge clouds, star formation around Cloud C appears
more compactly clustered during the last 10 Myr, i.e., from the older
cluster NGC\,796 to young clusters of UV stars and YSOs found in two
compact bright \hii\ regions.  A clustered mode of massive star
formation is usually found in high-mass molecular clouds; Cloud C does
provide a preferred environment owing to its high mass, i.e.,
$7\times10^3 M_\odot$ which is the highest among all eight clouds
\citep{MNetal06}, as well as its location in a region with the
highest N(\hi ) in the Bridge.  There appears a sequence of star
formation from north to south, i.e., from the $\sim$ 10-Myr-old
NGC\,796, to the south-central \hii\ region containing two YSOs
$\lesssim 0.1$ Myr.  Although at present this south-central \hii\
region has UV stars and YSOs with the lowest masses among the three
clusters, it is near the CO peak of Cloud C revealed by ASTE
observations, and may possibly develop a massive cluster similar to
its northern neighbors.

\subsubsection{Molecular Cloud D}

\begin{figure}
\epsscale{0.8}
\plotone{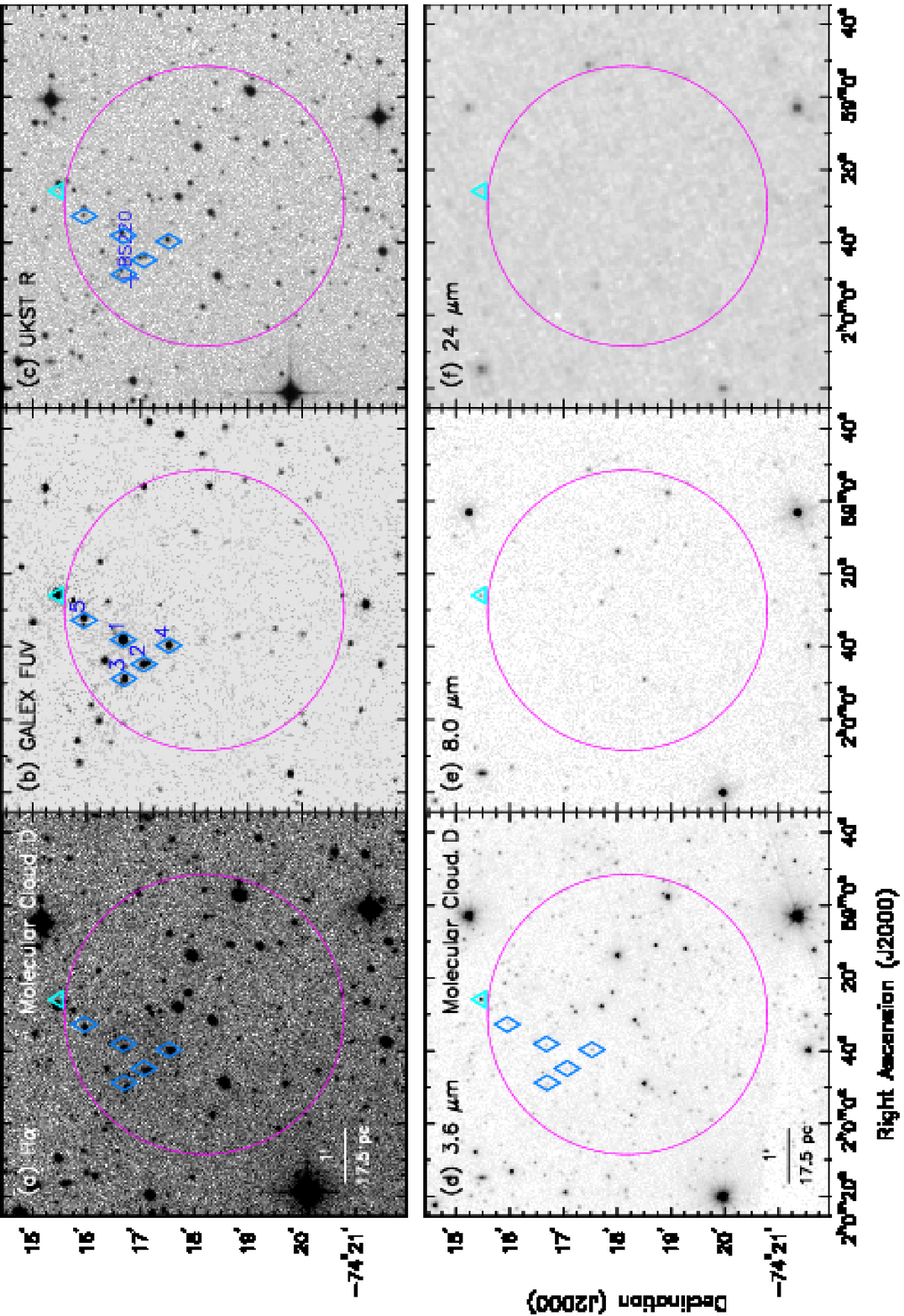}
\caption{Molecular Cloud D, with the same images and annotations as Fig~\ref{fig:yso_coa}.
 \label{fig:yso_cod}}
\end{figure}

Molecular Cloud D overlaps with a stellar association BS220
(Figure~\ref{fig:yso_cod}), and shows no sign of massive star
formation in the last 10 Myr due to the lack of \hii\ regions.
Examination on {\it GALEX} UV and {\it Spitzer} MIR data finds 
neither blue stars nor YSOs in Cloud D.  
This cloud appears to be quiescent in star formation, even though its mass, 
$1\times10^3 M_\odot$, is comparable to other clouds in the Bridge 
\citep{MNetal06}.  
It is possible that Cloud D is just formed and star formation has not yet begun.  
Alternatively, it is 
also possibly a false CO detection since it has the lowest peak main-beam
temperature, 10 mK, barely higher than the rms noise temperature of 9 mK
\citep{MNetal06}.  Deep CO observations of Cloud D are needed to
verify the detection.

\subsubsection{Molecular Cloud E}

\begin{figure}
\epsscale{0.8}
\plotone{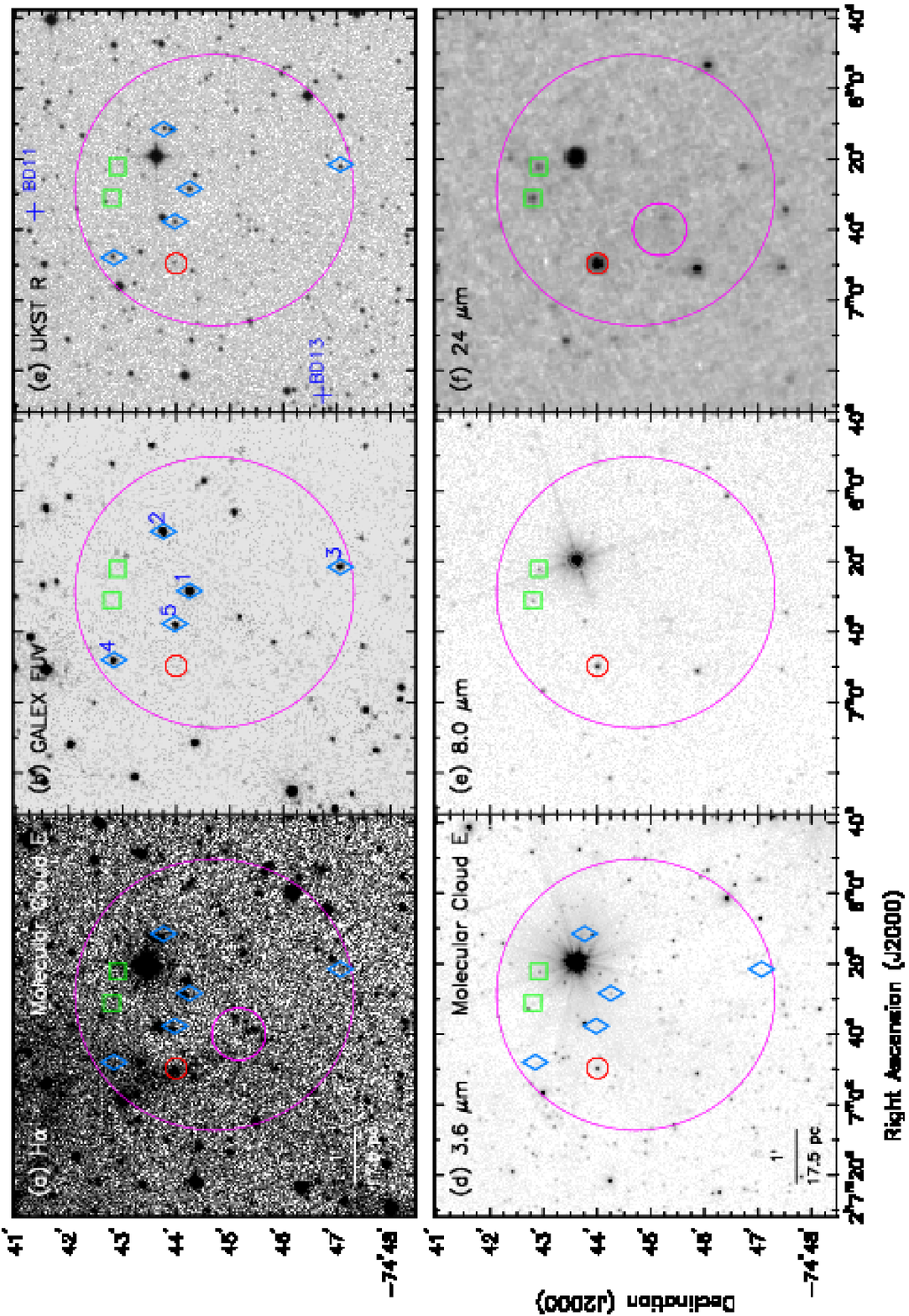}
\caption{Molecular Cloud E, with the same images and annotations as Fig~\ref{fig:yso_coa}.
 \label{fig:yso_coe}}
\end{figure}

Three large stellar associations are located to the east of Molecular
Cloud E, with two of them, BD11 and BD13 \citep{BD00}, overlapping on
the Cloud's north and east edge (Figure~\ref{fig:yso_coe}c).  Evidence
of fairly recent (5-10 Myr old) star formation is shown by the \hii\
shell DEM\,S\,171 that abuts the east side of Cloud E.  DEM\,S\,171 is
one of the few large \ha\ structures in the Bridge \citep{MJ86,MP07},
with part of the shell rims shown in Figure~\ref{fig:yso_coe}.  It has
a size of $\sim 8'\times8'$, or $130\times130$ pc$^2$, and the shell
expansion is suggested to be driven by winds of a Wolf-Rayet star or
supernova explosion \citep{Gretal01}.  Compared to the plausible
Wolf-Rayet star responsible for DEM\,S\,171, the brightest and bluest
UV stars in Cloud E have $FUV=15.64$, and 15.90, and
($FUV-NUV$)$=-0.14$ and $-0.32$ respectively, consistent with B2-3V
stars.  These two stars show no identifiable \hii\ regions but
only some faint diffuse \ha\ emission in their surroundings
(Figure~\ref{fig:yso_coe}a), suggesting that they have dispersed their
\hii\ regions, and are relatively old.

Current star formation in Molecular Cloud E is represented by one
embedded YSO and two fainter YSOs (Figure~\ref{fig:yso_coe}).  
The three YSOs, \ysoei , \ysoefi, and \ysoefii , have mass estimates 
$\sim 5-6 M_\odot$ (Tables \ref{sedfits} and \ref{fysofits}), not as
massive as the $\sim 8-10 M_\odot$ found in other molecular clouds in
the Bridge.  YSO \ysoei\ is resolved into a compact multiple that is
most clearly visible in the B-band image.  
This compact multiple consists of only embedded sources, unlike YSOs in 
Clouds A-C that are frequently found in pairs with UV-bright early B stars.
YSOs in Cloud E are all found near diffuse \ha\ emission.
YSO \ysoei\ and fainter YSO \ysoefii\ are located at the edges of the 
\hii\ shell DEM\,S\,171,  so is the peak of the ASTE CO emission
(Figure~\ref{fig:yso_coe}a); their formation might have
been triggered by the shell expansion.  The remaining fainter YSO
\ysoefi\ is near the edge of faint diffuse \ha\ emission, though it is
difficult to distinguish if this diffuse emission is associated with the 
UV-bright star or from the PSF of the nearby foreground star that is 
saturated in R-band but not detected in UV images (Figure~\ref{fig:yso_coe}).

Our examination of massive star and YSO populations in Molecular Cloud
E shows that a sparse distribution of a few early-to-mid B (B2-3 and
later) stars have formed in the recent past $\gtrsim 10$ Myr, as well
as a few more recently.  Given that YSOs are on the edge of the \hii\
shells and not as massive as the ionizing star, the current star
formation is consistent with a triggered star formation scenario.  In
addition, these YSOs are the dominant sources in their groups, in
contrast to pairs of high-mass stars and YSOs found in other clouds,
suggesting that massive star formation may not have been active in
Cloud E until triggered by shell expansion into the molecular cloud or
gas accumulation in the shell \citep[e.g., the collect and collapse
scenario,][]{El98}.  Given Cloud E's low mass of $1\times10^3 M_\odot$
\citep{MNetal06} and spread-out distribution of YSOs along the shell
rims, the star formation will most likely only ever amount to
distributed small groups of early-to-mid B stars.

\subsubsection{Molecular Cloud F}

\begin{figure}
\epsscale{0.8}
\plotone{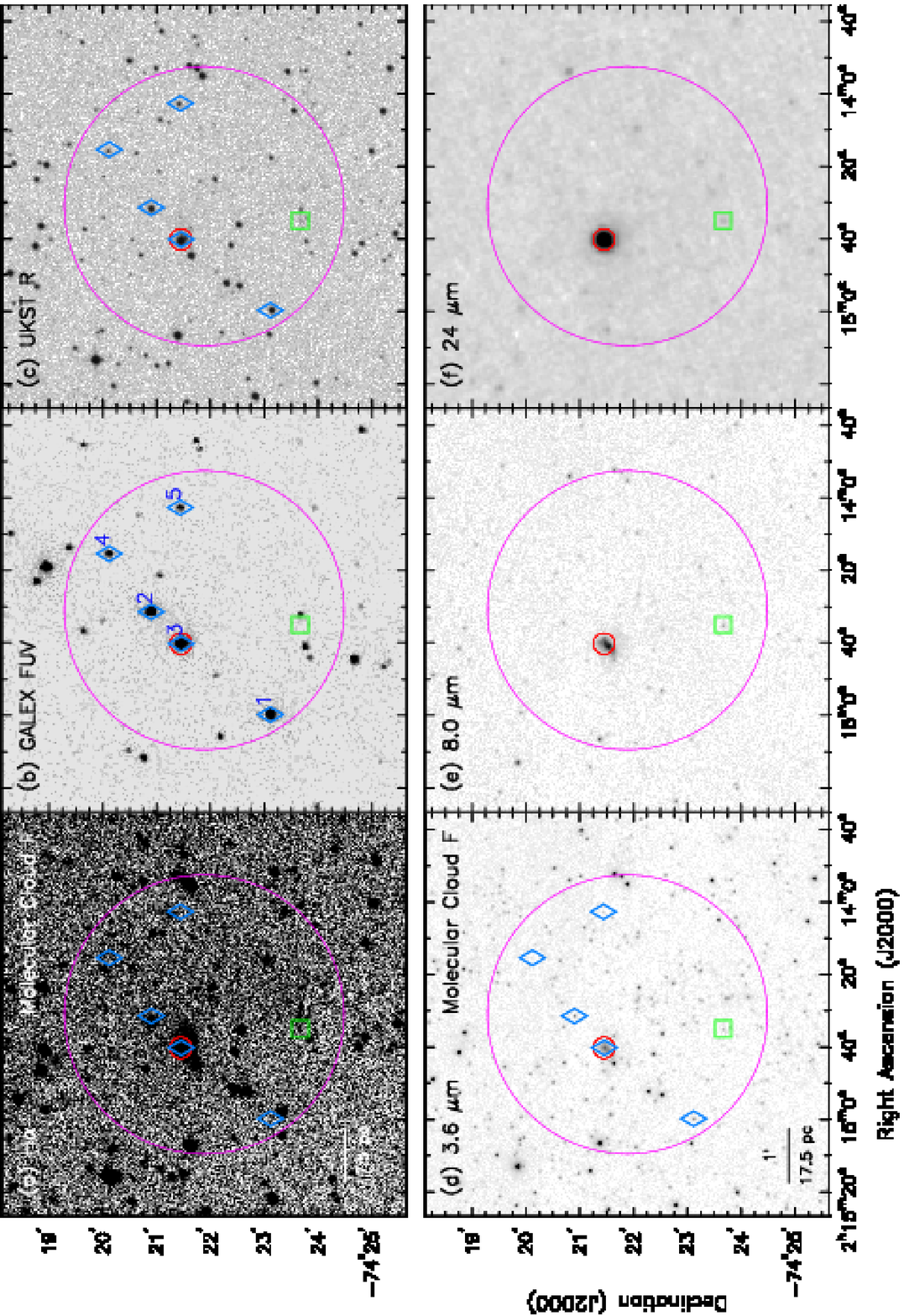}
\caption{Molecular Cloud F, with the same images and annotations as Fig~\ref{fig:yso_coa}.
 \label{fig:yso_cof}}
\end{figure}

Molecular Cloud F appears to be in a relatively quiescent environment
as there are no cataloged stellar associations or clusters.  Evidence
of relatively recent massive stars is shown by diffuse, faint \ha\
emission within the cloud in the \ha\ image
(Figure~\ref{fig:yso_cof}a).  This \ha\ emission surrounds three
bright blue stars revealed by {\it GALEX} UV images
(Figure~\ref{fig:yso_cof}b).  They have $FUV = 14.50$, 14.51, and
14.53, and ($FUV-NUV$) $= -0.16$, $-0.11$, and $-0.14$, respectively,
consistent with B1-2V stars.  Some or all three blue stars are likely
the ionizing sources this diffuse, low-brightness \ha\ emission, and
they would be $\gtrsim 10$ Myr old.

The current star formation in Cloud F is represented by one embedded
YSO and one fainter YSO (Figure~\ref{fig:yso_cof}).  YSO \ysofi\ has
an estimated mass $\gtrsim 10 M_\odot$, among  the most massive YSOs 
in the Bridge (Table~\ref{sedfits}).  It is in a compact system of
massive stars at different evolutionary stages, with the other massive
source being the second brightest UV star $\sim$ B1-2V, corresponding
to $\sim 13-10 M_\odot$.  The fainter YSO \ysoffi\ has a mass $\sim 4
M_\odot$ (Table~\ref{fysofits}), and appears as a single source.

The examination of massive star and YSO populations in Molecular Cloud
F shows that distributed early B (B1-2) stars have formed in the
recent past ($\gtrsim 10$ Myr) as well as in the current.  There is no
dense clustering of higher mass stars/YSOs, not unexpected from Cloud
F's low mass of $1\times10^3 M_\odot$ \citep{MNetal06}.  The formation
of fainter YSO \ysoffi\ might have been affected by the stellar energy
feedback as it is within the diffuse, low-brightness \ha\ emission, 
though higher resolution of \ha\ and CO maps are needed to assess 
such a possibility.

\subsubsection{Molecular Clouds H$+$G}

\begin{figure}
\epsscale{0.8}
\plotone{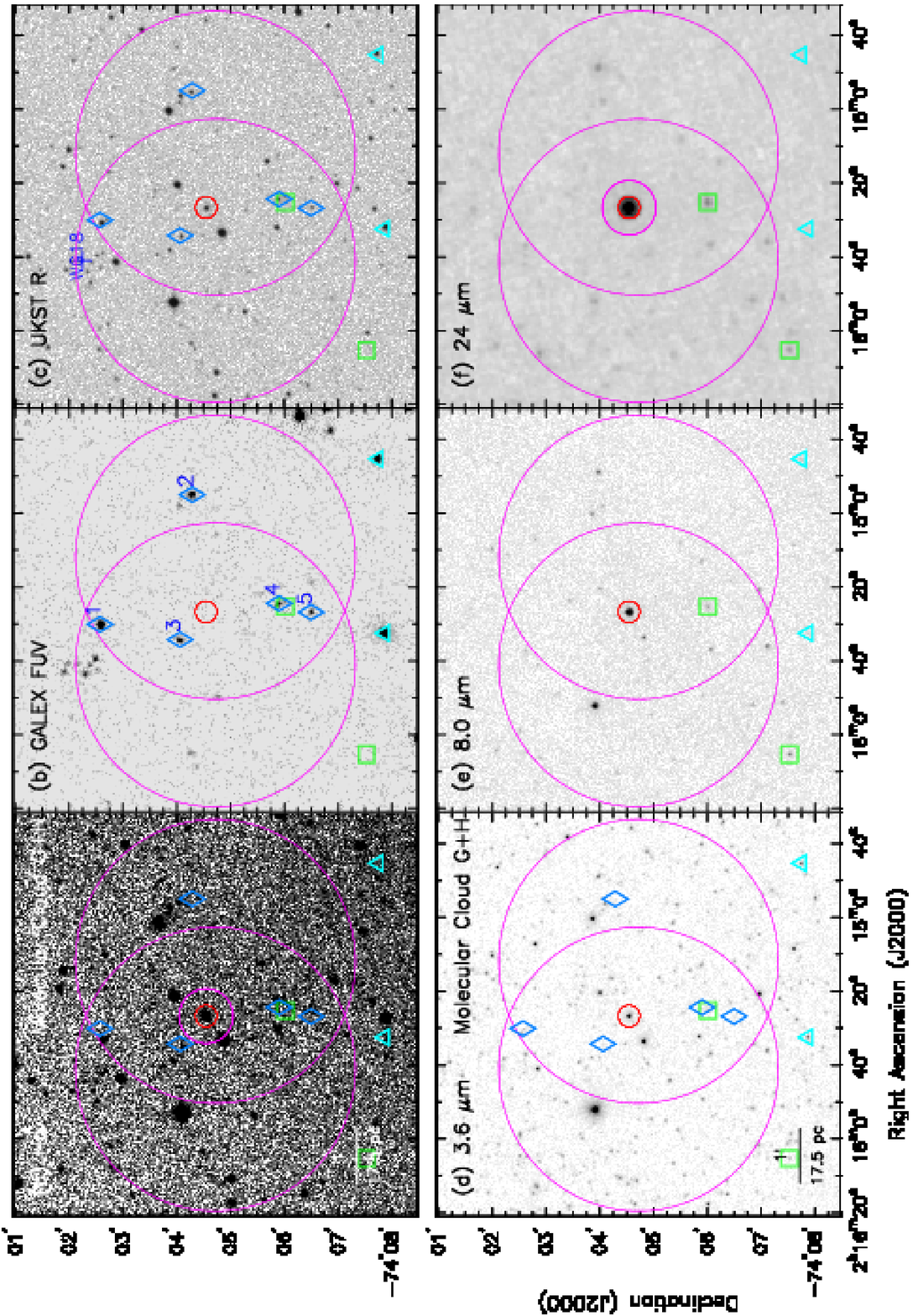}
\caption{Molecular Cloud G, with the same images and annotations as Fig~\ref{fig:yso_coa}.
 \label{fig:yso_cog}}
\end{figure}

Molecular Cloud H is part of Cloud G and hence included in 
its discussion. 
Molecular Cloud G abuts two stellar associations, a large one BD33 on
the west and a small one WG18 on the north east.  The \ha\ image shows
diffuse, low-brightness emission within the cloud and slightly
brighter emission to the southeast (Figure~\ref{fig:yso_cog}).  The
two brightest blue stars in the cloud revealed by {\it GALEX} have
$FUV = 16.28$ and 16.94, and ($FUV-NUV$)=$-0.01$ and 0.03,
respectively, consistent with $\sim$ B3V stars.  They are plausibly the
ionizing sources of this diffuse \ha\ emission, and would be
$\gtrsim 10$ Myr old.

The current star formation in Cloud G is revealed by one YSO and one
fainter YSO (Figure~\ref{fig:yso_cog}).  YSO \ysogi\ has an estimated
mass $\sim 9 M_\odot$ and an age $\le 0.1$ Myr
(Table~\ref{sedfits}).  It is a single isolated source, unlike the
majority of YSOs in other clouds are found in pairs with UV-bright
early B stars; the massive star formation in Cloud G appears to be just
beginning.  The fainter YSO \ysogfi\ has a mass $\sim 5 M_\odot$.
This YSO is within the diffuse faint \ha\ emission, but association is
unclear.

The massive star and YSO contents of Molecular Cloud G show that
distributed early-to-mid B ($\sim$ B3) stars have formed in the recent past
$\gtrsim 10$ Myr as well as are currently forming.  Although at
present there is only one $\sim 9 M_\odot$ YSO, it is possibly the
first newly formed massive star in a cluster as it is at the ASTE CO
peak and Cloud G has the second highest mass of $5\times10^3 M_\odot$
in the Bridge \citep{MNetal06}.  Massive star formation appears to
be just beginning in Cloud G.

\subsection{Star Formation Efficiency and Rate}

The relation between gas surface density and SFR is one of the most critical
links between star formation and galaxy evolution, and also the most widely
used relation in extragalactic astronomy.
The observed relation, the ``Schmidt-Kennicutt (S-K) law'',  is tight
when properties are averaged on kpc scales 
\citep{Ke89,Ke98,Keetal09}, 
but appears to break down at scales $\le$ a few hundred pc, as shown 
in recent high-resolution studies of M\,33 
\citep{Onetal10,Scetal10}.
It is conceivable that the SFR and GMC content averaged over too small 
a surface area do not adequately sample GMCs at different evolutionary 
stages, and thus do not show a good relation.
As demonstrated in our LMC study, while individual GMCs show different 
evolutionary stages (based on their association with different advancement 
of star formation activities) and have different SFRs, the S-K relation is 
observed when averaging over several GMCs in large \hii\ complexes  
of sizes $\sim 200$ pc \citep{CCetal10}.
On the other hand, the SFR of a GMC also depends on SFE.
Much lower SFRs than expected from the S-K relation have been observed
in the outer disks of spiral and dwarf galaxies, implying that 
star formation may be affected by environmental factors \citep{Bietal10}.
The proximity of the Bridge provides an excellent laboratory to use 
resolved stellar and gas contents to critically examine environmental 
effects, specifically metallicity and tidal effects, on star formation.

Applying the same method as our study of massive YSOs in LMC \hii\
regions \citep{CCetal10}, we assess the ``instantaneous'' SFE, 
SFE$_{\rm YSO}$, of the molecular clouds in the Bridge
using the known massive YSO content.
The total mass of the current
star formation, $M^{\rm total}_{\rm YSO}$, can be estimated using the
number of YSOs within different mass bins and assuming a
Salpeter's stellar initial mass function (IMF).  The number of YSOs in individual
molecular clouds is too small to infer $M^{\rm total}_{\rm YSO}$ with
reasonable uncertainties, however, the total number of YSOs in all 7
molecular clouds can be used to assess an average SFE$_{\rm YSO}$ for
the collection of clouds, albeit averaging over different evolutionary
stages (recall Section 5.2).  $M^{\rm total}_{\rm YSO}$ of the 7
molecular clouds is calculated by integrating the IMF from lower to 
upper mass limits, $M_l$ and $M_u$.
We use the highest mass of YSO observed for $M_u$ and 1 $M_\odot$
for $M_l$; the adoption of 1 $M_\odot$ is to facilitate comparisons
with other work as it is commonly used. 
$M^{\rm total}_{\rm YSO}$ is estimated
excluding fainter YSOs, since except for the two confirmed by IRS
spectra, their identification as YSOs is less certain.  The $M^{\rm
  total}_{\rm YSO}$ is then divided by the sum of cloud masses to
obtain SFE$_{\rm YSO}$, and these three quantities are listed in
Table~\ref{sfr}.  
Also listed are uncertainties in $M^{\rm total}_{\rm YSO}$; they are 
directly related to the uncertainties in mass estimates for individual 
massive YSOs used in number counts.  
Such uncertainties are thus estimated using the largest and smallest 
mass ranges covered by these YSOs.  
This exercise yields a dimensionless efficiency (mass
ratio) of 0.015$^{+0.009}_{-0.004}$.

To determine the current star formation {\em rate}, SFR$_{\rm YSO}$,
from $M^{\rm total}_{\rm YSO}$ requires a timescale.  While the age
of each massive YSO can be constrained using its evolutionary stage,
uncertainties of high-mass YSO accretion models make it difficult to
set this age more precisely than $\lesssim 1$ Myr.  On the other hand,
if we presume that all high- and intermediate-mass YSOs within
molecular clouds are currently forming in a burst, the formation
timescale for a cluster or association may be more relevant to the 
{\em cloud} efficiency than the age of any individual protostar.
Carrying on with this assumption, we adopt a cluster formation time
of $\sim 1$ Myr \citep[e.g.,][]{BBV03} and derive SFR$_{\rm YSO}$ of
2.9$\times$10$^{-4}$ $M_\odot$~yr$^{-1}$ and SFR$_{\rm YSO}$ per
surface area, $\Sigma_{\rm SFR YSO}$, of 7.6$\times$10$^{-3}$
$M_\odot$~yr$^{-1}$~kpc$^{-2}$ (Table~\ref{sfr}).

For comparison, we have also estimated the SFR averaged over the last
$\sim 10$ Myr,  commonly derived from \ha\ and/or 24 \um\ fluxes 
calibrated on an ensemble of \hii\ regions \citep[e.g.,][]{Ke98}.
The average SFR of the Bridge is estimated using the integrated 24 \um\ 
flux with the prescription of \citet{Caetal07}:
\begin{equation}
SFR_{24}(M_\odot~{\rm yr}^{-1}) = 
1.27\times10^{-38}[L(24~\mu m)]^{0.8850},
\end{equation}
where $L$(24 \um ) is the 24 \um\ luminosity in ergs~s$^{-1}$.  To
measure the 24 \um\ luminosity, we use an aperture size of 5\farcm2,
twice the beam size of the NANTEN CO observations since the molecular
clouds were not resolved; the largest uncertainties come from flux
calibration, $\sim 10$\% (MIPS Data Handbook).  The 24 \um\
luminosity, SFR$_{24}$, and SFR$_{24}$ per surface area, $\Sigma_{\rm
  SFR 24}$, are given in Table~\ref{sfr}.  The Bridge has $\Sigma_{\rm
  SFR 24}=5.2\times10^{-4} M_\odot$~yr$^{-1}$~kpc$^{-2}$, much
  lower than $\Sigma_{\rm SFR YSO}=7.6\times10^{-3}$ by a factor of 15.  
This discrepancy is consistent with our previous analysis of LMC regions, 
in which we found that below a threshold gas
surface density of $\sim$200$M_\odot$~pc$^{-2}$, star formation rates
derived from detailed YSO analysis 
frequently exceed the rates
calculated from integrated 24 \um\ and H$\alpha$ emission
\citep{CCetal10,Inetal08}.  We attributed this to the ability of
detailed YSO analysis to better account for variations in cloud
evolutionary state and stellar mass function than the integrated
measures.

\begin{figure}
\plotone{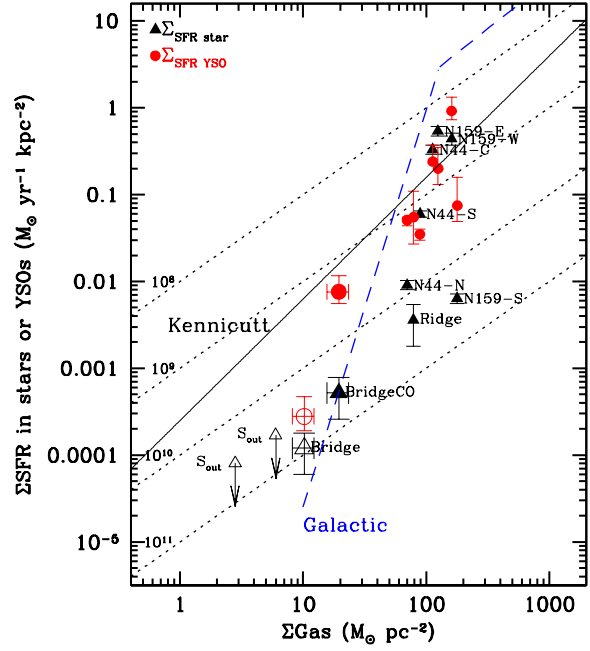}
\caption{
Relation between the SFR per unit area and gas density of 
 all molecular clouds (i.e. locations of CO J=1-0 detection; solid symbols
 with a label ``BridgeCO'') 
 and the entire studied area (open symbols with a label ``Bridge'') in the Bridge. 
 For comparison with the Bridge molecular clouds, also plotted are
 SFRs estimated for the molecular ridge (Ridge) and six
 individual GMCs in \hii\ complexes in the LMC \citep{Inetal08,CCetal10}.
 Each region has two estimated SFRs using different tracers, $\Sigma_{\rm SFR YSO}$ 
 (circles) and $\Sigma_{{\rm SFR star}}$ (triangles labeled by the 
 region name); in the Bridge, $\Sigma_{{\rm SFR star}} = \Sigma_{{\rm SFR 24}}$ while
 for other regions $\Sigma_{{\rm SFR star}}$ are estimated using integrated \ha +24 \um\
 fluxes.
 The solid line is the S-K relation, the dotted lines correspond to
 gas depletion timescales at constant SFRs from 10$^8$ to 10$^{11}$ yr,
 and the dashed lines mark the broken power law derived from Galactic
 star-forming regions \citep{Heidermanetal10}.
 GMCs without prominent \hii\ regions, i.e., the Bridge, Ridge, and two 
 LMC GMCs N\,44-N and N\,159-S have 
 $\Sigma_{{\rm SFR star}}$
 11--56 times smaller than expected from the S-K relation, but their
 $\Sigma_{\rm SFR YSO}$ are in better agreement with the S-K relation.
 Finally, the SFR estimated with FUV and
 N(\hi ) for the outer disks of 17 nearby spiral galaxies \citep{Bietal10}
 are plotted and labeled as S$_{\rm out}$.
 \label{fig:sfr}}
\end{figure}

Within the limitations of the sparse CO measurements, we can determine
where the Bridge clouds lies relative to the S-K 
relation.
The SFR per surface area expected from the S-K relation is 
\begin{equation}
\Sigma_{\rm SFR Gas}(M_\odot~{\rm yr}^{-1}~{\rm kpc}^{-2}) =
 2.5\times10^{-4}(\frac{\Sigma_{Gas}}{M_\odot~{\rm pc}^{-2}})^{1.4}
\end{equation}
where $\Sigma_{\rm Gas}$ is the sum of molecular and atomic surface densities, 
$\Sigma_{\rm H2} + \Sigma_{\sc HI}$ \citep{Ke98}.
The average $\Sigma_{\rm H2}$ for the Bridge clouds is estimated from 
the CO intensities given in \citet{MNetal06} and a CO-to-H$_2$ 
conversion factor $X_{\rm CO} = 1.4\times10^{21}$ cm$^{-2}$
(K km~s$^{-1}$)$^{-1}$ used in that paper.
The average $\Sigma_{\sc HI}$ is measured from \hi\ maps \citep{Muetal03b}.
Then the expected SFR from the total gas surface density is given
in Table~\ref{sfr}. 
The largest uncertainties come from the range of $X_{\rm CO} = 
1.3^{+1.6}_{-0.8}~10^{21}$ cm$^{-2}$ (K km~s$^{-1}$)$^{-1}$ found 
in the Bridge \citep{Muetal10}; since $\Sigma_{\rm H2}$ comprises 
5--17\% $\Sigma_{\rm Gas}$, the associated uncertainty in 
$\Sigma_{\rm SFR Gas}$ is thus $< 20$\%.
As shown in Figure~\ref{fig:sfr}, the Bridge has $\Sigma_{\rm SFR Gas}
\sim 31$ times $\Sigma_{\rm SFR 24}$, while $\Sigma_{\rm SFR YSO}$ 
is in agreement with $\Sigma_{\rm SFR Gas}$ within a factor of 2.
If $\Sigma_{\rm H2}$ found in the seven molecular clouds is 
representative for all clouds associated with YSOs in the Bridge,
using the whole sample of YSOs shows a similar trend that SFRs
estimated with a comprehensive YSO inventory agree with SFRs expected 
from the S-K relation and they are much higher than inferred from 
integrated 24 \um\ luminosities.
This discrepency between $\Sigma_{\rm SFR 24}$ and $\Sigma_{\rm SFR YSO}$ 
is also seen in some GMCs in the LMC, namely those without bright \hii\ 
regions \citep{Inetal08,CCetal10}.
For such GMCs in the LMC we attributed the derivation of lower 
$\Sigma_{\rm SFR 24}$ to their lower luminosity-to-mass ($L/M$) ratios 
as the star formation mostly occurred in lower-mass or less rich clusters 
that do not fully sample the high-mass end (with high $L/M$) of 
the stellar IMF, while the prescription in \citet[][i.e., 
Equation (1)]{Caetal07} is based on rich clusters with high $L/M$.
It is not surprising that the Bridge follows the same trend, since
its star formation is dominated by small, low-mass clusters 
evidenced by the scarcity of bright, large \hii\ regions \citep{MP07} 
and spread-out distribution of intermediate- and high-mass YSOs.
It is worth noting that at lower $\Sigma_{\rm Gas}$, 
$\Sigma_{\rm SFR YSO}$ of molecular clouds in the Bridge is 
still in agreement with the S-K relation.

Finally, we examine current star formation and estimate SFE and SFR 
for the Bridge as a whole.
The total mass of current star formation can be estimated using
the YSO content.  
The lowest mass bin used for counting depends on the photometric
completeness of infrared sources in the Bridge.
We have constructed a mass function of the YSOs and found moderate
incompleteness at masses $< 6 M_\odot$ (Figure~\ref{fig:haebe_hist}a); 
thus, only masses $\ge 6 M_\odot$ are used in the counting.
$M^{\rm total}_{\rm YSO}$, SFR$_{\rm YSO}$, and $\Sigma_{\rm SFR YSO}$
are estimated for the entire Bridge and listed in Table~\ref{sfr}.
For comparison, $\Sigma_{\rm SFR 24}$ and $\Sigma_{\rm SFR Gas}$ are 
also estimated using $L$(24 \um ) and $\Sigma_{\sc HI}$ of the entire
Bridge and given in Table~\ref{sfr}.
As shown in Figure~\ref{fig:sfr}, $\Sigma_{\rm SFR YSO}$ of the 
whole Bridge drops by a factor of 27 compared to 
$\Sigma_{\rm SFR YSO}$ of {\it all molecular clouds} in the Bridge,
while there is less than a factor of 3 difference in the predicted
$\Sigma_{\rm SFR Gas}$ from $\Sigma_{\rm Gas}$ in all clouds and
compared to the entire Bridge area.
If all YSOs in the Bridge are formed in molecular clouds, this 
much lower $\Sigma_{\rm SFR YSO}$ of the whole area would indicate
a much lower fraction of converting \hi\ to molecular gas, consistent
with that expected for low-metallicity environment \citep{KMT09b}. 
We have further examined if SFRs in molecular clouds in the Bridge 
are extremely low by comparing them to the $\Sigma_{\rm Gas}$-$\Sigma_{\rm SFR}$ 
relation derived using resolved YSOs and massive clumps in star-forming 
regions in the Galaxy \citep{Heidermanetal10}.
The \citet{Heidermanetal10} study demonstrated a much higher 
$\Sigma_{\rm SFR}$ than that expected from the S-K relation
(as can be seen in Figure~\ref{fig:sfr}).
Despite the very low $\Sigma_{\rm Gas}$ that is not covered in the 
Galactic sample, molecular clouds in the Bridge show a higher 
$\Sigma_{\rm SFR}$ than expected from the Galactic relation, favoring 
that the low $\Sigma_{\rm SFR}$ in the entire Bridge is more likely 
attributed to a low efficiency in cloud formation.
 
A similarly low SFR is found in the recent study of the outer disks 
of spiral and dwarf galaxies, using FUV as SFR tracer
and \hi\ for gas density \citep{Bietal10}.
For comparison, we have used their data of spiral galaxies to estimate 
the average $\Sigma_{\rm SFR FUV}$ for different N(\hi ) and plotted 
in Figure~\ref{fig:sfr}, though only upper limits can be obtained 
since data with S/N $< 3 \sigma$ are not included in their table.
These $\Sigma_{\rm SFR FUV}$ of the outer disks of spirals and 
$\Sigma_{\rm SFR YSO}$ of the whole area of the Bridge lie near
that expected for long gas depletion time of $\sim 10^{10-11}$ yr.
If SFRs in these outer disks have been at similar level in the 
last 100 Myr, their low $\Sigma_{\rm SFR FUV}$ is likely a result 
of the inefficiency of molecular cloud formation such as in the 
Bridge.

Lastly, the YSO content in the Bridge also indicates a possibly 
different star formation mode.
Extrapolating from the total number of YSOs found in the 6--10.4
$M_\odot$ (Table~\ref{sfr}), the expected number of YSOs with $M_\star=$
11--50 $M_\odot$ is 15$^{+11}_{-5}$ for a Salpeter IMF, and even 
7$^{+5}_{-2}$ should be O-type (17.5--50 $M_\odot$) YSOs.
However no YSOs with $M_\star > 10 M_\odot$ were found in the Bridge.
It is unlikely that higher-mass YSOs are at even earlier evolutionary 
stages since our far-IR {\it Herschel} survey of the Bridge show no 
such embedded high-mass YSOs \citep{Meixneretal13}.
It is not clear if the higher-mass YSOs are simply not yet formed or 
such YSOs are less likely to be formed in the Bridge's environment.
Alternatively, the total number of YSOs in the 6--10.4 $M_\odot$
in the Bridge might be overpopulated as a result of preferentially 
lower-mass star formation expected in the colliding flow scenario 
(recall Section 5.1).
To investigate if star formation mode is different in the Bridge,
a more comprehensive census of low-mass YSOs is needed.

\section{Summary}

We have analyzed the distribution of young stellar objects (YSOs) in
the Magellanic Bridge, the nearest tidally disturbed low-metallicity
system.  We selected YSO candidates from {\em Spitzer} photometry
obtained for the SAGE-SMC Legacy program \citep{GKetal09}, 
using a
combination of color and SED cuts,
building on the techniques of \citet{WBetal08} and \citet{GC09}.  We
present a list of high-confidence brighter massive YSOs, and a list of
somewhat fainter YSO candidates which may suffer from some
contamination by unresolved background galaxies.  We fit each source's
SED with dust radiative transfer models from \citet{RTetal06} to
constrain physical properties including central mass, envelope mass,
and accretion rate.  We find only weak correlation between the
evolutionary ``Stage'' implied by the SED fit (high accretion rate
normalized to central source mass) and our ``Type'' classification, in
which sources with redder SEDs and fainter parsec-scale interstellar
diffuse emission are presumed to be less evolved.  The differences are
likely due to a combination of a not extremely tight relation between
circumstellar environment and protostellar evolutionary stage (which
affects the ``Type''), and the fact that not all warm circumstellar
dust contributing to the FIR SED may actually accrete onto the
protostar (which affects the ``Stage'').  We also analyzed {\em
Spitzer} photometry for a list of NIR-derived candidate HAeBe stars
\citep{Naetal05} in the Bridge, and find that about half of the HAeBe
candidates show no significant MIR excess, and the other half we model
as Stage II, or more evolved, YSOs.  All of these findings are
consistent with our similar analysis of LMC star formation regions.

YSOs in the Bridge do show one particular contrast to those with
similar mass in the higher metallicity LMC.  Sources at the same
evolutionary stage (derived primarily from the FIR flux and derived
circumstellar dust mass) are brighter at optical wavelengths in the
Bridge compared to the LMC.  One explanation is that the circumstellar
envelopes are more permeable or clumpy at low metallicity, allowing
more short-wavelength radiation to escape for the same mass of warm
clumps.  
Another particularity of the Bridge arises from the comparison of
MIR-selected YSOs and NIR-selected HAeBe candidates -- the most
massive YSOs are less massive than the most massive HAeBe candidates.
This is consistent with the star formation in the Bridge becoming less
vigorous with time, i.e. forming less massive stars than in the past.

Of particular interest is star formation activity at the locations
where CO 1-0 has been detected by \citet{MNetal06}.  All CO 1-0 clouds
except their lowest signal-to-noise detection are associated with MIR
YSOs.  Many of the clouds contain multiple YSOs, and several show
evidence of multiple generations of star formation and possible
triggering, for example a large HII region with MIR YSOs on its rim.
Many of the YSOs in clouds are also closely associated with UV-bright
main-sequence intermediate to massive stars, further supporting
heterogeneity and non-coeval formation.  

We compared the overall star formation rate and efficiency in CO-detected
molecular clouds to the atomic and molecular gas column.
Global measures of star formation (integrated 24$\mu$m luminosities)
fall below the rates predicted by the total gas column density and the
S-K relation.  However, detailed analysis of the YSO
content brings the measured rates in closer agreement with gas-based
predictions -- this agrees with results in LMC regions and may be a
result of incomplete sampling of the stellar mass function, which
systematically reduces integrated H$\alpha$ and 24$\mu$ because of the
steep stellar mass-luminosity relation.  
When the entire Bridge area is considered (in contrast to only the
molecular clouds), the star formation rate predicted from the
relatively high HI mass is significantly higher than the total star
formation that we detect.  This is consistent with very inefficient
molecular cloud formation efficiency in this environment, but
relatively efficient star formation in molecular clouds that can form.

\acknowledgments
This work was supported through JPL grants 1282653 and 
1288328, and also partly by the European Research Council 
for the ERC Advanced Grant {\it GLOSTAR} under constract \# 247078.
M.S. acknowledges financial support from the NASA ADAP award NNX11AG50G.
M.M. and J.P.S. acknowledge financial support from the NASA {\it Herschel} 
Science Center, JPL contracts \# 1381522 and 1381650.
This study made use of data products of the Two Micron All Sky Survey, 
which is a joint project of the University of Massachusetts and 
the Infrared Processing and Analysis Center/California Institute 
of Technology, funded by the National Aeronautics and the Space 
Administration and the National Science Foundation.



\clearpage
\begin{figure}
\epsscale{1.0}
\plotone{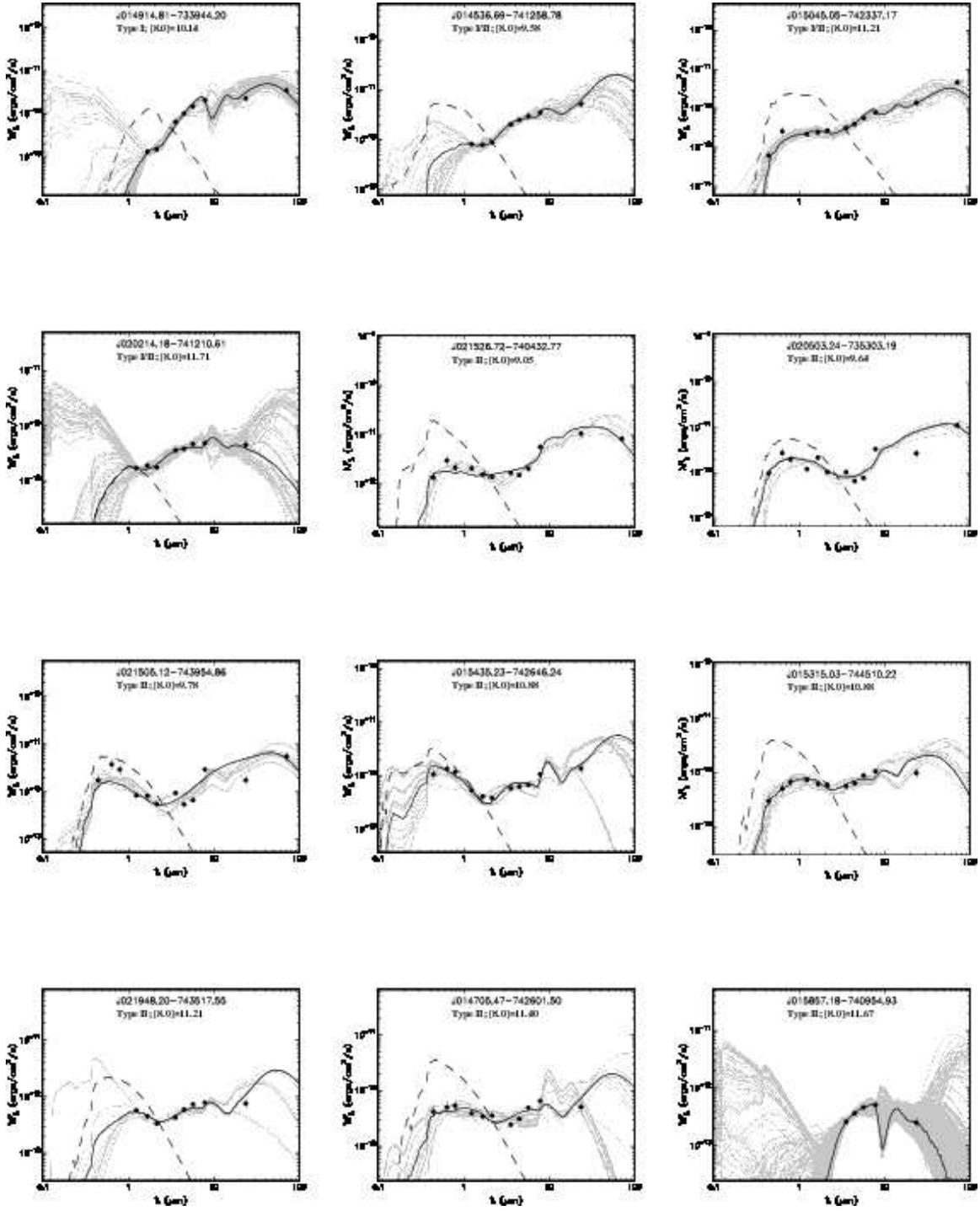}
\caption{SEDs of 26 YSOs identified in the Bridge.
 Filled circles are the flux values converted from magnitudes listed in Tables~\ref{ysoclass}
 and \ref{ysoclassb}.
 The source names are labeled at the top of the plot.
 Triangles are upper limits.  Error bars are shown if larger than the data
 points. The solid black line shows the best-fit model, and the dashed black
 line illustrates the radiation from the central star reddened by the 
 total line-of-sight
 $A_V$.  The gray lines  show all acceptable models.  
\label{fig:fit}}.
\end{figure}

\clearpage
{\plotone{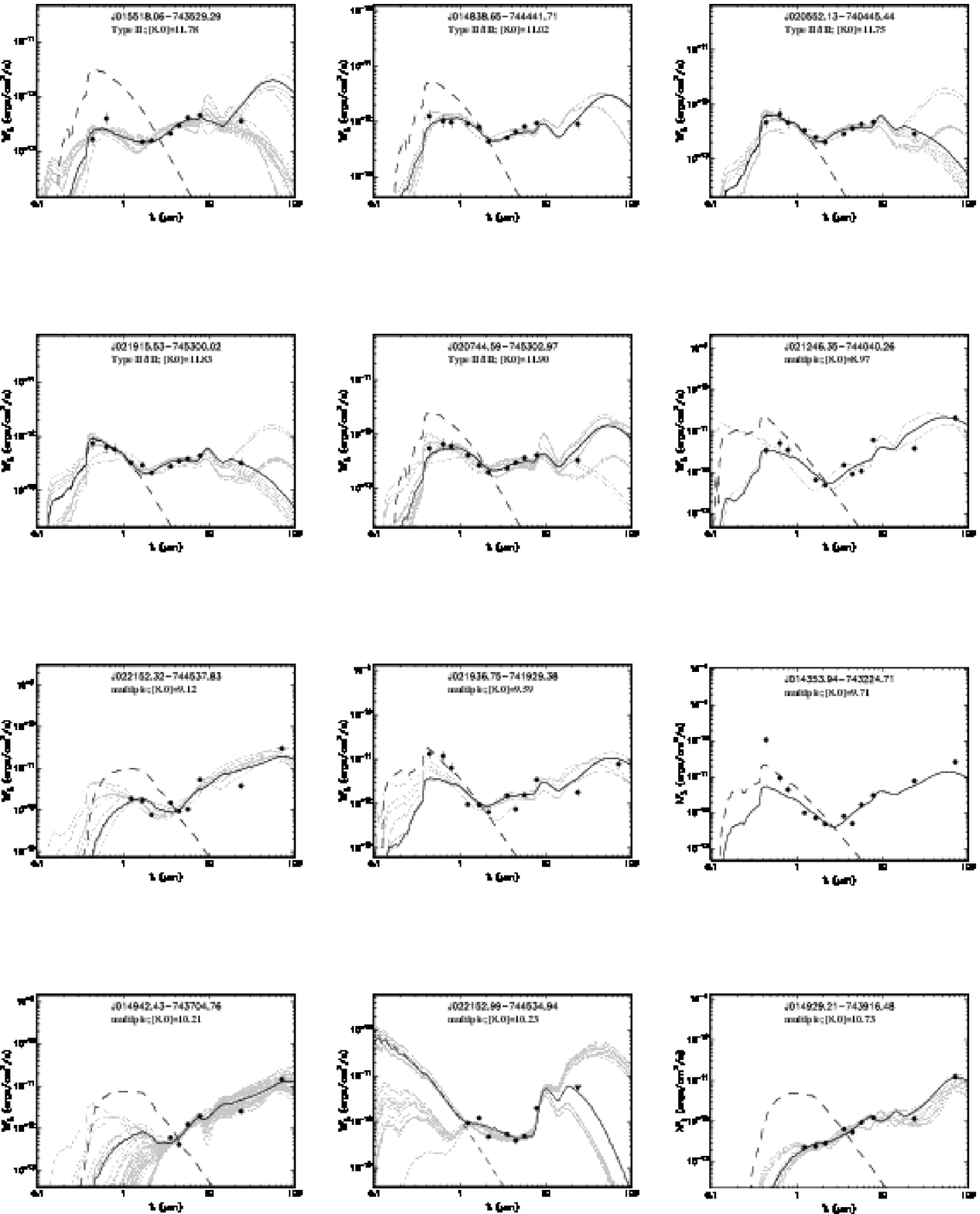}}\\[2mm]
\centerline{Figure~\ref{fig:fit} --- Continued.}

\clearpage
{\epsscale{0.65}
\plotone{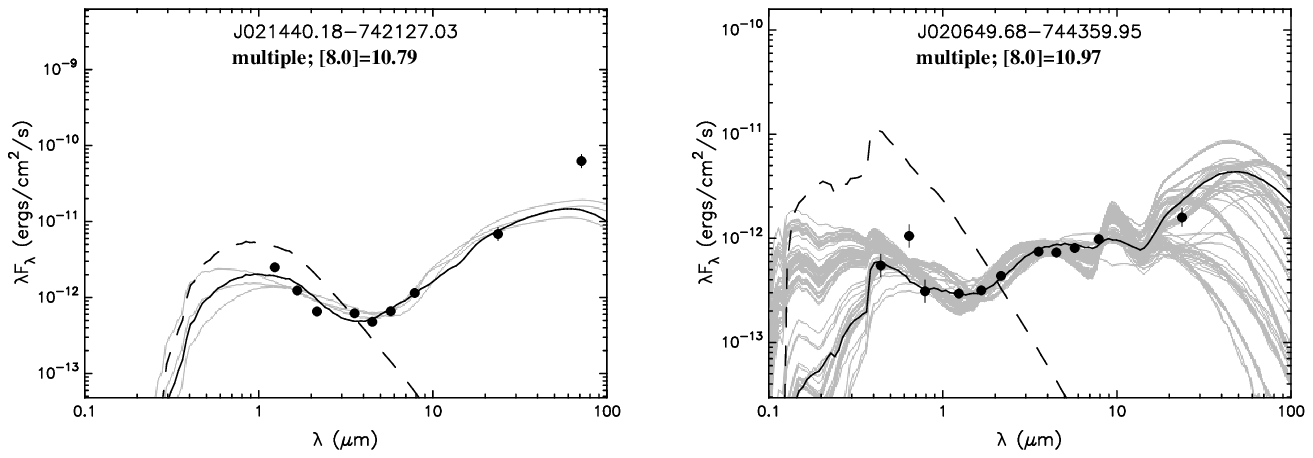}}\\[2mm]
\centerline{Figure~\ref{fig:fit} --- Continued.}

\clearpage
\begin{figure}
\epsscale{0.75}
\plotone{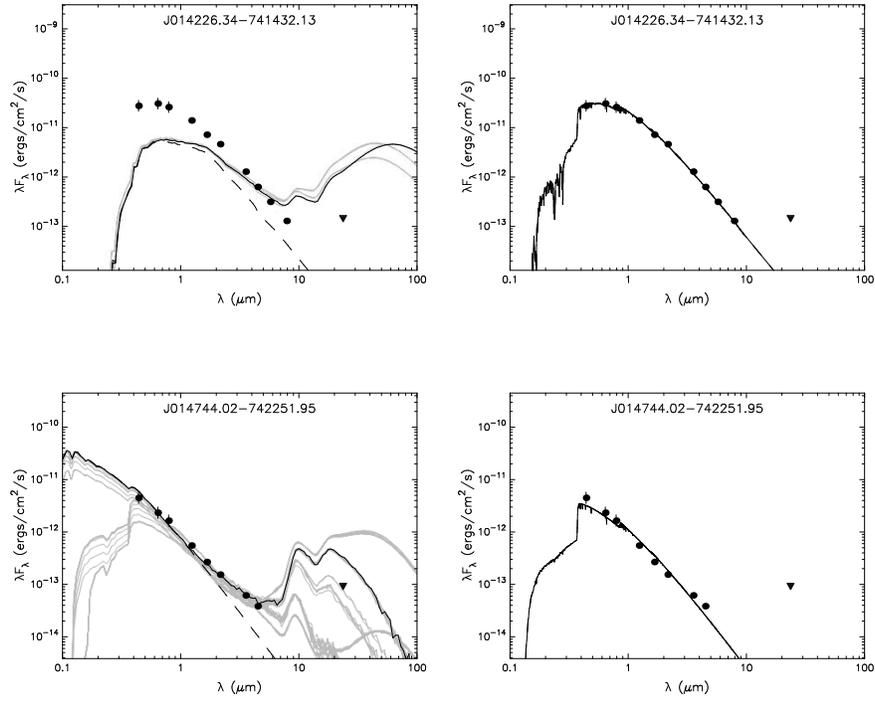}
\caption{Upper: SEDs (filled circles and triangles) of HAeBe candidate
 J014226.34$-$741432.13 overlaid with best and acceptable fits by YSO 
 models (left) and single best fit by stellar models (right).  Symbols are the
 same as Figure~\ref{fig:fit}. 
 Lower: the same setup of HAeBe candidate J014744.02$-$742551.95.
 \label{fig:haebe}}.
\end{figure}

\epsscale{.75}

\epsscale{.95}

\epsscale{1.}

\clearpage

\epsscale{0.5}

\clearpage
\newpage

\begin{deluxetable}{rrrrrrrrll}
\tablecolumns{10}
\tablecaption{Multi-wavelength Photometry for $\lambda > 3$ \um\ of YSO Candidates Selected from CMD
 Criteria \label{ysoclass}}
\tablewidth{0pc}
\tablehead{
 \colhead{Name} &  \colhead{No}  &
 \colhead{[3.6]} & 
 \colhead{[4.5]} & 
 \colhead{[5.8]} & 
 \colhead{[8.0]} & 
 \colhead{[24]}  & 
 \colhead{[70]}  &
 \colhead{Class.}  &
 \colhead{Remarks} \\
 \colhead{(1)} &  \colhead{(2)}  &
 \colhead{(3)} & 
 \colhead{(4)} & 
 \colhead{(5)} & 
 \colhead{(6)} & 
 \colhead{(7)} & 
 \colhead{(8)} & 
 \colhead{(9)} & \colhead{(10)} 
}

\startdata

J013928.99-744839.06 &  12 & 13.80  0.01 & 13.64  0.01 & 12.76  0.03 & 10.09  0.01 &  6.64  0.02 &   1.24  0.03 &  G &  \\ 
J013958.84-744902.13 &  18 & 13.35  0.01 & 13.27  0.01 & 12.69  0.03 & 10.32  0.01 &  7.42  0.03 &   2.17  0.04 &  G &  \\ 
J014025.67-741014.16 &  43 & 13.89  0.01 & 13.77  0.01 & 13.35  0.04 & 11.25  0.02 &  8.35  0.04 &  \nodata &  G &  \\ 
J014036.87-741245.39 &  36 & 13.98  0.01 & 13.94  0.01 & \nodata & 10.96  0.01 &  8.82  0.06 &  \nodata &  G &  \\ 
J014114.46-744823.14 &  38 & 13.85  0.01 & 13.82  0.02 & 13.26  0.04 & 10.99  0.01 &  8.06  0.04 &   2.60  0.05 &  G &  \\ 
J014121.42-734508.53 &  45 & 14.28  0.01 & 13.91  0.01 & 13.82  0.05 & 11.45  0.01 &  8.46  0.04 &  \nodata &  G &  \\ 
J014208.96-735202.92 &  21 & 13.88  0.01 & 13.78  0.01 & 13.18  0.04 & 10.37  0.01 &  7.79  0.02 &   2.54  0.06 &  G &  \\ 
J014353.94-743224.71 &   9 & 13.66  0.01 & 13.41  0.01 & 11.39  0.01 &  9.72  0.01 &  5.11  0.01 &   0.22  0.02 &  II &  mul\\ 
J014402.46-743333.15 &  47 & 14.35  0.01 & 13.94  0.01 & 13.73  0.06 & 11.53  0.02 &  8.73  0.05 &  \nodata &  G &  \\ 
J014536.69-741258.78 &   5 & 12.64  0.01 & 11.70  0.01 & 10.75  0.01 &  9.58  0.00 &  5.55  0.01 &  \nodata &  I/II &  \\ 
J014705.47-742601.50 &  44 & 14.93  0.01 & 13.94  0.01 & 12.71  0.02 & 11.40  0.01 &  8.06  0.03 &  \nodata &  II &  \\ 
J014813.20-734532.97 &  24 & 13.94  0.01 & 13.88  0.02 & 13.09  0.04 & 10.46  0.01 &  7.91  0.03 &   2.59  0.05 &  G &  \\ 
J014838.65-744441.71 &  39 & 14.14  0.01 & 13.14  0.01 & 12.18  0.02 & 11.01  0.01 &  7.45  0.02 &  \nodata &  II/III &  \\ 
J014914.81-733944.20 &  13 & 13.91  0.01 & 12.65  0.01 & 11.52  0.01 & 10.14  0.01 &  6.44  0.01 &   2.38  0.05 &  I &  \\ 
J014927.10-740011.41 &  11 & 13.45  0.01 & 13.38  0.01 & 12.67  0.03 &  9.95  0.01 &  7.67  0.02 &   2.27  0.04 &  G &  \\ 
J014929.21-743916.48 &  27 & 13.90  0.02 & 13.37  0.01 & 12.05  0.02 & 10.73  0.03 &  7.22  0.03 &   1.03  0.02 &  I/II &  mul\\ 
J014942.43-743704.76 &  15 & 13.98  0.01 & 13.64  0.01 & 11.73  0.02 & 10.21  0.01 &  6.33  0.01 &   0.87  0.02 &  I &  mul\\ 
J015005.67-734714.96 &  25 & 13.68  0.01 & 13.40  0.01 & 12.57  0.02 & 10.46  0.01 &  7.28  0.02 &  \nodata &  G &  \\ 
J015039.92-735041.67 &  26 & 13.87  0.01 & 13.75  0.01 & 13.09  0.03 & 10.62  0.01 &  7.25  0.02 &   2.09  0.04 &  G &  \\ 
J015045.05-742337.17 &  41 & 14.65  0.01 & 13.69  0.01 & 12.57  0.02 & 11.20  0.01 &  6.99  0.02 &   2.10  0.04 &  I/II &  \\ 
J015103.81-745306.07 &  20 & 13.53  0.01 & 13.44  0.01 & \nodata & 10.34  0.01 &  7.41  0.02 &   2.21  0.04 &  G &  \\ 
J015113.35-740308.28 &  50 & 14.21  0.01 & 13.89  0.01 & 13.78  0.04 & 11.61  0.01 &  8.78  0.05 &  \nodata &  G &  \\ 
J015135.01-735425.52 &  46 & 13.96  0.01 & 13.92  0.02 & \nodata & 11.45  0.01 &  9.51  0.15 &  \nodata &  G &  \\ 
J015138.67-743000.61 &  23 & 13.87  0.01 & 13.65  0.01 & 13.17  0.03 & 10.44  0.01 &  7.26  0.02 &   1.99  0.04 &  G &  \\ 
J015148.77-745014.92 &  48 & 14.12  0.02 & 13.97  0.02 & 13.71  0.07 & 11.54  0.02 &  8.42  0.04 &  \nodata &  G &  \\ 
J015217.87-744755.06 &  58 & 14.22  0.01 & 13.98  0.02 & \nodata & 11.83  0.02 &  9.60  0.11 &  \nodata &  G &  \\ 
J015315.03-744510.22 &  33 & \nodata & 13.17  0.01 & 12.09  0.01 & 10.87  0.01 &  7.36  0.02 &  \nodata &  II &  ext\\ 
J015435.23-742646.24 &  34 & 14.00  0.01 & 13.19  0.01 & 12.36  0.02 & 10.87  0.01 &  7.02  0.02 &  \nodata &  II &  \\ 
J015518.06-743529.29 &  56 & 15.09  0.02 & 13.98  0.01 & 12.90  0.03 & 11.78  0.02 &  8.45  0.04 &  \nodata &  II &  \\ 
J015535.26-734110.17 &  49 & 14.10  0.01 & 13.66  0.01 & 13.64  0.05 & 11.59  0.01 &  8.18  0.03 &  \nodata &  G &  \\ 
J015717.36-741104.70 &   6 & 13.54  0.01 & 13.42  0.01 & 12.32  0.02 &  9.59  0.00 &  7.39  0.02 &   1.68  0.03 &  G &  \\ 
J015809.18-740955.36 &  40 & 14.04  0.03 & 13.95  0.03 & 13.41  0.06 & 11.16  0.03 &  8.11  0.03 &   2.66  0.06 &  G &  \\ 
J015857.18-740954.93 &  52 & 14.87  0.01 & 13.78  0.01 & 12.78  0.02 & 11.66  0.01 &  8.84  0.06 &  \nodata &  II &  \\ 
J015929.09-742214.05 &  51 & 14.38  0.01 & 13.82  0.01 & 13.23  0.04 & 11.63  0.01 &  7.37  0.02 &  \nodata &  G &  \\ 
J020116.73-735926.19 &  14 & 13.67  0.01 & 13.50  0.01 & \nodata & 10.17  0.01 &  7.55  0.02 &   1.83  0.03 &  G &  \\ 
J020159.23-740621.09 &  28 & 13.71  0.04 & 13.68  0.04 & \nodata & 10.75  0.07 & \nodata &   1.57  0.02 &  G &  \\ 
J020214.18-741210.61 &  53 & 14.52  0.01 & 13.72  0.01 & 12.77  0.03 & 11.71  0.02 &  8.21  0.04 &  \nodata &  I/II &  ext\\ 
J020237.42-735549.07 &  29 & 13.29  0.01 & 13.28  0.01 & 12.84  0.03 & 10.77  0.01 &  8.61  0.05 &  \nodata &  G &  \\ 
J020440.85-735746.87 &  55 & 14.28  0.01 & 13.90  0.01 & 13.82  0.06 & 11.76  0.02 &  9.10  0.07 &  \nodata &  G &  \\ 
J020503.24-735303.19 &   8 & \nodata & 13.14  0.01 & 12.24  0.02 &  9.64  0.00 &  6.28  0.01 &   1.15  0.02 &  II &  \\ 
J020552.13-740445.44 &  54 & 14.78  0.01 & 13.79  0.01 & 12.86  0.03 & 11.75  0.02 &  8.70  0.05 &  \nodata &  II/III &  \\ 
J020649.68-744359.95 &  37 & 13.74  0.01 & 13.03  0.01 & 12.18  0.02 & 10.96  0.01 &  6.85  0.01 &  \nodata &  II &  mul\\ 
J020744.59-745302.97 &  59 & 14.98  0.02 & 13.99  0.01 & 13.03  0.03 & 11.90  0.02 &  8.54  0.04 &  \nodata &  II/III &  \\ 
J021246.35-744040.26 &   1 & 12.96  0.01 & 12.78  0.01 & 11.86  0.02 &  8.96  0.00 &  5.89  0.01 &   0.47  0.02 &  II &  mul\\ 
J021249.62-740848.22 &  32 & \nodata & 12.94  0.01 & 12.54  0.03 & 10.87  0.02 &  8.58  0.05 &   2.46  0.05 &  G &  \\ 
J021440.18-742127.03 &  30 & 13.94  0.03 & 13.49  0.03 & 12.41  0.05 & 10.79  0.05 &  5.26  0.01 &  -0.72  0.01 &  I &  mul\\ 
J021505.20-743954.86 &  10 & \nodata & 13.34  0.01 & 12.39  0.02 &  9.78  0.01 &  6.74  0.02 &   1.89  0.04 &  II &  \\ 
J021526.72-740432.73 &   2 & 12.86  0.01 & 12.24  0.01 & 11.19  0.01 &  9.05  0.00 &  4.79  0.01 &   1.47  0.03 &  II &  ext\\ 
J021648.92-742412.56 &  22 & 12.98  0.03 & 13.03  0.03 & 12.12  0.06 & 10.37  0.07 &  6.13  0.01 &   0.61  0.02 &  G &  \\ 
J021654.37-743558.30 &   4 & \nodata & 12.98  0.01 & 11.98  0.02 &  9.34  0.01 &  6.49  0.01 &   1.22  0.02 &  G &  \\ 
J021654.69-743940.85 &  60 & \nodata & 14.00  0.02 & 13.89  0.07 & 11.98  0.02 &  8.51  0.04 &  \nodata &  G &  \\ 
J021732.44-744527.82 &  17 & 13.54  0.02 & 13.52  0.03 & \nodata & 10.29  0.04 &  8.22  0.05 &   1.69  0.03 &  G &  \\ 
J021822.89-742336.31 &  35 & 13.54  0.01 & 13.47  0.01 & 13.03  0.04 & 10.89  0.01 &  8.57  0.04 &  \nodata &  G &  \\ 
J021845.29-734541.11 &  31 & 14.13  0.01 & 13.95  0.02 & 13.82  0.06 & 10.85  0.01 &  8.54  0.07 &  \nodata &  G &  \\ 
J021915.53-745300.02 &  57 & 14.79  0.02 & 13.84  0.01 & 12.98  0.03 & 11.83  0.02 &  8.57  0.04 &  \nodata &  II/III &  \\ 
J021936.75-741929.38 &   7 & 12.96  0.01 & 12.98  0.01 & 11.47  0.01 &  9.59  0.01 &  6.71  0.01 &   1.53  0.03 &  II &  mul\\ 
J021948.20-743517.55 &  42 & 14.34  0.01 & 13.26  0.01 & 12.29  0.02 & 11.21  0.01 &  7.64  0.02 &  \nodata &  II &  \\ 
J022030.32-744654.37 &  19 & 13.52  0.01 & 13.30  0.01 & 13.04  0.03 & 10.34  0.01 &  7.63  0.02 &   2.02  0.04 &  G &  \\ 
J022152.32-744537.83 &   3 & 12.86  0.01 & 12.66  0.01 & 11.79  0.01 &  9.06  0.01 &  5.48  0.01 &   0.07  0.02 &  II &  mul\\ 
J022152.99-744534.94 &  16 & 14.10  0.10 & 13.71  0.06 & 12.70  0.05 & 10.23  0.04 & \nodata &  \nodata &  II &  mul\\ 

\enddata
\tablecomments{Column 1: source name. Column 2: Ranking of the brightness at
 8 \um. Columns 3-8: photometric measurements in 3.6, 4.5, 5.8, 8.0, 24, and
 70 \um\ bands in magnitudes. 
 Measurements with uncertainties of $99.9$ are the upper brightness limits
 as they include fluxes from neighbors or backgrounds.
 The uncertainties listed here are only errors in measurements and do not
 include errors in flux calibration, i.e., 5\% in 3.6, 4.5, 5.8, and
 8.0 \um, 10\% in 24 \um, and 20\% in 70 \um.  Thus, the total uncertainty
 of a flux is the quadratic sum of the measurement error and the calibration
 error.
 Columns 9 and 10: classification and remarks: D -- diffuse emission,
 ext -- extended source, G -- background galaxy,
 I/II/III -- Type I/II/III YSO, mul -- multiple, S -- star.}
\end{deluxetable}

\begin{deluxetable}{rrrrrrrrll}
\tablecolumns{10}
\tablecaption{Multi-wavelength Photometry for $\lambda < 3$ \um\ of YSO Candidates Selected from CMD
 Criteria \label{ysoclassb}}
\tablewidth{0pc}
\tablehead{
 \colhead{Name} &  \colhead{No}  &
 \colhead{$B$} & 
 \colhead{$R$} & 
 \colhead{$I$} & 
 \colhead{$J$} & 
 \colhead{$H$} & 
 \colhead{$K_s$} & 
 \colhead{Class.}  &
 \colhead{Remarks} \\
 \colhead{(1)} &  \colhead{(2)}  &
 \colhead{(3)} & 
 \colhead{(4)} & 
 \colhead{(5)} & 
 \colhead{(6)} & 
 \colhead{(7)} & 
 \colhead{(8)} & 
 \colhead{(9)} & \colhead{(10)} 
}

\startdata

J013928.99-744839.06 &  12 & 18.49  0.15 & 16.35  0.15 & 15.92  0.15 & 17.05  0.09 & 16.49  0.12 &  15.70  0.17 &  G & \\ 
J013958.84-744902.13 &  18 & 17.38  0.15 & 15.57  0.15 & 15.15  0.15 & 16.68  0.06 & 16.00  0.07 &  15.48  0.16 &  G & \\ 
J014025.67-741014.16 &  43 & 17.61  0.15 & 16.33  0.15 & 16.54  0.15 & 16.94  0.12 & 16.16  0.09 &  15.46  0.05 &  G & \\ 
J014036.87-741245.39 &  36 & 17.98  0.15 & 16.52  0.15 & 16.43  0.15 & 16.41  0.12 & 16.32  0.07 &  15.60  0.10 &  G & \\ 
J014114.46-744823.14 &  38 & 17.81  0.15 & 16.24  0.15 & 16.42  0.15 & 16.52  0.05 & 15.92  0.10 &  15.49  0.14 &  G & \\ 
J014121.42-734508.53 &  45 & 19.16  0.15 & 17.37  0.15 & 17.05  0.15 & \nodata & \nodata &  \nodata &  G & \\ 
J014208.96-735202.92 &  21 & 17.90  0.15 & 16.51  0.15 & 16.59  0.15 & 17.31  0.07 & 16.52  0.08 &  16.00  0.12 &  G & \\ 
J014353.94-743224.71 &   9 & 13.48  0.15 & 15.42  0.15 & 15.75  0.15 & 16.45  0.05 & 16.00  0.05 &  15.67  0.05 &  II & mul\\ 
J014402.46-743333.15 &  47 & 20.85  0.15 & 18.00  0.15 & 17.76  0.15 & 17.32  0.08 & 16.44  0.07 &  15.74  0.06 &  G & \\ 
J014536.69-741258.78 &   5 & \nodata & \nodata & \nodata & 16.68  0.06 & 15.92  0.06 &  15.03  0.04 &  I/II & \\ 
J014705.47-742601.50 &  44 & 19.49  0.15 & 18.59  0.15 & 18.05  0.15 & 17.48  0.02 & 16.80  0.04 &  16.01  0.04 &  II & \\ 
J014813.20-734532.97 &  24 & \nodata & \nodata & \nodata & 17.19  0.08 & 16.45  0.07 &  16.11  0.13 &  G & \\ 
J014838.65-744441.71 &  39 & 18.31  0.15 & 17.83  0.15 & 17.40  0.15 & 16.55  0.02 & 15.91  0.02 &  15.81  0.06 &  II/III & \\ 
J014914.81-733944.20 &  13 & \nodata & \nodata & \nodata & \nodata & 17.86  0.11 &  16.97  0.11 &  I & \\ 
J014927.10-740011.41 &  11 & \nodata & \nodata & \nodata & 16.77  0.11 & 15.65  0.05 &  15.23  0.05 &  G & \\ 
J014929.21-743916.48 &  27 & \nodata & \nodata & \nodata & 18.05  0.21 & 17.19  0.18 &  16.26  0.10 &  I/II & mul\\ 
J014942.43-743704.76 &  15 & \nodata & \nodata & \nodata & \nodata & \nodata &  \nodata &  I & mul\\ 
J015005.67-734714.96 &  25 & 17.48  0.15 & 15.88  0.15 & 15.48  0.15 & 16.16  0.13 & 16.23  0.14 &  15.88  0.21 &  G & \\ 
J015039.92-735041.67 &  26 & 18.26  0.15 & 16.69  0.15 & 16.65  0.15 & \nodata & \nodata &  16.14  0.15 &  G & \\ 
J015045.05-742337.17 &  41 & 21.57  0.15 & 19.30  0.15 & 100.00  0.15 & 18.12  0.06 & 17.20  0.10 &  16.32  0.06 &  I/II & \\ 
J015103.81-745306.07 &  20 & \nodata & \nodata & \nodata & 16.76  0.05 & 16.17  0.06 &  15.63  0.11 &  G & \\ 
J015113.35-740308.28 &  50 & 20.09  0.15 & 17.83  0.15 & 17.34  0.15 & 17.30  0.15 & 15.89  0.07 &  15.52  0.14 &  G & \\ 
J015135.01-735425.52 &  46 & 17.94  0.15 & 16.38  0.15 & 16.51  0.15 & 16.98  0.20 & 16.11  0.06 &  15.23  0.11 &  G & \\ 
J015138.67-743000.61 &  23 & 18.39  0.15 & 16.72  0.15 & 16.65  0.15 & 16.71  0.07 & 16.11  0.07 &  15.37  0.07 &  G & \\ 
J015148.77-745014.92 &  48 & 17.81  0.15 & 16.27  0.15 & 16.24  0.15 & 17.37  0.12 & 16.50  0.09 &  15.75  0.08 &  G & \\ 
J015217.87-744755.06 &  58 & \nodata & \nodata & \nodata & 17.36  0.06 & 16.54  0.11 &  15.75  0.08 &  G & \\ 
J015315.03-744510.22 &  33 & 19.86  0.15 & 18.60  0.15 & 17.80  0.15 & 16.79  0.02 & 16.19  0.02 &  15.47  0.03 &  II & ext\\ 
J015435.23-742646.24 &  34 & 18.48  0.15 & 17.52  0.15 & 17.21  0.15 & 17.14  0.04 & 16.61  0.05 &  15.95  0.05 &  II & \\ 
J015518.06-743529.29 &  56 & 20.50  0.15 & 18.85  0.15 & 100.00  0.15 & \nodata & 17.72  0.08 &  16.92  0.09 &  II & \\ 
J015535.26-734110.17 &  49 & 19.15  0.15 & 17.08  0.15 & 17.07  0.15 & 17.65  0.10 & 16.16  0.10 &  15.56  0.08 &  G & \\ 
J015717.36-741104.70 &   6 & 19.00  0.15 & 17.30  0.15 & 17.19  0.15 & 16.84  0.07 & 16.02  0.06 &  15.35  0.10 &  G & \\ 
J015809.18-740955.36 &  40 & \nodata & \nodata & \nodata & 16.97  0.05 & 16.43  0.07 &  15.70  0.11 &  G & \\ 
J015857.18-740954.93 &  52 & \nodata & \nodata & \nodata & \nodata & \nodata &  \nodata &  II & \\ 
J015929.09-742214.05 &  51 & 21.16  0.15 & 18.29  0.15 & 18.10  0.15 & 17.23  0.07 & 16.84  0.15 &  15.88  0.09 &  G & \\ 
J020116.73-735926.19 &  14 & \nodata & \nodata & \nodata & 17.11  0.07 & 16.16  0.07 &  15.74  0.12 &  G & \\ 
J020159.23-740621.09 &  28 & 15.36  0.15 & 14.66  0.15 & 14.88  0.15 & 16.63  0.04 & 16.12  0.06 &  15.77  0.12 &  G & \\ 
J020214.18-741210.61 &  53 & \nodata & \nodata & \nodata & 18.38  0.05 & 17.47  0.07 &  16.79  0.10 &  I/II & ext\\ 
J020237.42-735549.07 &  29 & 17.39  0.15 & 15.87  0.15 & 15.75  0.15 & 15.73  0.02 & 15.55  0.05 &  14.85  0.05 &  G & \\ 
J020440.85-735746.87 &  55 & 19.56  0.15 & 17.45  0.15 & 17.23  0.15 & 16.87  0.09 & 15.89  0.05 &  15.07  0.05 &  G & \\ 
J020503.24-735303.19 &   8 & 18.59  0.15 & 16.77  0.15 & 16.62  0.15 & 16.25  0.03 & 14.82  0.04 &  14.86  0.09 &  II & \\ 
J020552.13-740445.44 &  54 & 19.38  0.15 & 18.32  0.15 & 18.22  0.15 & 17.66  0.04 & 17.18  0.06 &  16.66  0.07 &  II/III & \\ 
J020649.68-744359.95 &  37 & 19.21  0.15 & 17.80  0.15 & 18.64  0.15 & 17.79  0.05 & 16.91  0.04 &  15.81  0.04 &  II & mul\\ 
J020744.59-745302.97 &  59 & 19.21  0.15 & 18.31  0.15 & 17.93  0.15 & 17.43  0.02 & 17.08  0.04 &  16.67  0.06 &  II/III & \\ 
J021246.35-744040.26 &   1 & 17.22  0.15 & 16.09  0.15 & 16.02  0.15 & \nodata & 16.11  0.05 &  15.68  0.15 &  II & mul\\ 
J021249.62-740848.22 &  32 & 16.82  0.15 & 15.41  0.15 & 15.26  0.15 & 15.90  0.05 & 15.21  0.04 &  14.69  0.06 &  G & \\ 
J021440.18-742127.03 &  30 & \nodata & \nodata & \nodata & 15.47  0.02 & 15.43  0.03 &  15.37  0.04 &  I & mul\\ 
J021505.20-743954.86 &  10 & 17.94  0.15 & 16.42  0.15 & 16.20  0.15 & 16.66  0.05 & 15.88  0.06 &  15.55  0.10 &  II & \\ 
J021526.72-740432.73 &   2 & \nodata & \nodata & \nodata & 15.67  0.02 & 15.19  0.02 &  14.54  0.03 &  II & ext\\ 
J021648.92-742412.56 &  22 & 10.62  0.15 & 13.72  0.15 & 13.96  0.15 & 15.85  0.06 & 15.35  0.12 &  14.68  0.05 &  G & \\ 
J021654.37-743558.30 &   4 & 16.85  0.15 & 15.65  0.15 & 15.60  0.15 & 16.45  0.05 & 15.58  0.05 &  15.23  0.05 &  G & \\ 
J021654.69-743940.85 &  60 & \nodata & \nodata & \nodata & 17.31  0.06 & 16.60  0.05 &  16.06  0.12 &  G & \\ 
J021732.44-744527.82 &  17 & 13.36  0.15 & 15.18  0.15 & 15.30  0.15 & 16.70  0.06 & 16.25  0.08 &  15.99  0.14 &  G & \\ 
J021822.89-742336.31 &  35 & 17.95  0.15 & 16.38  0.15 & 16.28  0.15 & 16.92  0.06 & 15.84  0.16 &  15.07  0.10 &  G & \\ 
J021845.29-734541.11 &  31 & 18.17  0.15 & 16.80  0.15 & 16.40  0.15 & 17.25  0.05 & 16.72  0.10 &  16.34  0.14 &  G & \\ 
J021915.53-745300.02 &  57 & 18.87  0.15 & 18.35  0.15 & 17.96  0.15 & 17.69  0.04 & 16.98  0.05 &  16.59  0.06 &  II/III & \\ 
J021936.75-741929.38 &   7 & 15.76  0.15 & 15.17  0.15 & 15.36  0.15 & 16.51  0.05 & 15.72  0.06 &  15.39  0.09 &  II & mul\\ 
J021948.20-743517.55 &  42 & \nodata & \nodata & \nodata & 17.06  0.03 & 16.53  0.05 &  16.10  0.09 &  II & \\ 
J022030.32-744654.37 &  19 & 18.55  0.15 & 16.72  0.15 & 16.55  0.15 & \nodata & \nodata &  15.11  0.08 &  G & \\ 
J022152.32-744537.83 &   3 & \nodata & \nodata & \nodata & 15.76  0.03 & 15.11  0.03 &  15.21  0.14 &  II & mul\\ 
J022152.99-744534.94 &  16 & \nodata & \nodata & \nodata & 16.52  0.05 & 15.44  0.07 &  15.73  0.18 &  II & mul\\ 

\enddata
\tablecomments{Column 1: source name. Column 2: Ranking of the brightness at
 8 \um. Columns 3-8: $BRIJHK_s$ photometric measurements in magnitudes. 
 Measurements with uncertainties of $99.9$ are the upper brightness limits
 as they include fluxes from neighbors or backgrounds.
 The uncertainties listed here are only errors in measurements and do not
 include errors in flux calibration, i.e., 10\% in $BRIJHK_s$. Thus, 
 the total uncertainty of a flux is the quadratic sum
 of the measurement error and the calibration error.
 Column 10
 and 11: classification and remarks: D -- diffuse emission,
 ext -- extended source, G -- background galaxy,
 I/II/III -- Type I/II/III YSO, mul -- multiple, S -- star.}
\end{deluxetable}

\clearpage
\begin{turnpage}
\begin{deluxetable}{lrrrrrrrrrr}
\tablecolumns{11}
\tablecaption{Inferred Physical Parameters from SED Fits to YSOs \label{sedfits}}
\tablewidth{0pc}
\tablehead{
  \colhead{} & 
 \colhead{[8.0]} & 
 \colhead{} &
 \colhead{Stage} &
 \colhead{$\bar{M}_{\ast}$} &
 \colhead{$\bar{L}_{\rm tot}$} &
 \colhead{$\bar{\dot{M}}_{\rm env} $} & 
 \colhead{$\bar{M}_{\rm disk}$} & 
 \colhead{$\bar{\tau_{\ast}}$} &
 \colhead{$\bar{A_V}$} &
 \colhead{$\bar{incl.}$} \\
 \colhead{Source Name} & 
 \colhead{(mag)} &
 \colhead{Type} &
 \colhead{Range} & 
 \colhead{($M_\odot$)} & 
 \colhead{($L_\odot$)} &
 \colhead{($M_\odot$/yr)} & 
 \colhead{($R_\odot$)} & 
 \colhead{(yr)} & 
 \colhead{(mag)} &
 \colhead{($\circ$)} 
}

\startdata

J014914.81-733944.20 & 10.14 & I &  1.5$\pm$ 0.5 &  6.6$\pm$ 1.2 & 1.3E+03$\pm$9.7E+02 & 3.2E-04$\pm$3.1E-04 & 4.5E-01$\pm$3.5E-01 & 4.3E+05$\pm$4.3E+05 &  13.1$\pm$  9.3 & 38.8$\pm$21.8 \\ 
J014536.69-741258.78 &  9.58 & I/II &  1.6$\pm$ 0.5 &  9.6$\pm$ 2.6 & 6.4E+03$\pm$5.6E+03 & 1.7E-04$\pm$2.2E-04 & 3.3E-01$\pm$2.7E-01 & 9.6E+05$\pm$8.0E+05 &  17.2$\pm$ 12.8 & 45.2$\pm$19.8 \\ 
J015045.05-742337.17 & 11.21 & I/II &  1.1$\pm$ 0.3 &  6.2$\pm$ 1.0 & 5.1E+02$\pm$2.8E+02 & 2.2E-03$\pm$1.6E-03 & 4.5E-01$\pm$3.7E-01 & 1.2E+05$\pm$2.0E+05 &   0.9$\pm$  1.0 & 23.4$\pm$15.1 \\ 
J020214.18-741210.61 & 11.71 & I/II &  1.6$\pm$ 0.5 &  5.1$\pm$ 1.0 & 4.7E+02$\pm$2.0E+02 & 6.8E-04$\pm$6.6E-04 & 1.4E-01$\pm$1.1E-01 & 1.6E+06$\pm$2.0E+06 &   1.3$\pm$  1.8 & 40.9$\pm$21.7 \\ 
J021526.72-740432.77 &  9.05 & II &  1.0$\pm$ 0.0 &  9.1$\pm$ 0.8 & 2.0E+03$\pm$7.3E+02 & 5.2E-04$\pm$3.9E-04 & 2.1E-01$\pm$1.6E-01 & 4.0E+04$\pm$2.9E+04 &   1.5$\pm$  1.6 & 29.6$\pm$16.6 \\ 
J020503.24-735303.19 &  9.64 & II &  1.0$\pm$ 0.0 &  7.9$\pm$ 0.8 & 1.5E+03$\pm$5.6E+02 & 9.5E-04$\pm$7.4E-04 & 2.3E-01$\pm$1.8E-01 & 9.6E+04$\pm$4.0E+04 &   1.1$\pm$  0.8 & 27.4$\pm$11.5 \\ 
J021505.12-743954.86 &  9.78 & II &  1.0$\pm$ 0.0 &  7.5$\pm$ 0.4 & 1.3E+03$\pm$2.5E+02 & 5.9E-04$\pm$4.9E-04 & 2.1E-01$\pm$1.7E-01 & 1.4E+05$\pm$5.4E+04 &   0.6$\pm$  0.6 & 37.1$\pm$16.6 \\ 
J015435.23-742646.24 & 10.88 & II &  1.7$\pm$ 0.5 &  6.2$\pm$ 1.0 & 1.1E+03$\pm$8.4E+02 & 2.4E-05$\pm$2.2E-05 & 9.8E-02$\pm$1.1E-01 & 9.7E+05$\pm$9.8E+05 &   0.6$\pm$  0.5 & 48.7$\pm$19.1 \\ 
J015315.03-744510.22 & 10.88 & II &  1.0$\pm$ 0.0 &  6.2$\pm$ 0.9 & 5.1E+02$\pm$2.0E+02 & 1.4E-04$\pm$1.2E-04 & 5.9E-01$\pm$4.5E-01 & 1.8E+05$\pm$1.0E+05 &   0.8$\pm$  0.9 & 40.2$\pm$17.8 \\ 
J021948.20-743517.55 & 11.21 & II &  1.1$\pm$ 0.3 &  6.0$\pm$ 0.8 & 5.0E+02$\pm$1.2E+02 & 8.6E-05$\pm$6.2E-05 & 7.1E-01$\pm$5.2E-01 & 3.4E+05$\pm$2.2E+05 &   1.7$\pm$  1.4 & 48.0$\pm$15.0 \\ 
J014705.47-742601.50 & 11.40 & II &  2.2$\pm$ 1.0 &  8.0$\pm$ 2.2 & 3.5E+03$\pm$3.0E+03 & 2.0E-04$\pm$2.9E-04 & 1.7E-01$\pm$1.5E-01 & 1.8E+06$\pm$1.4E+06 &   1.6$\pm$  1.0 & 53.4$\pm$22.4 \\ 
J015857.18-740954.93 & 11.67 & II &  2.0$\pm$ 0.1 &  8.2$\pm$ 1.8 & 3.8E+03$\pm$3.1E+03 & 8.0E-05$\pm$1.9E-04 & 3.9E-02$\pm$4.3E-02 & 2.3E+06$\pm$1.0E+06 &  36.5$\pm$ 12.0 & 56.6$\pm$20.7 \\ 
J015518.06-743529.29 & 11.78 & II &  1.9$\pm$ 0.3 &  5.0$\pm$ 0.7 & 4.7E+02$\pm$1.6E+02 & 4.7E-05$\pm$1.3E-04 & 1.1E-01$\pm$8.9E-02 & 2.4E+06$\pm$2.0E+06 &   1.6$\pm$  0.5 & 48.9$\pm$18.9 \\ 
J014838.65-744441.71 & 11.02 & II/III &  1.0$\pm$ 0.0 &  6.2$\pm$ 0.5 & 5.6E+02$\pm$1.5E+02 & 9.9E-05$\pm$7.2E-05 & 6.6E-01$\pm$4.9E-01 & 2.7E+05$\pm$6.4E+04 &   0.6$\pm$  0.5 & 41.9$\pm$11.7 \\ 
J020552.13-740445.44 & 11.75 & II/III &  1.9$\pm$ 0.3 &  4.4$\pm$ 0.5 & 3.0E+02$\pm$6.4E+01 & 1.1E-05$\pm$1.2E-05 & 1.5E-01$\pm$1.2E-01 & 9.2E+05$\pm$2.1E+05 &   1.2$\pm$  0.5 & 46.8$\pm$17.4 \\ 
J021915.53-745300.02 & 11.83 & II/III &  1.9$\pm$ 0.3 &  4.4$\pm$ 0.5 & 3.0E+02$\pm$6.8E+01 & 1.3E-05$\pm$1.3E-05 & 1.5E-01$\pm$1.2E-01 & 9.1E+05$\pm$2.3E+05 &   0.9$\pm$  0.4 & 43.6$\pm$16.9 \\ 
J020744.59-745302.97 & 11.90 & II/III &  1.2$\pm$ 0.4 &  5.2$\pm$ 0.6 & 4.4E+02$\pm$1.1E+02 & 5.1E-05$\pm$3.8E-05 & 2.0E-01$\pm$1.6E-01 & 5.6E+05$\pm$2.6E+05 &   1.2$\pm$  0.7 & 43.6$\pm$18.3 \\ 
J021246.35-744040.26 &  8.97 & mul &  1.0$\pm$ 0.0 &  8.4$\pm$ 0.9 & 1.8E+03$\pm$5.8E+02 & 1.2E-03$\pm$1.0E-03 & 3.8E-01$\pm$2.9E-01 & 8.2E+04$\pm$4.0E+04 &   0.3$\pm$  0.3 & 32.0$\pm$13.0 \\ 
J022152.32-744537.83 &  9.12 & mul &  1.0$\pm$ 0.0 &  9.5$\pm$ 1.0 & 2.3E+03$\pm$7.2E+02 & 5.5E-03$\pm$4.1E-03 & 4.3E-01$\pm$3.4E-01 & 4.4E+04$\pm$4.4E+04 &   1.1$\pm$  1.0 & 19.5$\pm$ 4.0 \\ 
J021936.75-741929.38 &  9.59 & mul &  1.0$\pm$ 0.0 &  7.5$\pm$ 0.4 & 1.4E+03$\pm$2.4E+02 & 5.5E-04$\pm$4.6E-04 & 1.0E-01$\pm$8.7E-02 & 1.4E+05$\pm$5.4E+04 &   0.1$\pm$  0.2 & 29.2$\pm$ 9.9 \\ 
J014353.94-743224.71 &  9.71 & mul &  1.0$\pm$ 0.0 &  8.1$\pm$ 1.0 & 1.7E+03$\pm$5.4E+02 & 2.6E-03$\pm$2.1E-03 & 6.3E-01$\pm$5.2E-01 & 1.1E+05$\pm$5.8E+04 &   0.0$\pm$  0.0 & 18.2$\pm$ 0.0 \\ 
J014942.43-743704.76 & 10.21 & mul &  1.0$\pm$ 0.1 &  8.2$\pm$ 1.7 & 1.5E+03$\pm$8.5E+02 & 5.7E-03$\pm$4.3E-03 & 3.7E-01$\pm$3.1E-01 & 8.2E+04$\pm$1.8E+05 &   7.9$\pm$  8.4 & 22.3$\pm$11.1 \\ 
J022152.99-744734.94 & 10.23 & mul &  2.3$\pm$ 0.8 & 10.0$\pm$ 0.9 & 6.5E+03$\pm$1.9E+03 & 2.0E-05$\pm$1.9E-05 & 1.0E-01$\pm$9.6E-02 & 1.4E+06$\pm$8.8E+05 &   0.1$\pm$  0.1 & 69.5$\pm$22.1 \\ 
J014929.21-743916.48 & 10.73 & mul &  1.2$\pm$ 0.4 &  7.8$\pm$ 1.1 & 1.3E+03$\pm$4.6E+02 & 4.3E-03$\pm$3.2E-03 & 3.0E-01$\pm$2.5E-01 & 1.4E+05$\pm$2.9E+05 &  10.9$\pm$  8.9 & 26.3$\pm$16.7 \\ 
J021440.18-742127.03 & 10.79 & mul &  1.8$\pm$ 0.4 & 10.4$\pm$ 1.1 & 6.4E+03$\pm$2.3E+03 & 1.5E-03$\pm$1.6E-03 & 8.3E-02$\pm$1.0E-01 & 6.9E+05$\pm$3.2E+05 &  35.6$\pm$  9.5 & 50.2$\pm$23.1 \\ 
J020649.68-744359.95 & 10.97 & mul &  1.7$\pm$ 0.4 &  6.1$\pm$ 0.6 & 1.1E+03$\pm$3.8E+02 & 1.8E-05$\pm$1.6E-05 & 6.2E-02$\pm$7.3E-02 & 1.8E+06$\pm$1.7E+06 &   1.0$\pm$  0.6 & 48.2$\pm$19.9 \\ 

\enddata
\end{deluxetable}
\end{turnpage}
\clearpage

\clearpage
\begin{turnpage}
\begin{deluxetable}{lrrrrrrrr}
\tablecolumns{9}
\tablecaption{Inferred Physical Parameters from SED Fits to Fainter YSO Candidates \label{fysofits}}
\tablewidth{0pc}
\tablehead{
 \colhead{} &
 \colhead{$\bar{M}_{\ast}$ } &
 \colhead{$\bar{L}_{\rm tot}$} &
 \colhead{$\bar{\dot{M}}_{\rm env} $} & 
 \colhead{$\bar{M}_{\rm disk}$} & 
 \colhead{Stage} &
 \colhead{$\bar{\tau_{\ast}}$} &
 \colhead{$\bar{A_V}$} &
 \colhead{$\bar{incl.}$} \\
 \colhead{Source Name} & 
 \colhead{($M_\odot$)} & 
 \colhead{($L_\odot$)} &
 \colhead{($M_\odot$/yr)} & 
 \colhead{($R_\odot$)} & 
 \colhead{Range} & 
 \colhead{(yr)} & 
 \colhead{(mag)} &
 \colhead{($\circ$)} 
}

\startdata

J013810.63-735606.90 &  5.6$\pm$ 1.5 & 1.0E+03$\pm$1.0E+03 & 5.5E-05$\pm$1.1E-04 & 5.6E-02$\pm$5.0E-02 &  1.9$\pm$ 0.3 & 2.8E+06$\pm$1.7E+06 &  34.7$\pm$ 13.8 & 49.2$\pm$22.3 \\ 
J014003.44-743845.81 &  5.9$\pm$ 1.6 & 1.2E+03$\pm$1.7E+03 & 1.8E-05$\pm$1.5E-05 & 5.4E-02$\pm$5.9E-02 &  1.7$\pm$ 0.7 & 8.7E+05$\pm$6.9E+05 &   1.0$\pm$  0.4 & 42.9$\pm$19.2 \\ 
J014011.01-740303.42 &  5.1$\pm$ 1.7 & 5.2E+02$\pm$5.0E+02 & 3.8E-04$\pm$4.3E-04 & 1.9E-01$\pm$1.5E-01 &  1.4$\pm$ 0.5 & 9.8E+05$\pm$1.4E+06 &   1.9$\pm$  2.4 & 40.7$\pm$21.0 \\ 
J014109.05-743704.11 &  3.5$\pm$ 1.7 & 1.8E+02$\pm$4.0E+02 & 1.9E-04$\pm$2.0E-04 & 1.1E-01$\pm$9.1E-02 &  1.2$\pm$ 0.4 & 5.0E+05$\pm$1.1E+06 &  20.0$\pm$ 14.9 & 42.4$\pm$19.2 \\ 
J014110.65-743748.07 &  5.1$\pm$ 1.2 & 5.9E+02$\pm$9.8E+02 & 5.9E-05$\pm$1.1E-04 & 5.9E-02$\pm$5.4E-02 &  1.7$\pm$ 0.5 & 2.0E+06$\pm$2.0E+06 &  18.8$\pm$ 10.6 & 50.0$\pm$20.9 \\ 

\enddata
\tablecomments{This table is available in its entirety in a machine-readable form in the online journal. A portion is shown here for guidance 
 regarding its form and content.}
\end{deluxetable}
\end{turnpage}
\clearpage

\clearpage
\begin{turnpage}
\begin{deluxetable}{lrrrrrrrr}
\tablecolumns{9}
\tablecaption{Inferred Physical Parameters from SED Fits to HAeBe Candidates \label{aebefits}}
\tablewidth{0pc}
\tablehead{
 \colhead{} &
 \colhead{$\bar{M}_{\ast}$ } &
 \colhead{$\bar{L}_{\rm tot}$} &
 \colhead{$\bar{\dot{M}}_{\rm env} $} & 
 \colhead{$\bar{M}_{\rm disk}$} & 
 \colhead{Stage} &
 \colhead{$\bar{\tau_{\ast}}$} &
 \colhead{$\bar{A_V}$} &
 \colhead{$\bar{incl.}$} \\
 \colhead{Source Name} & 
 \colhead{($M_\odot$)} & 
 \colhead{($L_\odot$)} &
 \colhead{($M_\odot$/yr)} & 
 \colhead{($R_\odot$)} & 
 \colhead{Range} & 
 \colhead{(yr)} & 
 \colhead{(mag)} &
 \colhead{($\circ$)} 
}

\startdata

J013926.65-735802.92 &  5.5$\pm$ 0.4 & 2.6E+02$\pm$6.3E+01 & 1.9E-05$\pm$1.4E-05 & 1.7E-03$\pm$1.5E-03 &  1.2$\pm$ 0.4 & 3.5E+05$\pm$1.2E+05 &   0.0$\pm$  0.0 & 36.3$\pm$20.6 \\ 
J014039.65-743243.22 &  4.6$\pm$ 0.4 & 1.4E+02$\pm$7.6E+01 & 9.4E-06$\pm$1.3E-05 & 5.1E-02$\pm$5.1E-02 &  1.5$\pm$ 0.5 & 6.0E+05$\pm$1.6E+05 &   0.2$\pm$  0.3 & 49.9$\pm$19.7 \\ 
J014050.82-741030.86 & 14.6$\pm$ 0.0 & 1.9E+04$\pm$0.0E+00 & 0.0E-00$\pm$0.0E-00 & 3.2E-08$\pm$2.3E-08 &  3.0$\pm$ 0.0 & 1.3E+06$\pm$4.7E+00 &   0.0$\pm$  0.0 & 57.6$\pm$21.0 \\ 
J014142.53-735528.92 & 14.6$\pm$ 0.0 & 1.9E+04$\pm$0.0E+00 & 0.0E-00$\pm$0.0E-00 & 3.2E-08$\pm$2.3E-08 &  3.0$\pm$ 0.0 & 1.3E+06$\pm$0.0E+00 &   0.0$\pm$  0.0 & 57.8$\pm$21.0 \\ 
J014221.85-743013.85 &  5.0$\pm$ 0.2 & 3.0E+02$\pm$6.8E+01 & 8.0E-06$\pm$6.1E-06 & 6.6E-02$\pm$5.3E-02 &  1.3$\pm$ 0.5 & 5.4E+05$\pm$1.1E+05 &   0.0$\pm$  0.0 & 40.6$\pm$16.3 \\ 

\enddata
\tablecomments{This table is available in its entirety in a machine-readable form in the online journal. A portion is shown here for guidance 
 regarding its form and content.}
\end{deluxetable}
\end{turnpage}
\clearpage

\begin{deluxetable}{l|cc}
\tablecolumns{4}
\tablecaption{Star Formation Properties in the Bridge\label{sfr}}
\tablewidth{0pc}
\tablehead{
 \multicolumn{1}{c}{Property} &
 \colhead{Molecular Clouds} & \colhead{Whole Area} 
 }

\startdata

Size & $7\times2$\farcm6-radius circles & $180'\times80'$ \\
Area (kpc$^{2}$) & 0.038 & 3.4 \\
$M_{\rm CO}$  ($10^3 M_\odot$) &  19 & \nodata \\
$N_{\rm YSO}$($M_{\rm u1}-M_{\rm u2}$)\tablenotemark{a}  & 6(10.4-6.1)& 21 (10.4-6.0) \\
$M_{\rm YSO}^{\rm total}$($M_{\rm u}-M_{\rm l}$)\tablenotemark{b}   ($M_\odot$) & 290$^{+170}_{-70}$(10.4-1) & 970$^{+630}_{-315}$(10.4-1) \\
SFE$_{\rm YSO}$       &  0.015$^{+0.009}_{-0.004}$ & \nodata \\
SFR$_{\rm YSO}$ ($M_\odot$~yr$^{-1}$) & 2.9E-4& 9.7E-4\\
$\Sigma_{\rm SFR YSO}$ ($M_\odot$~yr$^{-1}$~kpc$^{-2}$) & 7.6$^{+4.5}_{-1.8}$E-3& 2.8$^{+1.9}_{-0.9}$E-4 \\
log($L$(24 \um)[ergs~s$^{-1}$]) &  37.5 & 39.0\\
SFR$_{24}$  ($M_\odot$~yr$^{-1}$) & 2.0E-5 & 4.2E-4\\
$\Sigma_{\rm SFR 24}$ ($M_\odot$~yr$^{-1}$~kpc$^{-2}$) & 5.2$^{+2.6}_{-2.6}$E-4 & 1.2$^{+0.6}_{-0.6}$E-4 \\
$\Sigma_{\rm HI}$ ($M_\odot$~pc$^{-2}$) & 17.9 & 10.2 \\
$\Sigma_{\rm H_2}$ ($M_\odot$~pc$^{-2}$) & 1.6 & \nodata \\
$\Sigma_{\rm SFR Gas}$ ($M_\odot$~yr$^{-1}$~kpc$^{-2}$) & 0.016 & 6.0E-3 \\

\enddata
\tablenotetext{a}{Number of YSOs with $\bar{M}_{\ast}$ in the mass range (u1-u2).}
\tablenotetext{b}{Total mass of YSOs extrapolated for the mass range (u-l).}
\end{deluxetable}

\end{document}